\documentclass[preprint,12pt]{elsarticle}

\PassOptionsToPackage{hyphens}{url}\usepackage{hyperref}
\usepackage{graphicx}
\usepackage{subcaption}
\usepackage{amssymb}
\usepackage{amsthm}
\usepackage{amsmath}
\usepackage{multirow}
\usepackage[utf8]{inputenc}

\usepackage{xcolor}
\usepackage{xargs}
\usepackage[colorinlistoftodos,textsize=footnotesize]{todonotes}
\usepackage{lineno}
\usepackage{cleveref}

\usepackage{empheq}
\usepackage{multicol}
\usepackage{booktabs}
\usepackage{microtype}

\usepackage{parskip}
\usepackage[section]{placeins}
\usepackage{algorithm}
\usepackage[noend]{algpseudocode}
\usepackage{listings}
\usepackage{grffile}
\usepackage[export]{adjustbox}

\newcommandx{\unsure}[2][1=]{\todo[linecolor=red,backgroundcolor=red!25,bordercolor=red,#1]{#2}}
\newcommandx{\change}[2][1=]{\todo[linecolor=blue,backgroundcolor=blue!25,bordercolor=blue,#1]{#2}}
\newcommandx{\info}[2][1=]{\todo[linecolor=OliveGreen,backgroundcolor=OliveGreen!25,bordercolor=OliveGreen,#1]{#2}}

\newcommand{\ten}[1]{\ensuremath{\mathbf{#1}}}

\newcommand{\code}[1]{\lstinline{#1}}

\journal{Computer Physics Communications}

\begin{document}

\begin{frontmatter}

% \title{Applications of adaptive EDAC-SPH for moving bodies in 2D}
  \title{Parallel adaptive weakly-compressible SPH for complex moving
    geometries}
  \author[IITB]{Asmelash Haftu \corref{cor1}}
  \ead{asmelash.a@aero.iitb.ac.in}
  \author[IITB]{Abhinav Muta}
  \ead{abhinavm@aero.iitb.ac.in}
  \author[IITB]{Prabhu Ramachandran}
  \ead{prabhu@aero.iitb.ac.in}
\address[IITB]{Department of Aerospace Engineering, Indian Institute of
  Technology Bombay, Powai, Mumbai 400076}

\cortext[cor1]{Corresponding author}

\begin{abstract}
  The use of adaptive spatial resolution to simulate flows of practical interest
  using Smoothed Particle Hydrodynamics (SPH) is of considerable
  importance. Recently, \citet{muta2021efficient} have proposed an efficient
  adaptive SPH method which is capable of handling large changes in particle
  resolution.  This allows the authors to simulate problems with much fewer
  particles than was possible earlier. The method was not demonstrated or tested
  with moving bodies or multiple bodies. In addition, the original method
  employed a large number of background particles to determine the spatial
  resolution of the fluid particles. In the present work we establish the
  formulation's effectiveness for simulating flow around stationary and moving
  geometries. We eliminate the need for the background particles in order to
  specify the geometry-based or solution-based adaptivity and we discuss the
  algorithms employed in detail. We consider a variety of benchmark problems,
  including the flow past two stationary cylinders, flow past different NACA
  airfoils at a range of Reynolds numbers, a moving square at various Reynolds
  numbers, and the flow past an oscillating cylinder. We also demonstrate
  different types of motions using single and multiple bodies. The source code
  is made available under an open source license, and our results are
  reproducible.
\end{abstract}

\begin{keyword}
%% keywords here, in the form: keyword \sep keyword
  {Adaptive spatial resolution}, {Incompressible flow}, {Moving geometries}, {Multi body
    simulation}, {Smoothed Particle Hydrodynamics}. %{Particle shifting},

  %% MSC codes here, in the form: \MSC code \sep code or \MSC[2008] code \sep
  %% code (2000 is the default)

\end{keyword}

\end{frontmatter}

%\linenumbers

{\bf Program Summary}
  %Delete as appropriate.

\newenvironment{sloppypar*}
 {\sloppy\ignorespaces}
 {\par}

\begin{small}
  \noindent
  % {\em Program Title:} ASPH Motion                                        \\
  {\em Program Title:} Parallel adaptive EDAC-SPH \\
  {\em CPC Library link to program files:} (to be added by Technical Editor) \\
  {\em Developer's repository link:} \url{https://gitlab.com/pypr/asph_motion}. \\
  {\em Code Ocean capsule:} (to be added by Technical Editor)\\
  {\em Licensing provisions:} BSD 3-clause\\
  {\em Programming language:} Python \\
  {\em External routines/libraries:} PySPH
  (\url{https://github.com/pypr/pysph}), matplotlib
  (\url{https://pypi.org/project/matplotlib/}), NumPy
  (\url{https://pypi.org/project/numpy/}), automan
  (\url{https://pypi.org/project/automan/}), compyle
  (\url{https://pypi.org/project/compyle/}). \\
  {\em Nature of problem:} Simulating fluid flow around multiple moving bodies
  requires the local resolution to be automatically adapted in order to capture
  all the necessary flow features. If a single resolution is used throughout
  the domain, the number of particles would be excessively large. \\
  {\em Solution method:} We validate and demonstrate the accuracy of the
  adaptive particle refinement algorithm in simulating flow past moving multiple
  geometries. Our algorithm is fully parallel, and we provide an open-source
  implementation. \\
\end{small}

\section{Introduction}%
\label{sec:intro}

The study of fluid flows in complex scenarios is immensely important in
science and engineering. Problems involving multiple bodies, thin structures,
and moving or deforming boundaries are a few examples of interest. For the
simulation of such systems, the mesh-based approach of computational fluid
dynamics (CFD) faces issues due to mesh distortion
\cite{bui2021smoothed, Ting19} in addition to the expertise and the
considerable amount of time required for good quality meshing. An alternative
to mesh-based approaches is to use meshless particle methods, particularly
Smoothed Particle Hydrodynamics (SPH). SPH is a Lagrangian, mesh-free method
that employs particles to discretize the fluid, the combined motion of which
is analogous to the flow of gases or liquids \cite{lind2020review,
  vacondio2020grand}. SPH is suitable for complex dynamics with multi-body and
multi-motion simulation, innately allowing simplified geometry description for
such problems. Most importantly, it allows dynamic adaptive particle
resolution \cite{violeau2016smoothed, shadloo2016smoothed}. It
has so far shown a significant headway that its desirability is apparent from
the increasingly large number of published works, simulation packages, and
range of applications. Since the initial application in astrophysics
\cite{lucy1977numerical, monaghan-gingold-stars-mnras-77}, SPH has been widely
studied and broadly applied to problems in the areas of fluid mechanics
\cite{liu2003smoothed, monaghan-review:2005, Liu2010SmoothedPH,
  violeau2012fluid, monaghan2012smoothed}, solid mechanics
\cite{Libersky1993, belytschko1996meshless}, and
fluid-structure interactions \cite{manenti2008fluid, khayyer2009enhanced}.
The range of applications studied also cover fields in hydrodynamics
\cite{violeau2007numerical, huang2019kernel}, multiphase flows
\cite{yang2019adaptive, wang2016overview}, free-surface
flows \cite{monaghan1994simulating, gotoh2018state}, turbulence \cite{violeau2007numerical,
  koukouvinis2009turbulence, monaghan2011turbulence}, atomization in
combustion \cite{braun2019numerical, chaussonnet2020progress}, food industries
\cite{wieth2016smoothed}, and many other areas. The grand challenges (GC) in SPH
have been discussed in detail by \citet{vacondio2020grand} and
\citet{lind2020review}. In the present work, we consider the robust
application of adaptive particle refinement/resolution (GC3), and SPH range of
applicability (GC5).

\citet{martel1994adaptive} introduced adaptive particle resolution in SPH for
cosmological simulation and treatment of shocks. \citet{kitsionas2002smoothed}
simulated a collapse of self-gravitating fluid by splitting particles to
increase the resolution in high density localities.
\citet{lastiwka2005adaptive} discussed an SPH framework for inserting and
removing particles adaptively, and the accuracy achieved by the method was
studied using a shock tube test case. \citet{reyes2011dynamic} investigated a
dynamic refinement scheme applicable to free surface as well as non-cohesive
soil problems. In a separate study, \citet{lopez2013dynamic} presented a
dynamic particle refinement method for SPH in which they refine the particles
by replacing them with smaller daughter particles. In a paper by
\citet{spreng2014local}, a local adaptive discretization algorithm was
presented for improving the accuracy of SPH simulations while reducing both
time and cost of computation. This involves a combination of refinement and
coarsening, which increases particles adaptively for areas of high interest in
the domain and reduces in regions with lower interest. A particle refinement
and coarsening method is discussed by \citet{barcarolo2014adaptive} in which
one parent particle was split into four child particles of equal mass within
the refinement zone and recombined after leaving the refinement zone. The
total mass was not found conserved at all times during the operation.
\citet{Khorasanizade2016DynamicFP} illustrated the particle splitting
technique using a dynamic approach. Adaptive mesh refinement of finite volume
based CFD methods motivated \citet{garcia2014equalizing} to create adaptive
refinement of the particle in SPH by using guard particles to separate the
zones with different levels of refinement. More studies related to adaptive
particle refinement can be found in \cite{omidvar2012wave, hu2017consistent,
  chiron2018analysis}.

Feldman and Bonet \cite{feldman2007dynamic} illustrated a dynamic particle
refinement method and studied the error generated as a result of particle
splitting. Candidate particles are identified and the original particle is
split into seven smaller particles for which mass distribution is optimized
through nonlinear minimization. The daughter particles are moved with the
velocity of the original particle to conserve flow properties. Vacondio et al.
\cite{vacondio2013variable} presented a variable resolution approach for
splitting and merging SPH particles and simulating the Navier-Stokes (NS)
equations. The particle sizes are dynamically modified by splitting and
coalescing procedures of individual particles to achieve the required
resolution in targeted zones. The adaptive algorithm for particles of variable
smoothing lengths is implemented in an SPH scheme derived using a variational
method (\cite{bonet1999variational} and \cite{kulasegaram2004variational}) to
ensure the conservation of mass and momentum. In this method, one parent
particle was split into seven child particles during the refinement phase,
whereas while coarsening, two-child particles were merged into one parent
particle. The anisotropy and non-uniformity of particle distributions are
controlled and generalized using particle shifting variable mass particles.
The splitting accuracy was demonstrated for two-dimensional cases. Yang and
Kong \cite{yang2019adaptive} developed an adaptive particle refinement
technique for multiphase flows; defined criteria for the spacing of a
reference particle which changes dynamically along with the location of the
interface. The adaptive resolution was used with a smoothing length which also
varied during the simulation. The smoothing length is considered variable
because the spacing between particles is not uniform and it is averaged for
the neighboring particles to reduce numerical errors. Refining and coarsening
of fluid particles are carried out via splitting and merging techniques
respectively. Near the interface, the particles are refined whereas, far away
from the interface the particles are coarsened.

Many issues with adaptive SPH have been addressed in the recent work of
\citet{muta2021efficient}. Specifically, the method is accurate, efficient,
supports multiple bodies, supports moving geometries, is automatically adaptive,
and supports both geometry-based as well as solution-based adaptivity. In the
original paper, the authors did not demonstrate any moving geometry problems or
consider multiple bodies. The original method also employs background particles
in order to set the particle resolution of the fluid particles. The number of
particles used for the background is typically the same order as the fluid
particles. This increases the complexity of the method and also increases its
memory usage.

In this work, we demonstrate the applications of the adaptive SPH method of
\citet{muta2021efficient} for simulation of complex cases involving stationary
and moving geometries. We also eliminate the need for the background particles
and this reduces the complexity of the method, simplifies its implementation,
and reduces the memory usage of the method. The method is proposed for the
simulations of incompressible and weakly compressible fluid flows employing
the Entropically Damped Artificial Compressibility (EDAC) formulation
\cite{ramachandran2019entropically}. The adaptive
resolution is fully automatic and produces an optimal mass distribution of
particles and is accurate for the benchmark problems considered. It is based
on methods developed by \cite{yang2019adaptive, feldman2007dynamic,
  vacondio2013variable, vacondio2012accurate,vacondio2016variable}, with
additional improvements. The refinement method elegantly allows static regions
of refinement, geometry-based automatic refinement, and solution-based
automatic adaptivity. The SPH method has not been extensively applied to
problems that involve the motion of solids in a fluid medium at high
resolution. Fortunately, moving or stationary solid bodies can be easily
simulated at a range of Reynolds numbers using the aforementioned approach.

The main goals of this paper are to, (i) validate the adaptive EDAC-SPH method
without the use of background particle by testing on incompressible flows at a
range of Reynolds numbers, (ii) apply the method to moving geometries, (iii)
apply it to complex geometries. We first simulate flow past two stationary
cylinders at different gap spacing between the cylinders. We then perform flow
simulation over stationary airfoils of narrow cross-section at a range of
Reynolds numbers to illustrate the robustness of the method. Next, we show the
flow around a moving square and cylinder for Reynolds numbers $50 - 600$ and
compare the results with established methods. We also apply the present model
and simulate an oscillating single circular cylinder for different amplitudes
and frequencies of oscillation. At the end we demonstrate different complex
motions and multi-body simulations.

We provide an open-source implementation based on the PySPH framework
\cite{ramachandran2019pysph}. The source code can be
obtained from \url{https://gitlab.com/pypr/asph_motion}. The manuscript is
reproducible and each figure is automatically generated through the use of an
automation framework~\cite{ramachandran2017automan}.

\section{Smoothed particle hydrodynamics formulation}%
\label{sec:sph}

The governing equations for the viscous, weakly-compressible fluid flows using
the entropically damped artificial compressibility (EDAC)
formulation~\cite{Clausen2013,ramachandran2019entropically}, written in
transport velocity formulation~\cite{Adami2013}, including the corrections
terms of \cite{adepu2021} involves two equations, one for the pressure
evolution which is given as,
\begin{equation}
  \label{eq:edac-corr}
  \frac{\tilde{\mathrm{d}}p}{\mathrm{d}t} =
  -\rho c_s^2 \text{div}(\ten{u})
  + \nu_{\text{e}} \nabla^2 p
  + (\tilde{\ten{u}}
  - \ten{u}) \cdot \nabla p,
\end{equation}
where $p$ is the pressure, $t$ is the time, $\rho$ is the density, $c_s$ is
the artificial speed of sound, $\tilde{\ten{u}}$ is the transport velocity.
$\frac{\tilde{\mathrm{d}} (.)}{\mathrm{d} t} = \frac{\partial (.)}{\partial t}
+ \tilde{\ten{u}} \cdot \text{grad} (.)$ is the material time derivative of a
particle advecting with the transport velocity $\tilde{\ten{u}}$, $\ten{u}$ is
the fluid velocity, and $\nu_{\text{e}}$ is the EDAC viscosity parameter. The
second equation is the momentum equation, which is given as,
\begin{equation}
  \label{eq:mom-corr}
  \frac{\tilde{\text{d}} \ten{u}}{\text{d}t} =
  -\frac{1}{\rho} \nabla p
  + \nu \nabla^2 \ten{u} + \ten{f}
  + \frac{1}{\rho} \nabla \cdot \rho (\ten{u} \otimes (\tilde{\ten{u}} - \ten{u}))
  + \ten{u}\,\mathrm{div}(\tilde{\ten{u}}),
\end{equation}
where $\nu$ is the kinematic viscosity of the fluid, and $\ten{f}$ is the
external body force. The positions $\ten{r}$ of the Lagrangian fluid particles
are updated using,
\begin{equation}
  \label{eq:pos-ode:tvf}
  \frac{\mathrm{d} \ten{r}}{\mathrm{d} t} = \tilde{\ten{u}}.
\end{equation}

We discretize the above governing equations using the variable-$h$ SPH
formulation \cite{monaghan-review:2005,vacondio2012accurate}. In the following
the subscripts $i$ and $j$ denote the index of the particles in the
discretized domain and the summation is taken over all the indices in the
neighborhood of $i^{\text{th}}$ particle. For the computation of the density
$\rho_i$ of a particle we use the gather form of summation density
\cite{hernquist1989,vacondio2012accurate},
\begin{equation}
  \label{eq:sd-gather}
  \rho_i = \rho(\ten{r}_i) = \sum_j m_j W(r_{ij}, h_i),
\end{equation}
where $W(r_{ij}, h_i) = W(|\ten{r}_i - \ten{r}_j|, h_i)$ is the kernel function,
and $h_i$ the smoothing length. We use the quintic spline
kernel~\cite{Liu2010SmoothedPH} in all our simulations, which is given by,
\begin{equation}
  \label{eq:kernel}
  W(q) =
  \begin{cases}
    \sigma_2 [{(3 - q)}^5 - 6{(2 - q)}^5 + 15{(1 - q)}^5] \quad%
    &\text{if}\ 0 \le{} q < 1, \\
     \sigma_2 [{(3 - q)}^5 - 6{(2 - q)}^5] \quad
    &\text{if}\ 1 \le{} q < 2, \\
     \sigma_2 {(3 - q)}^5 \quad &\text{if}\ 2 \le{} q < 3, \\
     0 \quad &\text{if}\ q \ge{} 3, \\
  \end{cases}
\end{equation}
where $\sigma_2 = 7/(478 \pi {h(\ten{r})}^2)$, and $q = |\ten{r}|/h$. The EDAC
pressure evolution equation is given by,
\begin{equation}
  \label{eq:edac:sph}
\begin{split}
  \frac{\tilde{\mathrm{d}}p}{\mathrm{d} t}(\ten{r}_i) =
  \quad&
     \frac{\rho_0 c_s^2}{\beta_i}\sum_j
     \frac{m_j}{\rho_j} \ten{u}_{ij} \cdot \nabla W(r_{ij}, h_i) \\
  +&
     \frac{1}{\beta_i}\sum_j \frac{m_j}{\rho_j} \nu_{\text{e}, ij}  (p_i - p_j)
     (
       \ten{r}_{ij} \cdot \nabla W(r_{ij}, h_{ij})
     ) \\
  +&
     \sum_j m_j
     [(\tilde{\ten{u}}_{i} - \ten{u}_{i})
     \cdot (P_i \nabla W(r_{ij}, h_i) + P_j \nabla W(r_{ij}, h_j))],
   \end{split}
 \end{equation}
where $\rho_0$ is the reference density,
$\ten{u}_{ij} = \ten{u}_i - \ten{u}_j$, $\beta_i$ is the additional term
obtained while deriving the variable-$h$ formulation
\cite{vacondio2012accurate}, which in $d$ dimensions is given by,
\begin{equation}
  \label{eq:beta}
  \beta_i = - \frac{1}{\rho_i d} \sum_j m_j r_{ij}
  \frac{\mathrm{d} W(r_{ij}, h_i)}{\mathrm{d}r_{ij}},
\end{equation}
$P_i$ and $P_j$, employing the pressure reduction technique
\cite{sph:basa-etal-2009} by removing the averaged pressure
$p_{\text{avg}, i} = \sum_j p_j / N_i$, are given by,
\begin{equation}
  \label{eq:mom:pre}
  P_i = \frac{(p_i - p_{\text{avg}, i})}{\rho_i^2 \beta_i}, \quad
  P_j = \frac{(p_j - p_{\text{avg}, j})}{\rho_j^2 \beta_j},
\end{equation}
and the averaged gradient of the kernel is given by,
\begin{equation}
  \label{eq:dw-avg}
  \nabla W(r_{ij}, h_{ij}) =
  \left(\frac{\nabla W(r_{ij}, h_i) + \nabla W(r_{ij}, h_j)}{2}\right).
\end{equation}
For pressure diffusion term in the EDAC equation we use the formulation
of~\citet{cleary1999} to discretize the coefficient $\nu_{\text{e}, ij}$ as,
 \begin{equation}
   \label{eq:nu-cleary}
   \nu_{\text{e}, ij} = 4\frac{\nu_{\text{e}, i} \nu_{\text{e}, j}}
   {(\nu_{\text{e}, i} + \nu_{\text{e}, j})},
 \end{equation}
 where the EDAC viscosity $\nu_{\text{e}, i}$, which is varying in space due to
 the presence of $h_i$, is discretized as \cite{ramachandran2019entropically},
\begin{equation}
  \label{eq:edac-alpha}
  \nu_{\text{e}, i} = \frac{\alpha_{\text{e}} c_s h_i}{8},
\end{equation}
where $\alpha_{\text{e}} = 1.5$ is used in all our simulations.

The momentum equation in the variable-$h$ SPH discretization is given by,
\begin{equation}
  \label{eq:mom-sph}
  \begin{split}
  \frac{\tilde{\mathrm{d}}\ten{u}}{\mathrm{d} t}(\ten{r}_i, t) =
  -&\sum_j m_j
    \left((P_i + A_i) \nabla W(r_{ij}, h_i)
    + (P_j + A_j) \nabla W(r_{ij}, h_j)\right) \\
    +& \frac{1}{\beta_i}\sum_j m_j \frac{4 \nu}{(\rho_i + \rho_j)}
    \frac{\ten{r}_{ij} \cdot \nabla W(r_{ij}, h_{ij})}
    {(|\ten{r}_{ij}|^{2} + \eta)} \ten{u}_{ij} \\
    -&
    \frac{1}{\beta_i}\sum_{j} \frac{m_j}{\rho_j}
  [(\tilde{\ten{u}}_{ij} - \ten{u}_{ij}) \cdot \nabla W(r_{ij}, h_i)] \ten{u}_i,
  \end{split}
\end{equation}
where,
\begin{equation}
  \label{eq:astress}
{A}_{i} = \frac{1}{\rho_{i} \beta_i} \ten{u}_{i} \otimes(\tilde{\ten{u}}_{i} -
\ten{u}_{i}),
\quad
{A}_{j} = \frac{1}{\rho_{j} \beta_j} \ten{u}_{j} \otimes(\tilde{\ten{u}}_{j} - \ten{u}_{j}),
\end{equation}
and $\eta = 0.001 h_i^2$ is a small number added to ensure a non-zero
denominator in case when $i = j$. All our simulations are viscous and we do not
employ any artificial viscosity.

We employ the shifting technique of~\citet{diff_smoothing_sph:lind:jcp:2009},
additionally limiting the particle movement if it shifts by more than 25\% of
its smoothing length, to ensure uniformity, leading to stable and accurate
simulations
\cite{acc_stab_xu:jcp:2009,diff_smoothing_sph:lind:jcp:2009,oger_ale_sph_2016}.
Rather than shifting the particles directly, since we are using the transport
velocity formulation, we incorporate the shifting into the computation of the
transport velocity as,
\begin{equation}
  \label{eq:shift-tvf}
  \tilde{\ten{u}}_i = \ten{u}_i + \theta \frac{\delta \ten{r}_i}{\Delta t},
\end{equation}
where,
\begin{equation}%
  \label{eq:shift-deltar}
  \delta \ten{r}_{i} = - \frac{h^2_i}{2} \sum_{j}
  \frac{m_j}{\rho_0} \left (
    1 + 0.24 {\left(
        \frac{W(r_{ij}, h_{ij})}{W(\Delta x, \xi h_{ij})}
      \right)}^{4}
  \right)
  \nabla W_{ij},
\end{equation}
where $\xi = 0.759298480738450$ is the point of inflection of the quintic
spline kernel~\cite{crespo2008}, and $h_{ij} = (h_i + h_j)/2$; and,
\begin{equation}
  \label{eq:shift-limit}
  \theta =
  \begin{cases}
     \frac{0.25 h_i}{|\delta \ten{r}_{i}|}\quad &\text{if}\ |\delta \ten{r}_{i}|
     > 0.25 h_i, \\
     1 \quad &\text{otherwise}.
  \end{cases}
\end{equation}

\subsection{Boundary conditions}%
\label{sec:bc}

We use the dummy particle approach of~\citet{Adami2012} to enforce the boundary
conditions. We enforce no-penetration by using the wall velocity for the dummy
particles in the EDAC equation. We enforce the no-slip and free-slip boundary
condition by, first, extrapolating the fluid velocity onto the dummy particles
using Shepard interpolation:
\begin{equation}
  \label{eq:shepard}
  \hat{\ten{u}}_i = \frac{\sum_j \ten{u}_j W(r_{ij}, h_{ij})}
  {\sum_j W(r_{ij}, h_{ij})},
\end{equation}
where $j$ is the index of the fluid particles in the neighborhood of the
$i^{\text{th}}$ dummy particle; then, the velocity of the dummy particles is
computed from,
\begin{equation}
  \label{eq:bc-wall}
  \ten{u}_w = 2\ten{u}_i - \hat{\ten{u}}_i,
\end{equation}
where the subscript $w$ denotes the dummy wall particles, $\ten{u}_i$ is the
prescribed wall velocity. To impose the pressure gradient accurately near the
wall, the pressure on the dummy wall particles is computed by,
\begin{equation}
  \label{eq:bc-pre}
  p_w = \frac{\sum_f p_f W(r_{wf}, h_{wf}) +
    (\ten{g} - \ten{a}_w)\cdot
    \sum_f \rho_f \ten{r}_{wf} W(r_{wf}, h_{wf})}
  {\sum_f W(r_{wf}, h_{wf})},
\end{equation}
where the subscript $f$ denotes the fluid particles, $\ten{a}_w$ is the
acceleration of the wall, $g$ is the acceleration due to gravity (this is zero
in the present work), $r_{wf} = |\ten{r}_w - \ten{r}_f|$, and $h_{wf} = (h_w +
h_f)/2$.

We use the no-reflection boundary conditions on the inviscid wall of all
simulations. For this we use the approach of~\citet{lastiwka2009}, where we
compute the characteristic properties, $J_1, J_2$, and $J_3$, op. cit., from
the fluid and extrapolate these properties onto the dummy particles describing
the boundary using Shepard interpolation, finally, the fluid properties on the
dummy particles are set from these characteristic properties.

\subsection{Force computation}%
\label{sec:force-comp}

The forces due to the pressure and viscosity of the fluid on the solid are
computed by evaluating,
\begin{equation}
  \label{eq:force}
  \ten{f}^{\text{solid}} = m^{\text{solid}}\left(-\frac{1}{\rho}\nabla p
  + \nu \nabla \cdot \nabla \ten{u}\right),
\end{equation}
which in the variable-$h$ SPH discretization is written as,
\begin{equation}
  \label{eq:force-sph}
  \begin{split}
  \ten{f}^{\text{solid}}_{i} =
  -& \underbrace{m^{\text{solid}}_{i}\left[\sum_j m_j
  \left(P_i \nabla W(r_{ij}, h_i)
    + P_j \nabla W(r_{ij}, h_j)\right)\right]}_{\ten{f}_{i, \text{p}}}\\
+& \underbrace{m^{\text{solid}}_{i} \left[ \frac{1}{\beta_i}\sum_j m_j
  \frac{4 \nu}{(\rho_i + \rho_j)}
  \frac{\ten{r}_{ij} \cdot \nabla W(r_{ij}, h_{ij})}
  {(|\ten{r}_{ij}|^{2} + \eta)} \ten{u}_{ij}\right]}_{\ten{f}_{i, \text{visc}}}
  \end{split}
\end{equation}
where the summation index $j$ is over all the fluid particles in the
neighborhood of a $i^{\text{th}}$ dummy particle. The drag $C_d$, lift $C_l$,
and pressure $C_p$ coefficients used for comparison of our results with
reference data are expressed by:
\begin{equation}
  C_{d} =
  \frac{(\ten{f}_{\text{p}} + \ten{f}_{\text{visc}}) \cdot \ten{e}_x}
  {0.5 \rho_0 U^{2}_{\infty}D},
  \quad
  C_{l} =
  \frac{(\ten{f}_{\text{p}} + \ten{f}_{\text{visc}}) \cdot \ten{e}_y}
  {0.5 \rho_0 U^{2}_{\infty}D},
  \quad
  C_{p} = \frac{p}{0.5 \rho_0 U^{2}_{\infty}D},
\label{eq:cdcl}
\end{equation}
where $U_\infty$ is free-stream velocity of the fluid, $\ten{e}_x$ and
$\ten{e}_y$ are the unit vectors in the $x$ and $y$ directions, respectively,
$\ten{f}_{\text{p}} + \ten{f}_{\text{visc}} = \sum_j \ten{f}_{j, \text{p}} +
\ten{f}_{j, \text{visc}}$ is the sum of all forces of the dummy particles, and
$D$ is the characteristic length.

\subsection{Time integration}%
\label{sec:integration}

We use the Predict-Evaluate-Correct (PEC) integrator to integrate the position,
velocity, and pressure. First, we predict the properties at an intermediate time
$n + \frac{1}{2}$ by,
\begin{align}
  \label{eq:int-pre}
  \ten{u}^{n + \frac{1}{2}}_i &= \ten{u}^{n}_i + \frac{\Delta t}{2}
  \frac{\tilde{\mathrm{d}} \ten{u}_i^{n}}{\mathrm{d} t}, \\
  \tilde{\ten{u}}^{n + \frac{1}{2}}_i &= \ten{u}^{n+\frac{1}{2}}_i
                                        + \frac{\delta \ten{r}^n_i}{\Delta t}, \\
  \ten{r}^{n + \frac{1}{2}}_i &= \ten{r}^{n}_i + \frac{\Delta t}{2}
  \tilde{\ten{u}}_i^{n+\frac{1}{2}}, \\
  p^{n + \frac{1}{2}}_i &= p^{n}_i + \frac{\Delta t}{2}
  \frac{\tilde{\mathrm{d}}\ten{p}_i^{n}}{\mathrm{d} t},
\end{align}
next, we evaluate the rate of change of properties at $n + \frac{1}{2}$, then,
we correct the properties to get the corresponding values at the new time
$n + 1$ by,
\begin{align}
  \label{eq:int-cor}
  \ten{u}^{n + 1}_i &= \ten{u}^{n}_i + \Delta t
  \frac{\tilde{\mathrm{d}} \ten{u}_i^{n + \frac{1}{2}}}{\mathrm{d} t}, \\
  \tilde{\ten{u}}^{n + 1}_i &= \ten{u}^{n+1}_i
                              + \frac{\delta \ten{r}^{n+\frac{1}{2}}_i}{\Delta t}, \\
  \ten{r}^{n + 1}_i &= \ten{r}^{n}_i + \Delta t
  \tilde{\ten{u}}_i^{n+1}, \\
  p^{n + 1}_i &= p^{n}_i + \Delta t
                \frac{\tilde{\mathrm{d}}\ten{p}_i^{n+\frac{1}{2}}}{\mathrm{d} t}.
\end{align}
The time-step is determined by the highest resolution used in the domain, and
the minimum of the Courant–Friedrichs–Lewy (CFL) criterion and the viscous condition is taken:
\begin{align}
  \label{eq:timestep}
  \Delta t = \min\left(0.25 \left(\frac{h_{\min}}{U + c_s}\right),
  0.125 \left(\frac{h^2_{\min}}{\nu}\right)\right).
\end{align}

\section{Adaptive refinement}%
\label{sec:adaptive-res}

We discuss the adaptive refinement algorithm here. The present algorithm is
based on the work of \citet{muta2021efficient} in which the authors employ a
collection of background particles in order to determine the spatial
resolution of the fluid particles. The number of background particles is
typically the same size as the number of fluid particles and these do not
contribute significantly to the computation but do increase the memory
required by the algorithm and also increase the complexity of the
implementation. In the present work, we eliminate the background particles
altogether. The algorithms discussed here are thus slightly different from the
ones in \cite{muta2021efficient}.

Initially, we discretize all the solid bodies at the highest desired
resolution, and set a parameter $\Delta s_{\min}$ to this resolution. For
complex bodies a particle packing algorithms \cite{negi2019improved} can be
used to discretize at the desired resolution. We define two other parameters,
$\Delta s_{\max}$ the coarsest resolution in the domain, and $C_r =
\frac{\Delta s_{k+1}}{\Delta s_{k}}$, the particle spacing ratio
\cite{yang2019adaptive} between adjacent regions of refinement. The spacing
ratio $C_r$ controls the width of the refinement regions and is generally between 
1.05 and 1.2. %layers=regions?

We base our discussion on problems where solid bodies, stationary or moving,
are placed in a wind-tunnel-like setup with fluid streaming between the inlet
and outlet. Once the domain size is specified, we iterate over all the fluid
particles and mark the fluid particles which are in the neighborhood of the
particles defining the solid bodies; we use a simple integer mask to achieve
this. After fluid particles near the boundaries are identified, we iterate over
the rest of the particles and find the minimum $(\Delta s_{\min})$, maximum
$(\Delta s_{\max})$, and geometric mean $(\Delta s_{\text{avg}})$ of the
$\Delta s$ in the neighborhood of each particle. If
$\Delta s_{\max} / \Delta s_{\min} < C_r^3$ we set the value of $\Delta s$ to
$C_r \Delta s_{\min}$. If $\Delta s_{\max} / \Delta s_{\min} \ge C_r^3$, we set
$\Delta s = \Delta s_{\text{avg}}$. This immediately defines the reference
$m_{\text{ref}} = \rho (\Delta s)^d$, minimum $m_{\min} = 0.5 m_{\text{ref}}$
and maximum $m_{\max} = 1.05 m_{\text{ref}}$ mass. \Cref{alg:update-h} shows the
outline of this procedure.
\begin{algorithm}[!htp]
\caption{Update spacing of particles}\label{alg:update-h}
\begin{algorithmic}[1]
  \State{Define $\Delta s$ for all particles}
 \For{all $i$ of particles which are not fixed}
  \State{find neighbors of $i$}
  \State{find $\Delta s_{\min}$, and $\Delta s_{\max}$ in the neighborhood}
  \State{find average spacing $\Delta s_{\text{avg}}$ in the neighborhood}
  \If{${\frac{\Delta s_{\max}}{\Delta s_{\min}}} < {(C_r)}^3$}
  \State{$h_{\text{tmp}} \leftarrow \min(\Delta s_{\max}, C_r \Delta s_{\min})$}
  \Else{}
  \State{$h_{\text{tmp}} \leftarrow \Delta s_{\text{avg}}$}
  \EndIf{}
  \EndFor{}
  \For{all $i$ of particles which are not fixed}
  \State{$m_i \leftarrow \rho_i h^{2}_{\text{tmp}}$}
  \State{$m_{i, \max} \leftarrow 1.05 m_i \quad m_{i, \min} \leftarrow 0.5 m_i$}
  \EndFor{}
\end{algorithmic}
\end{algorithm}
\begin{figure}[htp]
  \centering
  \begin{subfigure}{0.45\textwidth}
    \includegraphics[width=\textwidth]{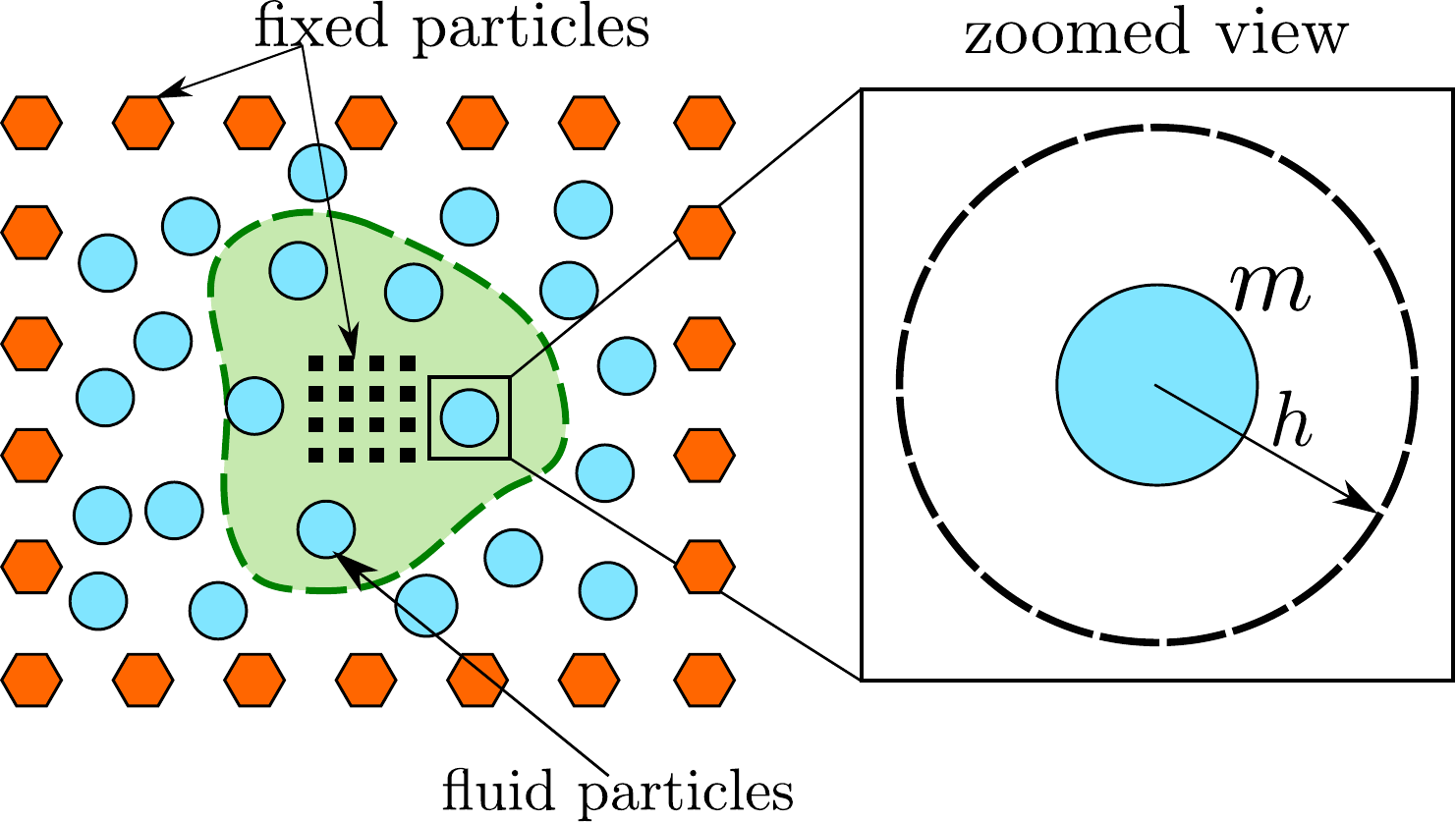}
      \subcaption{}%
      \label{fig:alg-2a}
  \end{subfigure}
  \begin{subfigure}{0.45\textwidth}
    \includegraphics[width=\textwidth]{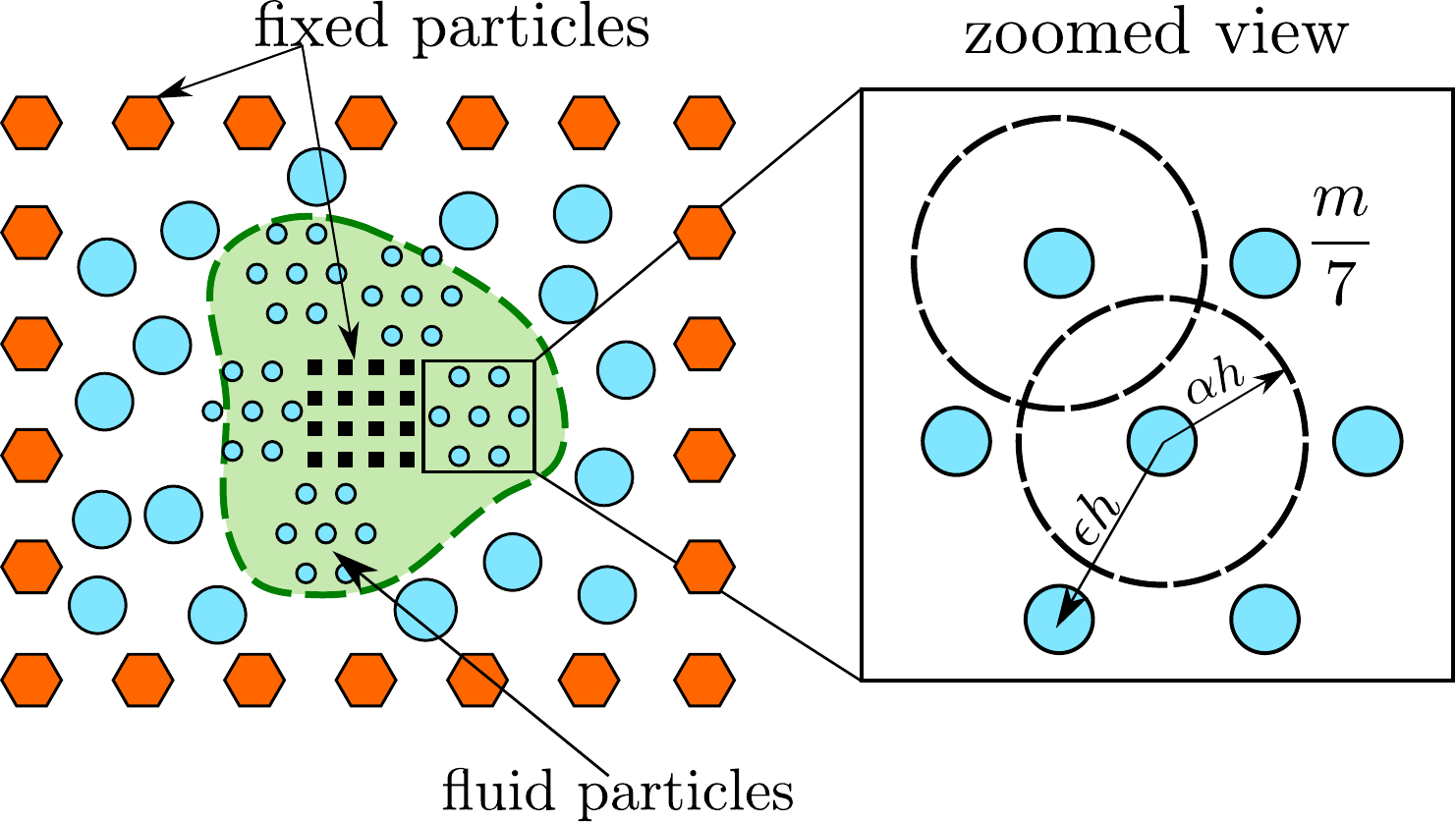}
      \subcaption{}%
      \label{fig:alg-2b}
  \end{subfigure}
  \caption{Illustration of the splitting algorithm (a) before the split, and (b)
    after the split. The particles outside the green zone have mass $m$ and the
    maximum mass is $m_{\max} = m$. All the particles inside the green region
    have maximum mass $m_{\max} = m/2$. After the split the particles undergo
    merging which is not illustrated.}%
  \label{fig:alg2}
\end{figure}

\Cref{alg:splitting} describes the splitting procedure which follows
\citet{feldman2007dynamic} and \citet{vacondio2012accurate}. \Cref{fig:alg2}
shows an illustration of the splitting algorithm. The orange hexagonal particles
indicate the outer boundary or the edge of the fixed internal domain, the black
square particles correspond to solid particles. Both the orange hexagonal and
the black square particles constitute the fixed particles and will not
subdivide. The blue particles correspond to the fluid particles and are
non-fixed as they may subdivide. The green zone indicates the region having
maximum mass $m_{\max} = m/2$. If a particle exceeds its maximum mass, we split
the particle into 7 \emph{daughter} particles, where we place 6 daughter
particles on the vertices of a hexagon, at a distance of $\epsilon = 0.4$ times
the smoothing length away from the original particle, centered around the
location of the original particle. We place the $7^{\text{th}}$ daughter
particle at the location of the original particle. The daughter particles'
smoothing length is set to $\alpha = 0.9$ times that of the original particle
smoothing length.
\begin{algorithm}[t]
  \caption{Splitting of the particles.}%
  \label{alg:splitting}
\begin{algorithmic}[1]
  \State{$\epsilon \leftarrow 0.4 \quad \alpha \leftarrow 0.9$}
  \For{all $i$ of particles which are not fixed}
  \If{$m_i > m_{i, \max}$}
  \State{split particle $i$ to 7 daughters}
  \For{$k = 0;\ k < 7;\ k\texttt{++}$}
  \State{$m_{i, k} \leftarrow \frac{m_i}{7}, \quad
  h_{i, k} \leftarrow \alpha h_i$}
  \State{$\ten{u}_{i, k} \leftarrow \ten{u}_i, \quad
    \tilde{\ten{u}}_{i, k} \leftarrow \tilde{\ten{u}}_i$}
  \State{$p_{i, k} \leftarrow p_i, \quad
    \rho_{i, k} \leftarrow \rho_i$}
  \EndFor{}
  \State{$\ten{x}_{i, 0} \leftarrow \ten{x}_i$}
  \For{$k = 1;\ k < 7;\ k\texttt{++}$}
  \State{$\ten{x}_{i, k} \leftarrow \ten{x}_i +
    \epsilon h_i \cos\left(\frac{k\pi}{3}\right)$}
  \EndFor{}
  \EndIf{}
  \EndFor{}
\end{algorithmic}
\end{algorithm}

The merging algorithm follows~\citet{vacondio2013variable} although with some
differences that allow us to perform this in parallel. In a single iteration,
a particle may merge only with one other particle. Two particles merge if the
distance between the two is less than the mean smoothing length of the two
$r_{ij} < (h_i + h_j)/2$, and the sum of mass of the two particles is less
than the maximum of the $m_{\max}$ of either of the particles $(m_i + m_j) <
\max{(m_{\max}[i], m_{\max}[j])}$, and $j$ is closest to particle $i$. This is
fully parallel, since the particle $i$ will only merge with particle $j$ if
particle $j$ is to merge with particle $i$.  We place the merged particle at,
\begin{equation}
  \label{eq:merged-pos}
  \ten{r}_m = \frac{m_i \ten{r}_i + m_j \ten{r}_j}{m_m},
\end{equation}
where $m_m = m_i + m_j$ is the mass of the merged particle. Set the velocity
using the mass-weighted mean as,
\begin{equation}
  \label{eq:merged-prop}
  \ten{u}_m = \frac{m_i \ten{u}_i + m_j \ten{u}_j}{m_m},
\end{equation}
similarly update other scalar properties, like, pressure, and transport
velocity. Set the smoothing radius using~\cite{vacondio2013variable},
\begin{equation}
  \label{eq:merged-h}
  h_{\text{merge}} = {\left( \frac{m_m W(\ten{0}, 1)}
      {m_i W(\ten{r}_m - \ten{r}_i, h_i) +
        m_j W(\ten{r}_m - \ten{r}_j, h_j)} \right)}^{1/d},
\end{equation}
where $W(\ten{x}, h)$ is the kernel function and $d$ is the number of spatial
dimensions. The pseudo-code for the algorithm is given in \cref{alg:merging}.
Note that when merging the particles, the particle with the smaller index is
the one that is retained.
\begin{algorithm}[!htp]
  \caption{Merging of particles.}%
  \label{alg:merging}
\begin{algorithmic}[1]
  \For{all $i$ which are not fixed}
  \State{find neighbors of $i$}
  \If{$m_i \le m_{i, \max}$}
  \For{$j$ in neighbor indices}
  \State{$m_{\text{merge}} \leftarrow m_i + m_j$}
  \State{$m_{\max, \min} \leftarrow \min(m_{i, \max}, m_{j, \max})$}
  \If{$m_{\text{merge}} < m_{\max, \min}$ \textbf{\&} $j$ is closest
    of all neighbors}
  \State{store index $j$ for merging}
  \EndIf{}
  \EndFor{}
  \EndIf{}
  \EndFor{}
  \For{all $i$ of particles which are to be merged}
  \If{merge pair of $i$ is $j$ and merge pair of $j$ is $i$ and $i < j$}
  \State{update $\ten{r}_i$ with its merged pair index using \cref{eq:merged-pos}}
  \State{update $\ten{u}_i$ with its merged pair index using \cref{eq:merged-prop}}
  \State{update $h_i$ with its merged pair index using \cref{eq:merged-h}}
  \Else
  \If{merge pairs match and $i > j$}
  \State{Delete particle $i$}
  \EndIf{}
  \EndIf{}
  \EndFor{}
\end{algorithmic}
\end{algorithm}

After finishing the splitting, we run the merging algorithm iteratively for three
times. After this, the smoothing length of all the particles is reset to the
value corresponding to the average mass of the neighbors,
\begin{equation}
  \label{eq:optimal-h}
  h_i = C {\left(\frac{1}{\rho_0} \frac{\sum_j m_j}{N_i}\right)}^{1/d},
\end{equation}
where $C$ is a constant corresponding to the value of smoothing length factor
used in the simulation. In our simulations we set this value to 1.2 for the
quintic spline kernel.

We employ shifting after our adaptive refinement procedure. The new position
$\ten{r}'_i$ of the particle is given by,
\begin{equation}%
  \label{eq:shift}
  \ten{r}'_{i} = \ten{r}_{i} + \theta \delta \ten{r}_{i},
\end{equation}
we correct the fluid properties by using a Taylor series approximation. Consider
a fluid property $\varphi_i$ the corrected value $\varphi'_i$ after shifting is
obtained by,
\begin{equation}%
  \label{eq:shift-correct}
  \varphi'_{i} = \varphi_i + {(\nabla \varphi)}_i\cdot \delta \ten{r}_{i}.
\end{equation}

\Cref{alg:apr} summarizes the adaptive particle refinement method.  First, we
set the $\Delta s$ of all the fluid particles in the neighborhood of a boundary
to the resolution of the boundary. Next, we set $\Delta s$ of all the particles
which satisfy specific criteria to the highest resolution. This is useful to
define solution adaptivity, where a particular flow feature is monitored, and
particles exceeding specific tolerance criteria are refined to the highest
resolution. After that, we run the update spacing algorithm \cref{alg:update-h}
which updates the $\Delta s$ of all the fluid particles. Then follows the
splitting and merging procedure as described above.
\begin{algorithm}[t]
  \caption{Adaptive particle refinement procedure.}%
  \label{alg:apr}
\begin{algorithmic}[1]
  \While{$t < t_{\text{final}}$}
  \For{all fluids $i$}
  \If{fluid particles closest to boundary}
  \State{$\Delta s_i \leftarrow \Delta s_{\min}$}
  \EndIf{}
  \If{fluid particles satisfy certain criteria for solution adaptivity}
  \State{$\Delta s_i \leftarrow \Delta s_{\min}$}
  \EndIf{}
  \EndFor{}
  \State{Update the spacing of particles (\cref{alg:update-h})}
  \State{Split the particles (\cref{alg:splitting})}
  \For{$i = 0; i < 3; i\texttt{++}$}
  \State{Merge the particles (\cref{alg:merging})}
  \EndFor{}
  \For{$i = 0; i < 3; i\texttt{++}$}
  \State{shift the particles using \cref{eq:shift}}
  \EndFor{}
  \State{correct the particle properties using \cref{eq:shift-correct}}
  \State{update the smoothing length using \cref{eq:optimal-h}}
  \State{begin the simulation using adaptive EDAC-SPH scheme (\cref{sec:sph})}
  \EndWhile{}
\end{algorithmic}
\end{algorithm}

The algorithms for the solution adaptivity are the same as that used in the
original work of \citet{muta2021efficient} with the difference that the
properties are directly set on the fluid particles and not on the background
particles. In the present work we use the vorticity to determine solution
adaptivity.

\subsection{Implementation in PySPH}%
\label{subsec:imp-pysph}

PySPH is an open-source, Python-based, multi-CPU, and GPU framework for SPH
simulations. We have used the Python based interfaces PySPH provides to
implement most of our adaptive algorithm. PySPH exposes Python-friendly
interface to the user and using internal mechanism generates a high
performance serial or parallel code based on the choice of the user. For an
overview of the PySPH design, see \citet{ramachandran2019pysph} and
\url{https://pysph.readthedocs.io} for a detailed reference.

We use the \code{Tool} interface to split and merge the particles.  The
\code{Tool} interface provides a way for a user defined callback to run before
or after the integration stages by overloading the \code{pre_stage} or
\code{post_stage} methods, respectively. It also provides a way to run a user
callback before the start of the integration step by overloading the
\code{pre_step} method. Listing \ref{lst:tool} shows a sample Python code used
in our adaptive particle refinement algorithm. The class
\code{AdaptiveResolution} overloads the \code{pre_step} method of \code{Tool} to
run a user callback \code{self.run} which in turn calls other specific functions
related to splitting, merging, and shifting.
\begin{lstlisting}[%
label={lst:tool},%
caption={Python code using the PySPH \code{Tool} interface.}%
]
from pysph.solver.tools import Tool

class AdaptiveResolution(Tool):
    def __init__(self):
        # ...

    def run(self):
        # ...
        self._add_remove_split_particles()
        if self.shifter is None:
            self.shifter = self.setup_shifting(
                self.fluid_name,
                [self.fluid_array],
                self.nnps, self.dim,
                rho0=self.rho0, iters=3
            )
            self.shifter.update()
            self.shifter.evaluate()

    def pre_step(self, solver=None):
        # ...
        self.count += 1
        if self._count % self.freq == 0:
            self.run()
\end{lstlisting}

These functionalities enable one to write the code without having to pay much
attention to the low-level high-performance complexities of implementing on
different platforms like multi-core CPUs, or GPUs. The respective splitting
and merging algorithms are all implemented in pure Python. For complete
details of our implementation one can see the source code provided in
\url{https://gitlab.com/pypr/asph_motion}.

In \cref{subsec:no-bg} we solve a case with a stationary cylinder and another
with a moving square to demonstrate that we do not require the background
particles. In \cref{subsec:perf} we demonstrate the parallel performance of our
algorithms by solving a problem and studying the speedup obtained as we increase
the number of threads.

\section{Source files description}

The algorithms mentioned in the manuscript are primarily contained in the
following files located in the \code{code/} directory: \code{edac.py},
\code{adapt.py}, \code{adaptv.py}, and \code{no_bg.py}. The
\code{hybrid_inlet_outlet.py} contains the code which implements the
no-reflection boundary conditions of \cite{lastiwka2009} and manages the
inlet/outlet particles.

We use \code{automan} \cite{ramachandran2017automan} Python package to automate
all of our results. The file \code{automate.py} contains the code to generate
all the results used in this paper. See \code{README.md} for more details
regarding the individual run time of the problems in \code{automate.py}. The
following files in the \code{code/} directory contain the test problems used in
the manuscript:

\begin{itemize}
	\item \code{fpc_auto.py}: Flow past a stationary cylinder.
	\item \code{moving_square.py}: Flow past a moving square (SPHERIC test 6).
	\item \code{ellipse_plunging.py}: Flow past a plunging ellipse at Re = 500.
	\item \code{ellipse_tranlation.py}: Flow past a translating ellipse at Re = 500.
	\item \code{fpc_bao_static.py}: Flow past two stationary cylinders at Re = 100.
	\item \code{fpc_bao_oscillation_1cylinder.py}: Flow past single oscillating
	cylinder at Re = 100.
	\item \code{airfoil_stationary_naca.py}: Flow past a stationary NACA5515
	airfoil.
	\item \code{s_shape.py}: Flow past a rotating S-shape at Re = 2000.
	\item \code{sun_airfoil08_aoa4.py}: Flow past a NACA0008 airfoil at $4^\circ$
	angle of attack and Re = 2000.
	\item \code{sun_airfoil12_aoa6.py}: Flow past a NACA0012 airfoil at $6^\circ$
	angle of attack and Re = 6000.
	\item \code{performance.py}: Code to measure the parallel performance of the
	algorithms.
\end{itemize}

\section{Results and Discussion}%
\label{sec:results}

\noindent
In this section we demonstrate practical applications of the adaptive smoothed
particle hydrodynamics method for selected problems in two dimensions. The
problems considered are fairly challenging to solve without the adaptive
refinement. We consider moving and stationary geometries specifically focusing
on the following problems:

\begin{enumerate}
\item flow around two stationary cylinders in a side-by-side configuration with
  different gap spacings;
\item flow over airfoils with different NACA profile at different Reynolds
  numbers and angles of attack;
\item flow around a moving square at different Reynolds numbers; and,
\item flow around a single oscillating cylinder with different oscillating
  frequencies.
\end{enumerate}

% \noindent
% The numerical analysis of flow around cylinders is presented for
% different computing arrangements. The layouts are: (i) flow around two
% stationary cylinders in a side-by-side configuration, and(ii) flow around a single
% oscillating cylinder. The airfoil simulations of NACA profiles are presented for
% stationary cases. The flow around a moving square problem has a square or
% cylinder geometry which moves across a static fluid and studies the
% incompressible viscous flow around the body.

\noindent
\subsection{Flow around two stationary cylinders in a side-by-side
  configuration}%
\label{subsec:fpc-2cyl-stationary}

\begin{figure}[h!]
  \centering \includegraphics[width=0.75\textwidth]{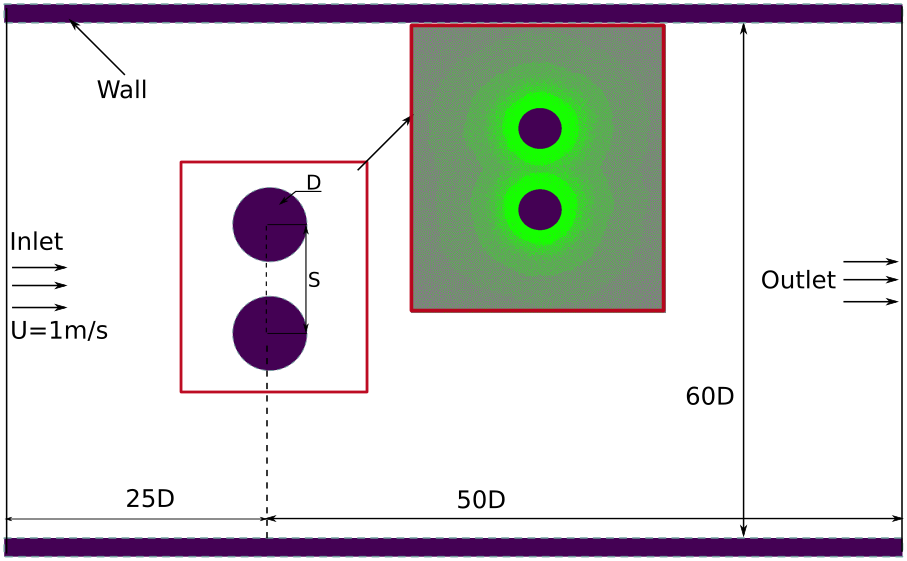}
  \caption{Schematic of the computational domain for the simulation of
    flow past two cylinders in a side-by-side arrangement.}%
  \label{fig:fpcdom}
\end{figure}

We study the flow around two stationary cylinders in a side-by-side
configuration at $Re = 100$. The flow produces distinct wake patterns at
different gap spacing between the cylinders. This problem is studied first
by~\citet{kang2003characteristics} and further used as a validation test case
by \citet{bao2013flow} but do not have SPH related studies. The diameter $D$
of each of the cylinders is $1 m$. The dimensions of the computational domain
are based on $D$, and the schematic diagram of the computational domain is
shown in \cref{fig:fpcdom}. For each inlet, outlet, and wall, we use $10$
layers of particles to get full kernel support. Three different cases of
center-to-center cylinder spacing $S = 1.2D, 1.5D$ and $4D$ are considered for
simulations. A constant velocity is prescribed at the inlet, $U_\infty =
1m/s$, and the kinematic viscosity $\nu$ is given by $U_{\infty} D / Re$.
Other geometric and numerical parameters are provided in Table
\ref{tab:table1}.
\begin{table}[h!]
  \begin{center}
    \caption{Simulation parameters for the problem described
      in~\ref{subsec:fpc-2cyl-stationary}.}%
    \label{tab:table1}
    \begin{tabular}[!htb]{cccc}
      \toprule
      Parameter & Value & Parameter & Value\\
      \midrule
      $D$ & $1 m$  & $U_\infty$ & $1 m/s$  \\
      $S$ & $D(1.2-4.0) m$ & $\rho$ & $1000 kg/m^3$  \\
      $\nu$ & $U_{\infty}D/Re$ & $\omega$ & $0$   \\
      \bottomrule
    \end{tabular}
  \end{center}
\end{table}
For all the cases, we deploy particles in the range of $75k$ to $88k$ on a $75$
m $\times\ 60$ m domain. The maximum spacing of fluid particles is
$D/\Delta x_{\max} = 2$ and the minimum resolution of the fluid particles, which
also matches the resolution of the boundary, is $D/\Delta x_{\min} = 100$; the
ratio of which is $50$ times smaller. Without an adaptive resolution, the
problem would need $45$ million particles to simulate at such accuracy using the
minimum resolution.

\begin{figure}[!ht]
  \centering
  \includegraphics[width=0.95\textwidth]{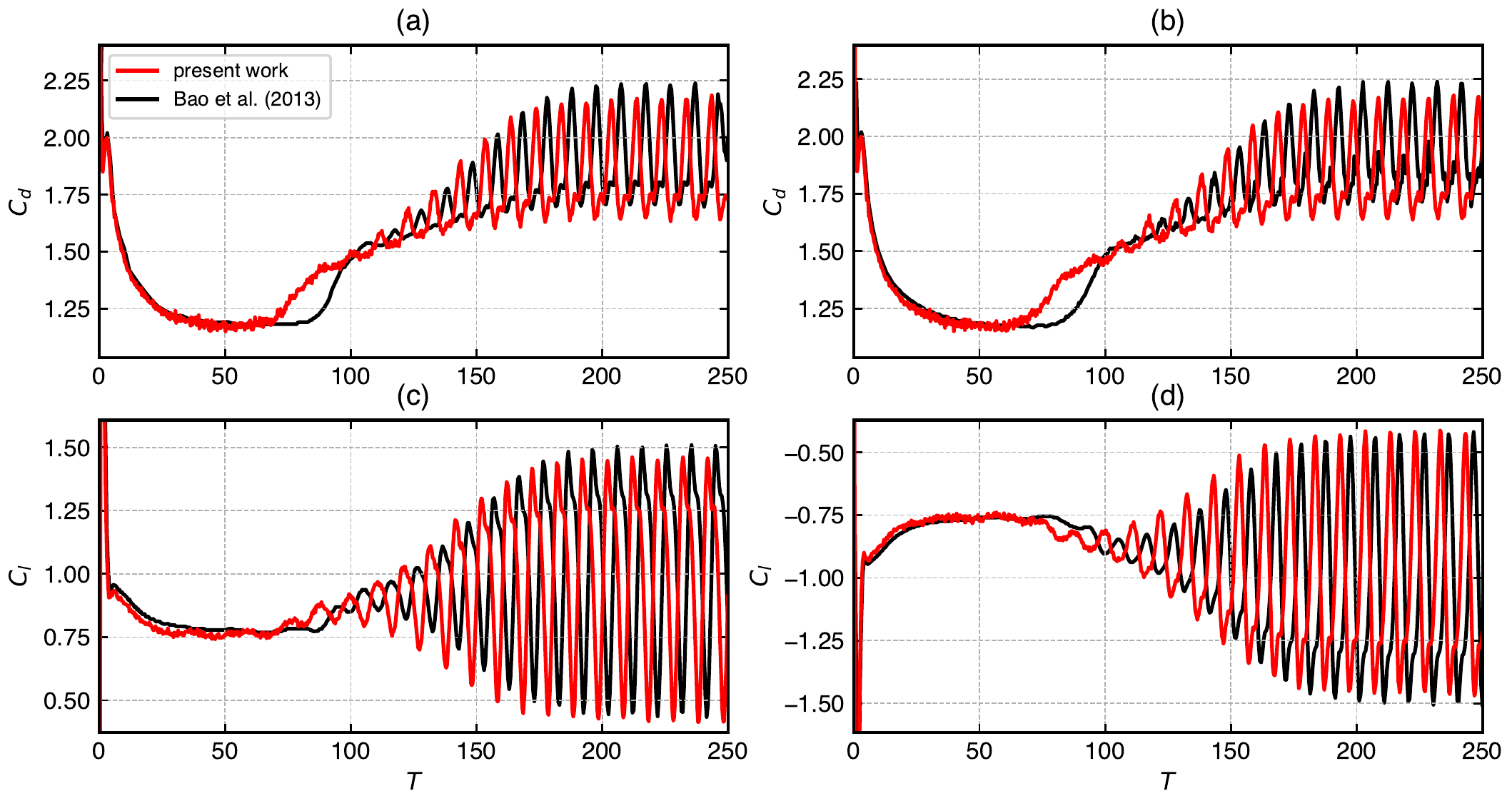}
  \caption{Time history of the drag and lift force coefficients for two stationary
    cylinders in a cross-flow at $Re=100$. Comparison of adaptive EDAC-SPH results
    with~\cite{bao2013flow} for gap spacing ratio of $S=1.2D$. (a) $C_d$ upper
    cylinder, (b) $C_d$ lower cylinder, (c) $C_l$ upper cylinder, (d) $C_l$
    lower cylinder.}%
  \label{fig:gap12}
\end{figure}
\begin{figure}[!ht]
  \centering
  \includegraphics[width=\textwidth]{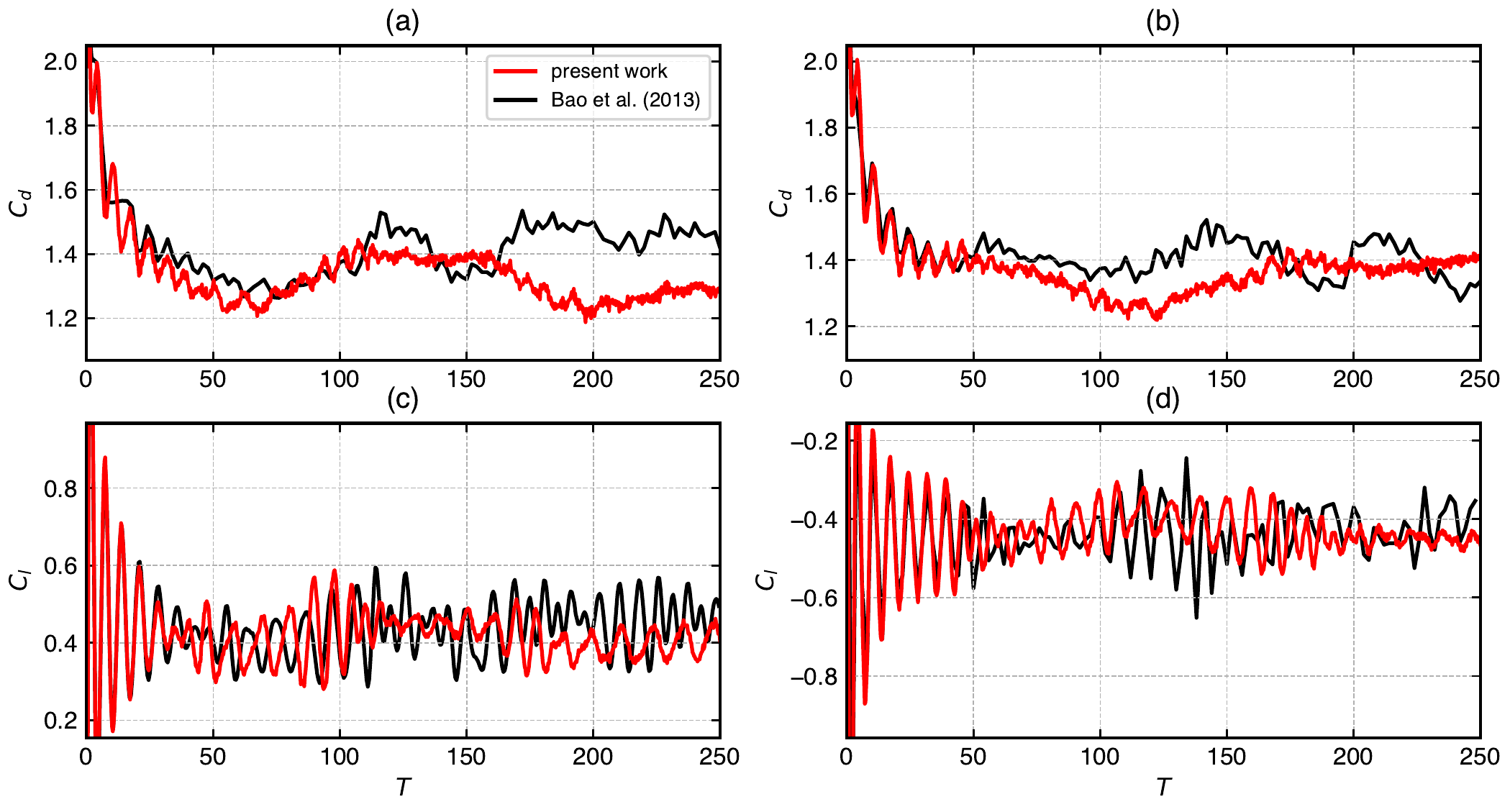}
  \caption {Time history of the drag and lift force coefficients for two stationary
    cylinders in a cross-flow at $Re=100$. Comparison of adaptive EDAC-SPH results
    with \cite{bao2013flow} for gap spacing ratio of $S=1.5D$. (a) $C_d$ upper
    cylinder, (b) $C_d$ lower cylinder, (c) $C_l$ upper cylinder, (d) $C_l$
    lower cylinder.}
  \label{fig:gap15}
\end{figure}
\begin{figure}[!ht]
  \centering
  \includegraphics[width=\textwidth]{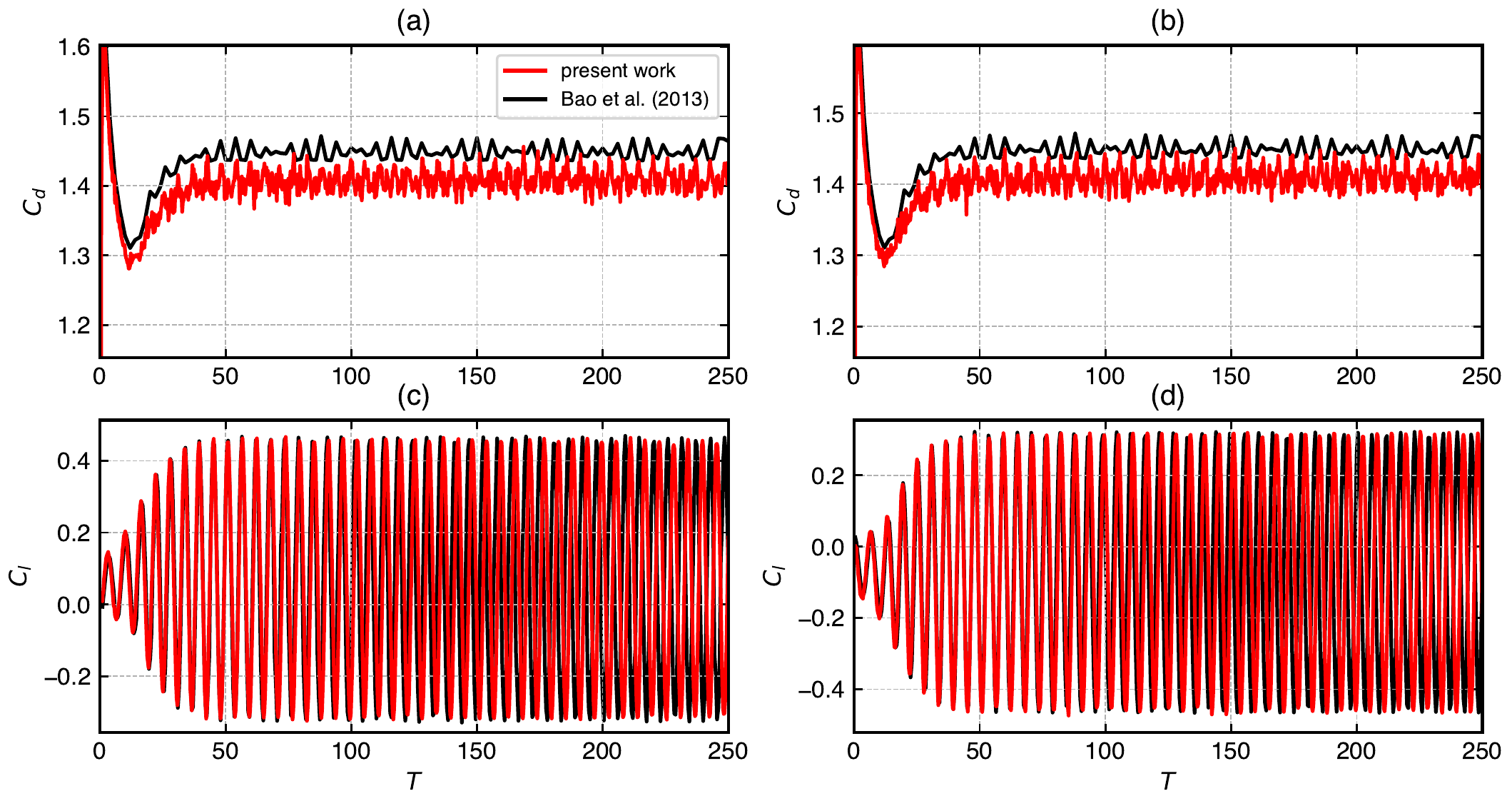}
  \caption {Time history of the drag and lift force coefficients for two stationary
    cylinders in a cross-flow at $Re=100$. Comparison of results of the present model
    with \cite{bao2013flow} for gap spacing ratio of $S=4.0D$. (a) $C_d$ upper
    cylinder, (b) $C_d$ lower cylinder, (c) $C_l$ upper cylinder, (d) $C_l$
    lower cylinder.}
  \label{fig:gap40}
\end{figure}

We are able to observe the three kinds of vortex shedding wake patterns that
occur for each configuration. The wake patterns are (i) single bluff body at
$S = 1.2D$, (ii) a random flip-flopping gap at $S = 1.5D$, and (iii) anti-phase
synchronized wake at $S = 4.0D$.  Time evolution of the simulated results of
total drag and lift force coefficients are presented in
\cref{fig:gap12,fig:gap15,fig:gap40}. The simulations took a fairly long
wall-clock time for $250$ seconds of simulation time, each case requiring an
average wall-clock time of $40$ hours on a server with $16$ cores and $32$
threads for the adopted resolution.

\Cref{fig:gap12} shows the comparison of time history of the coefficients of
total drag and lift for spacing $S = 1.2D$ up to $T = U_{\infty} t / D = 250$.
The figure illustrates the force characteristics of a single bluff body type
with entirely suppressed gap flow observed at $(S = 1.2D)$.
Our results show a phase difference which may be caused by the differences in
the nature of the solvers, time-stepping used, and boundary conditions. The
results are in good agreement with the reference data of \cite{bao2013flow}. The
trend, as well as the magnitude of the curves, is matching well.

\Cref{fig:gap15} shows the force characteristics for the spacing $S = 1.5D$.
The flow is complicated and belong to a randomly flip-flopping flow for which
the wake pattern looks irregular \cite{bao2013flow}. Our results compare very
well at the beginning up to around $75$ seconds of the flow. The results show
some discrepancy in the $C_d$ afterward but stay in the trend, matching the
amplitude and frequency. The $C_l$ results match better than the $C_d$ in all
the cases for both the top as well as bottom cylinders.

Figure \ref{fig:gap40} displays the comparison of results for the coefficients
drag and lift force at spacing $S = 4.0D$. The results are matching well with
the $C_d$ of the reference data in \cref{fig:gap40} (a) and \cref{fig:gap40} (b)
of the upper and lower cylinders with differences less than $5$ percent. The
results of $C_l$ are in a very good agreement as shown in \cref{fig:gap40} (c)
and \cref{fig:gap40} (d) for both the cylinders. The force characteristics of
the flow at this gap between the cylinders is anti-phase synchronized where the
vortex is shed behind each cylinder independently and symmetrical to the
centerline.

\begin{figure}[!ht]
  \centering
  \begin{subfigure}{0.8\textwidth}
    \includegraphics[width=\textwidth]{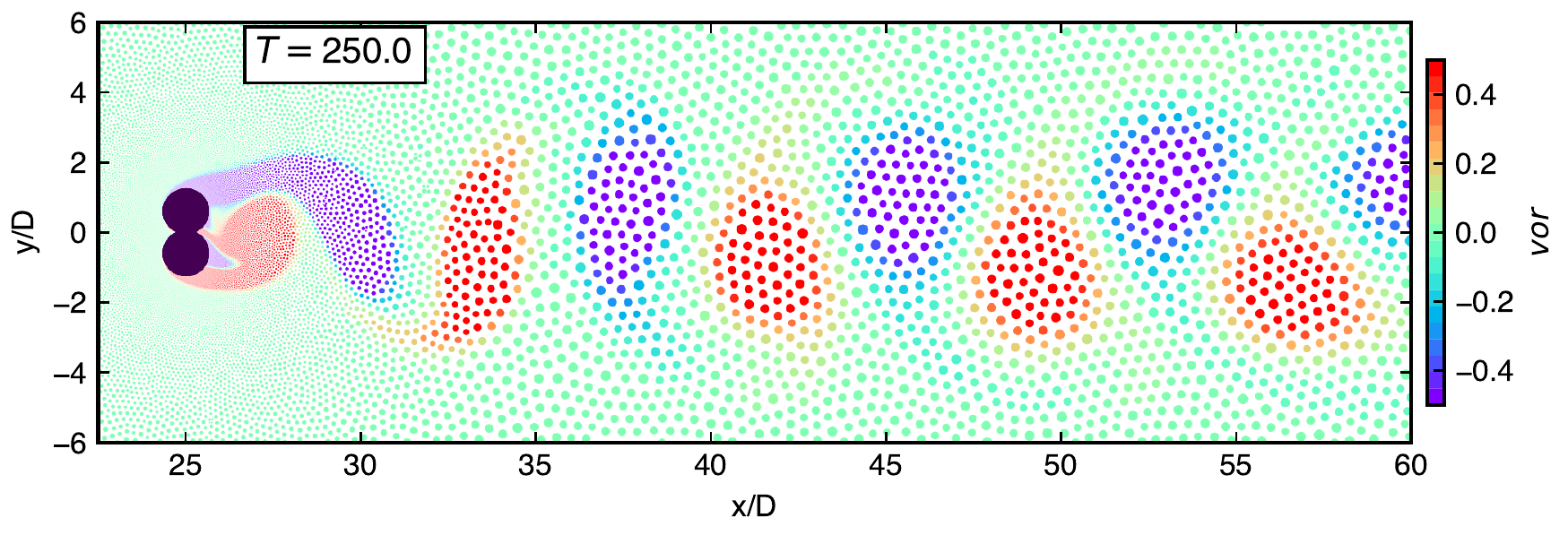}
      \subcaption{}
      \label{fig:vor-static-a}
  \end{subfigure}
  \\
  \begin{subfigure}{0.8\textwidth}
    \includegraphics[width=\textwidth]{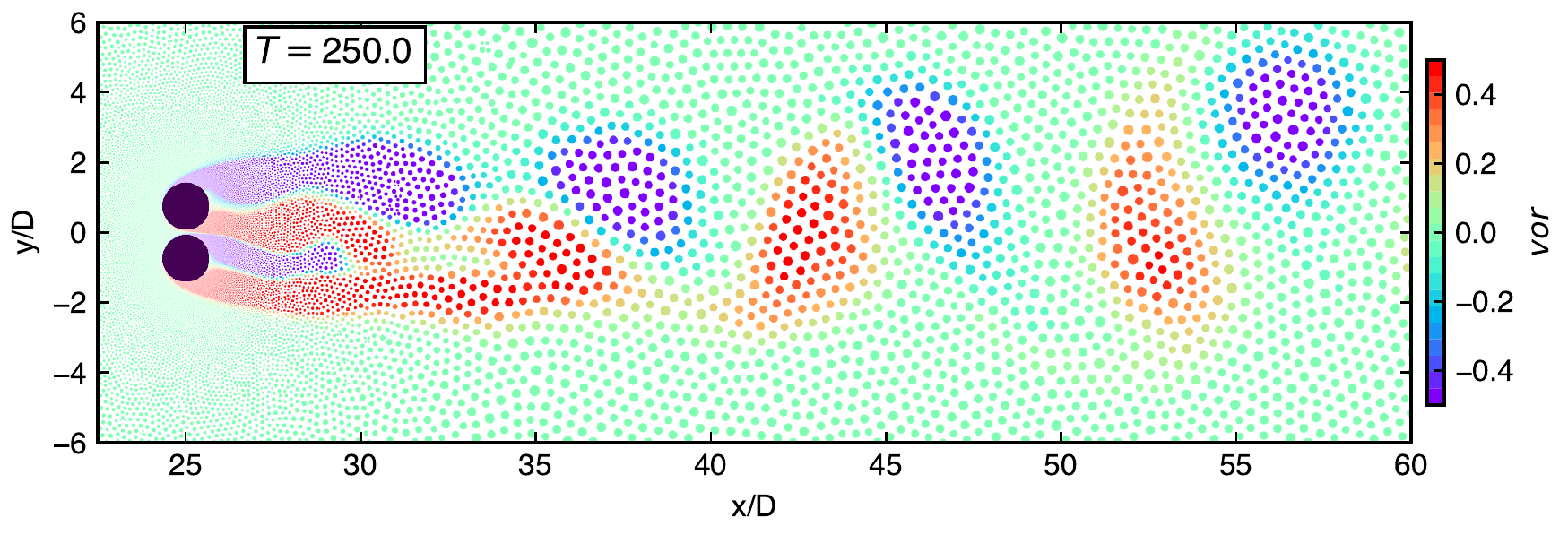}
      \subcaption{}
      \label{fig:vor-static-b}
  \end{subfigure}
  \\
  \begin{subfigure}{0.8\textwidth}
    \includegraphics[width=\textwidth]{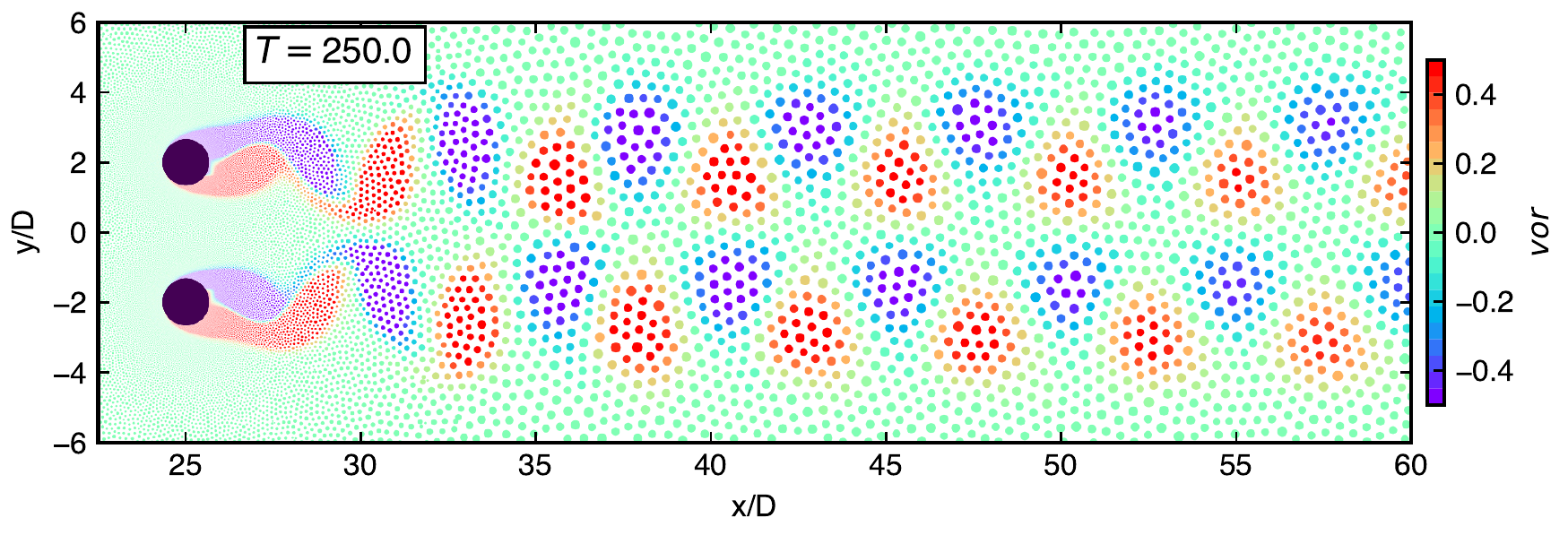}
      \subcaption{}
      \label{fig:vor-static-c}
  \end{subfigure}
  \caption{Vorticity distribution corresponding to flow past two stationary
    cylinders in a side-by-side arrangement at $Re=100$ with the gap spacing
    ratio of, (a) $S=1.2D$; (b) $S=1.5D$; (c) $S=4.0D$. The size of the points
    is proportional to the mass of the particles.}%
  \label{fig:vor-static}
\end{figure}
\Cref{fig:vor-static} exhibits the vorticity distribution corresponding to the three
spacing ratios. The results indicate that the flow characteristics and wake
structure explained in Figs. \ref{fig:gap12} to \ref{fig:gap40} are
justified. The flow structure strongly depends on the spacing gap for the given
Reynolds number. A suppressed flow, seemingly a single bluff-body wake pattern
is shown by the smallest spacing of $S = 1.2D$ in \cref{fig:vor-static-a}. In
\cref{fig:vor-static-b} the gap flow flip-flops for $S = 1.5D$ and seems
irregular. The wake pattern changes to a synchronized flow as the spacing
increases. As shown in \cref{fig:vor-static-c}, the instantaneous wake is
anti-phase synchronized; the vortex shedding occurs completely behind each
cylinder which is symmetric with the wake centerline.

The behaviour of the flows around two stationary cylinders and a single
oscillating cylinder, or both is helpful to understand the mechanism of flow
around two or more oscillating cylinders. The present SPH model is able to
simulate and accurately identify the wake characteristics of the flow. The force
quantities are in general in good agreement with the results available in the
literature. This accuracy is achieved with a proper particle resolution using a
reduced total number of particles.

\noindent
\subsection{Flow over airfoils with NACA profile}%
\label{subsec:fpa}

Flow around airfoil present difficulties to conduct high accuracy SPH
simulation. It can be more difficult when it is thinner, particle distribution
is irregular, or if it is moving. It is further challenging in complex motion
applications like in a rotational motion, elliptic motion, and motions which
combine heaving, oscillation, and pitching. One of the first applications of
SPH to model flow around an airfoil was conducted by
\citet{shadloo2011improved} and \cite{shadloo2012robust} in which they have
shown a comparative study between the weakly compressible and incompressible
SPH methods. However, their results have shown poor accuracy even when
employing large number of particles at moderate Reynolds number. With the help
of multi-resolution and particle shifting technique, \citet{sun2018multi}
indicated an improved success of airfoil SPH simulation for Reynolds number up
to $10,000$. \citet{huang2019kernel} illustrated an improved simulation of
viscous fluid flow around airfoils using SPH scheme possessing iterative
particle shifting technology (IPST) and which does not use kernel gradients.
The numerical study of airfoil motion is important for application like in
unmanned aerial vehicles (UAVs) for thrust generation and in vertical axis wind
turbines (VAWTs) for power optimization.

We discuss three stationary airfoil cases: (i) NACA5515 airfoil at
$\alpha=5^{\circ}$ and $Re=420$, (ii) NACA0008 at $\alpha=4^{\circ}$ and
$Re=2000$, (iii) NACA0008 at $\alpha=4^{\circ}$ and $Re=6000$. The computational
domain is $[-3L, 7L]\times[-3L, 3L]$, where $L = c = 1$ m is the characteristic length
(chord) with center at the airfoil midpoint. The algorithm of Negi and
Ramachandran \cite{negi2019improved} is used for packing the airfoils.

\subsubsection{NACA5515 airfoil at $\alpha=5^{\circ}$ and $Re=420$}%
\label{subsec:fpa-Re-420}
\begin{figure}[!ht]
  \centering
  \includegraphics[width=0.75\textwidth]{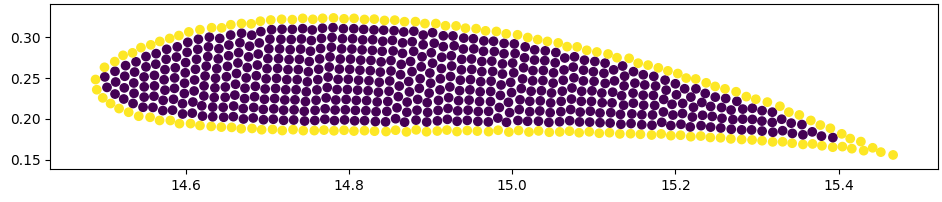}
  \caption{Airfoil with NACA5515 profile at $5^{\circ}$ angle of attack, yellow
    color indicates the surface particles. Particles are packed at a resolution
    $L/\Delta x = 80$ with the algorithm described
    in~\citet{negi2019improved}.}%
  \label{fig:fpa1}
\end{figure}
\Cref{fig:fpa1} depicts boundary points of NACA5515 marked in yellow color and
inner particles of the solid in violet. The particles on surface of the airfoil
(boundary points) are identified by computing divergence of the position
vector. All the surface points are identified correctly when the computed value
is lower than $d - 0.625$, $d$ is the dimension. The airfoil particles are
packed with the smallest spacing, $L/\Delta x_{\min} = 80$, which is equal to
the minimum resolution of fluid particles for mapping. We have simulated for
$T=12$ and compared the results with \cite{huang2019kernel}.
\begin{figure}[!ht]
  \centering
  \includegraphics[width=0.75\textwidth]{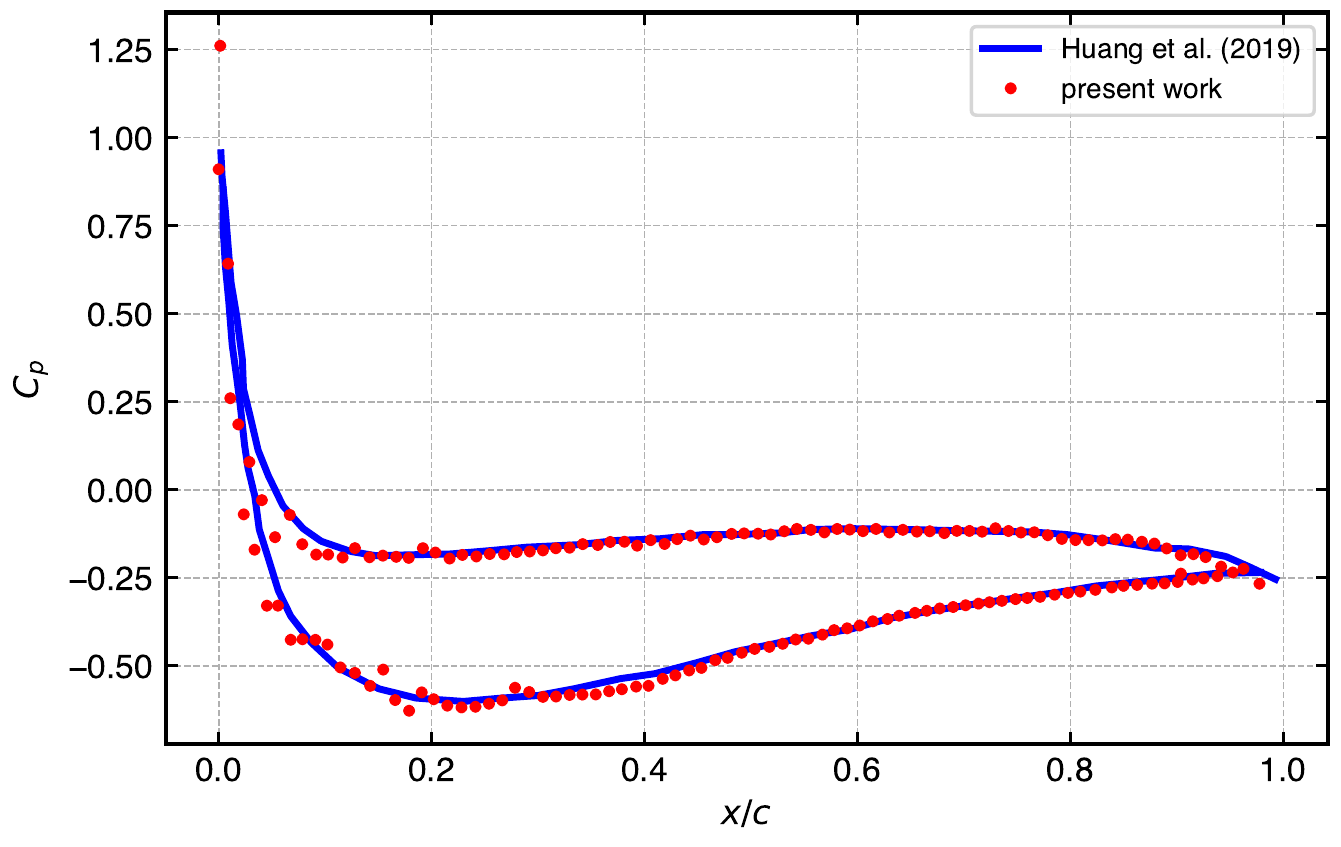}
  \caption{Comparison with \cite{huang2019kernel} of pressure coefficient $C_p$
    on the surface of NACA5515 airfoil at $T = 12$, with $\alpha=5^{\circ}$, and
    $Re=420$.}%
  \label{fig:fpa2}
\end{figure}

\Cref{fig:fpa2} shows the coefficient of pressure $C_p$ on the surface of an
airfoil with the NACA5515 profile. The airfoil is at an angle of attack
$\alpha=5^{\circ}$, $Re=420$ and uses particle spacing $L/\Delta
x_{\min}=80$. The simulated result of the present model agrees very well 
with \cite{huang2019kernel}. Based on the information provided in
\cite{huang2019kernel}, about $360k$ particles are employed to produce their
result. However, our solution require around $40k$ particles which could be
considered as a massive gain due to the adaptive resolution.

\subsubsection{NACA0008 airfoil at $\alpha=4^{\circ}$, $Re=2000$ and $Re=6000$}%
\label{subsec:fpa-Re-2k-6k}

The airfoil in NACA0008 profile has very thin cross-section which makes it a
challenging problem for numerical methods. This section discusses the flow
around the airfoil at an angle of attack $\alpha=4^{\circ}$. We simulate the
problem for two different Reynolds numbers which are $2000$ and $6000$. The
total drag and lift coefficient of the simulated results are compared with
Diffused Vortex Hydrodynamics (DVH) method of \citet{rossi2016simulating}, and
SPH method of \citet{sun2018multi}.

\begin{figure}[!ht]
	\centering
	\begin{subfigure}{0.48\textwidth}
		\includegraphics[width=\textwidth]{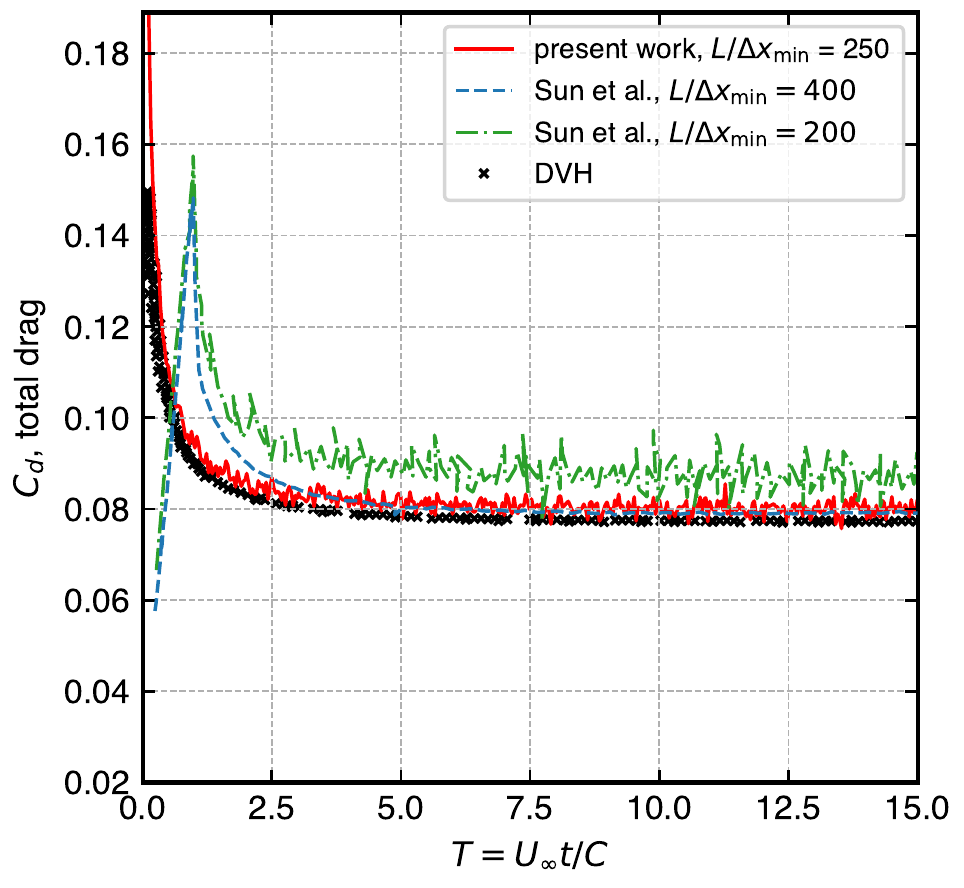}
	\end{subfigure}
	\begin{subfigure}{0.48\textwidth}
		\includegraphics[width=\textwidth]{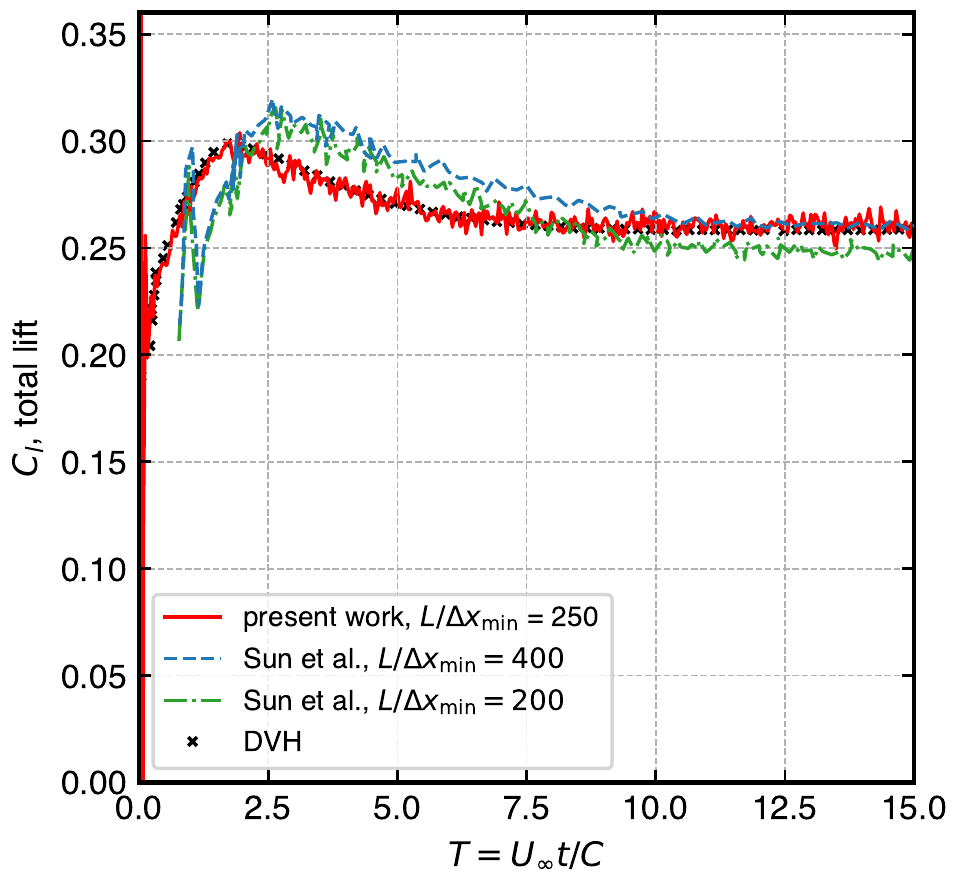}
	\end{subfigure}
	\\
	\begin{subfigure}{0.48\textwidth}
		\includegraphics[width=\textwidth]{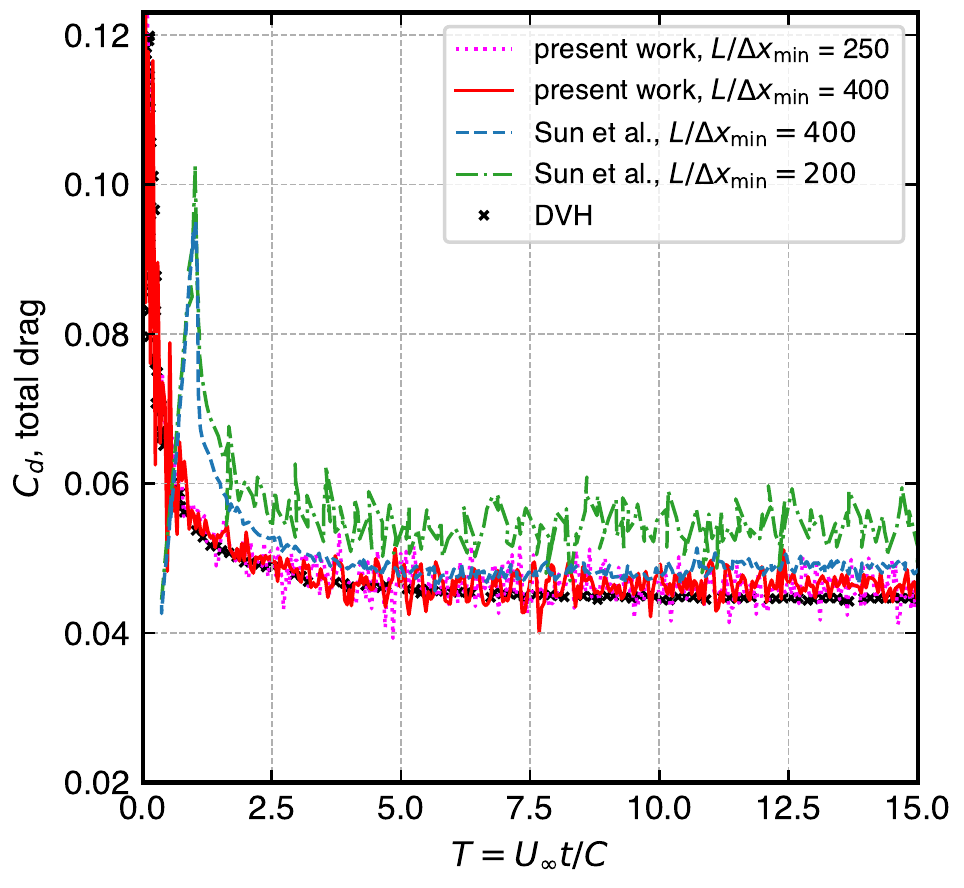}
	\end{subfigure}
	\begin{subfigure}{0.48\textwidth}
		\includegraphics[width=\textwidth]{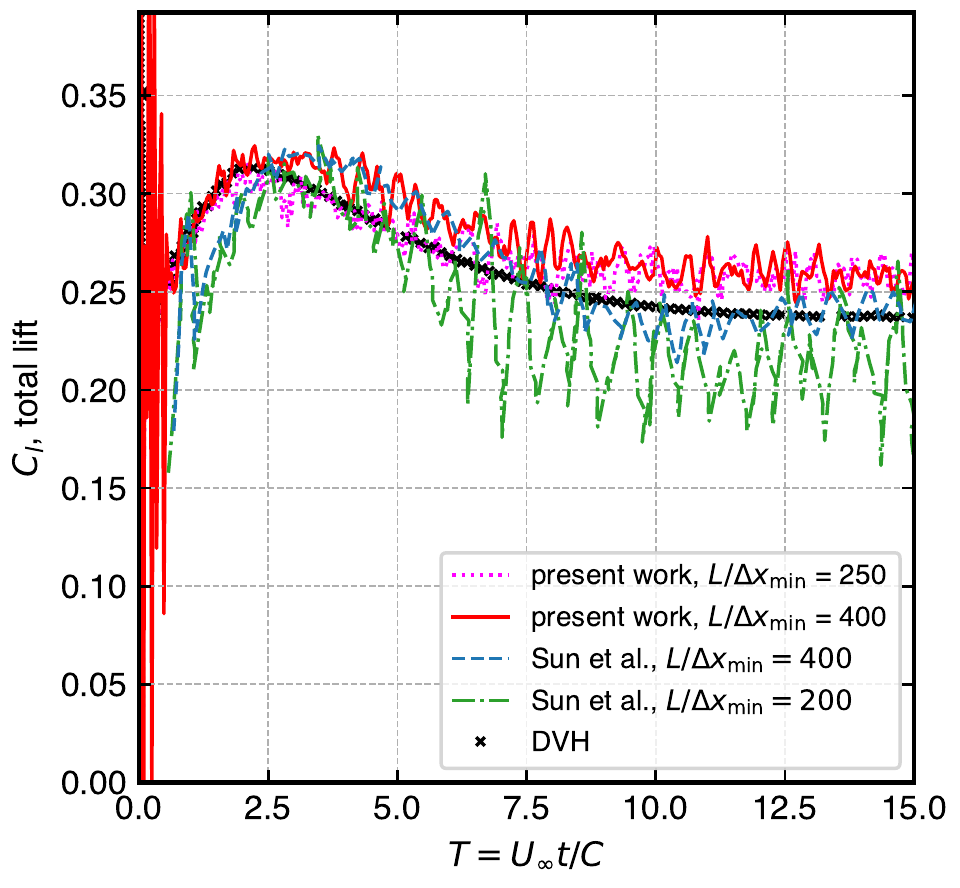}
	\end{subfigure}
	\caption{Flow over NACA0008 airfoil at $\alpha=4^{\circ}$ at $Re=2000$
          (upper) and $Re=6000$ (lower). Evolution of the drag (left) and lift
          (right) force coefficients. Comparison of the results with
          DVH~\cite{rossi2016simulating} and~\citet{sun2018multi}.}
	\label{fig:fpa3}
\end{figure}

Figure \ref{fig:fpa3} illustrates the evolution of force coefficients drag
(left) and lift (right). We simulate for $ T = 15$ using $L/\Delta x_{\min}=250$
for $Re = 2000$ (top) and $L/\Delta x_{\min}= 250, 400$ for $Re = 6000$
(bottom). At this resolution our simulations show very good comparison with DVH
\cite{rossi2016simulating} and have improved results compared with
\cite{sun2018multi}. The trend of the curve is captured very well except some
oscillations that can be minimized by increasing the resolution.  Sun et al.'s
result of comparable resolution has higher noise and misses some features of the
curve. Also, the number of particles (from Sun et al.'s Fig. 3) is more than
$250k$ particles for $L/\Delta x_{\min} = 400$, whereas our adaptive EDAC-SPH
simulation employs around $77k$ particles.

\Cref{tab:fpa-params} shows the number of particles used in the simulation for
two different Reynolds numbers. The number of particles is about $3.25$ times
less than \cite{sun2018multi}.
\begin{table}[!ht]
  \centering
  \begin{tabular}[!ht]{llll}
    \toprule
    Method & $Re=2000$  & $Re=6000$ & $L/\Delta x_{min}$ \\
    \midrule
    Sun et al. \cite{sun2018multi} & $250 k$ & $250 k $ & $400$ \\
    Adaptive EDAC-SPH & $76.5 k$ & $76.5 k$ & $250$ \\
    Adaptive EDAC-SPH (solution adaptivity) & --- & $260k$ & $250$ \\
    \bottomrule
  \end{tabular}
  \caption{Flow over NACA0008 airfoil at $\alpha=4^{\circ}$. Comparison between
    the approximate number of particles used in \cite{sun2018multi} and
    adaptive EDAC-SPH.}%
  \label{tab:fpa-params}
\end{table}

We ran the $Re = 6000$ case with and without solution adaptivity, where the
vorticity particles exceeding 5\% of the maximum vorticity in the domain are
refined to the highest resolution. The number of particles increases to
260k.
There is no significant difference in the lift or drag coefficient of the
airfoil between the solution adaptive and without the solution adaptive case
since the solid bodies are initially discretized at the highest resolution.
Our non-solution adaptivity case took about 18 hrs, whereas the solution
adaptivity case took about 41 hrs.
%There is no significant difference in the lift or drag coefficient of the
%airfoil between the solution adaptive and without solution adaptive case.
%
\begin{figure}[htp]
  \centering
  \includegraphics[width=0.95\textwidth]{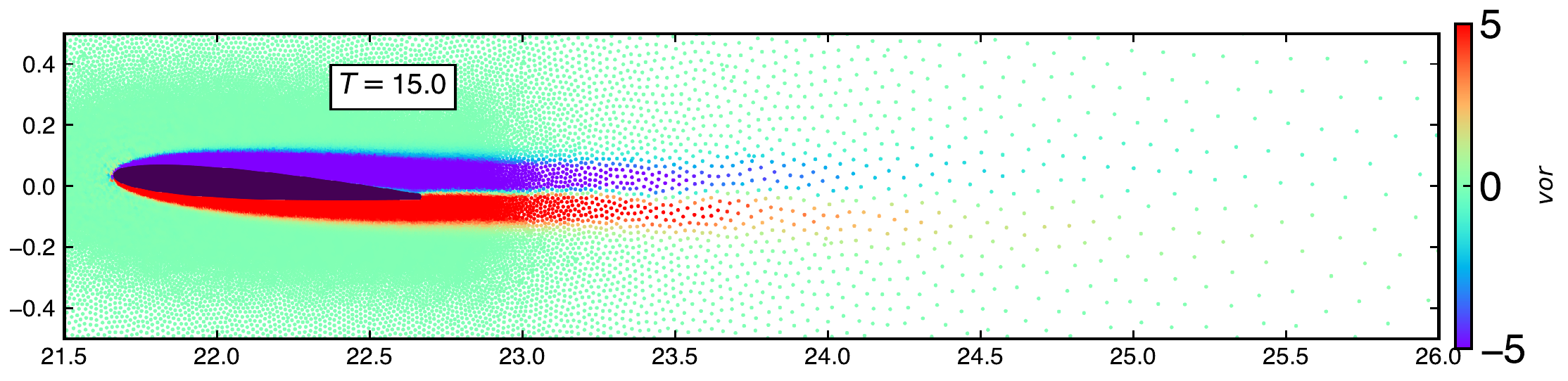}%
  \\
  \includegraphics[width=0.95\textwidth]{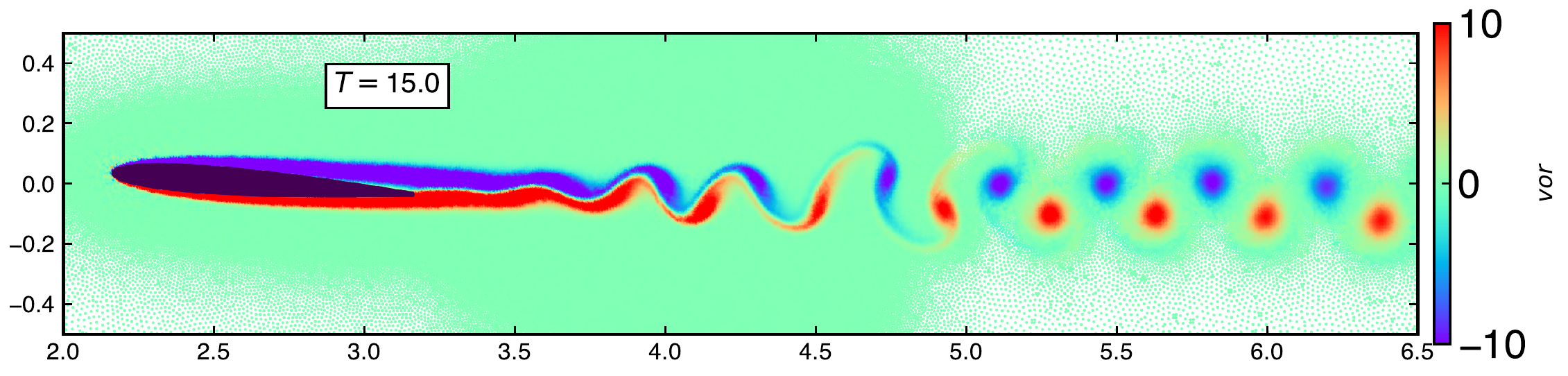}%
  \caption{Vorticity field around NACA0008 airfoil at of at $\alpha=4^{\circ}$.
    Simulated without solution adaptivity $Re=2000$ (top), and using solution
    adaptivity at $Re=6000$ (bottom).}%
  \label{fig:fpa4}
\end{figure}

In \cref{fig:fpa4}, we display the vorticity field around NACA0008 airfoil at an
angle of attack $\alpha=4^{\circ}$. The top figure is simulated without solution
adaptivity at $Re=2000$, and the bottom is simulated using solution adaptivity
at $Re=6000$. Due to the small angle of attack, the viscous flow remains
attached to the surface where a steady regime has reached immediately. The
selected airfoil configuration has practical application in micro-aerial
vehicles (MAVs) for the Reynolds numbers in the given range
\cite{rossi2016simulating}.

\subsection{Flow around a moving square}
\label{subsec:ms-mc}
The flow around a moving square represents a simple geometry but a flow
situation that can be very complex due to the sharp edges. The problem consists
of a square geometry moving through a stationary fluid kept inside a rectangular
box. The test can generate intense unsteady vorticity and has been a challenging
test case for SPH solvers.  This simulation is part of a validation test suite
compiled by the SPH rEsearch and engineeRing International Community
(SPHERIC)~\cite{spheric}. The benchmark results of this test case are simulated
using Finite Difference Navier-Stoked (FDNS) solver by
\cite{colicchio2006fluid}. In SPH \cite{vacondio2013variable}, and
\cite{marrone2013accurate} studied this problem. We study the time histories for
the drag force coefficients (pressure, viscous or total) and the generation and
diffusion of vorticity.

\begin{figure}[htp]
  \centering
  \begin{subfigure}{0.45\textwidth}
    \includegraphics[width=\textwidth]{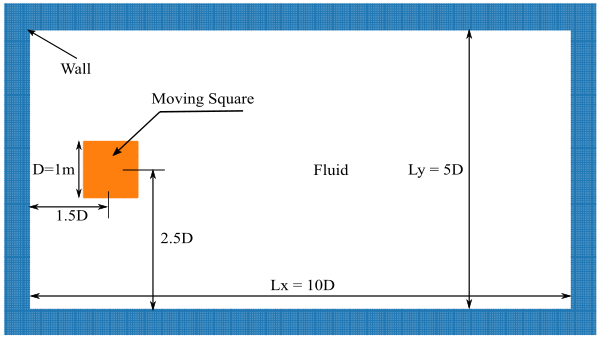}
      \subcaption{}
      \label{fig:geom_tl-a}
  \end{subfigure}
  \begin{subfigure}{0.45\textwidth}
    \includegraphics[width=\textwidth]{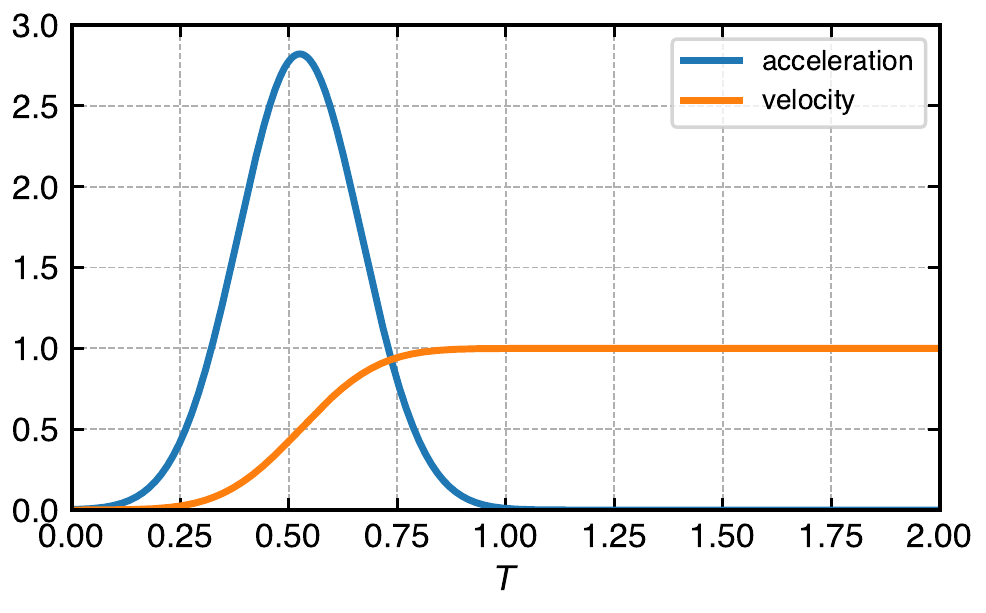}
      \subcaption{}
      \label{fig:geom_tl-b}
  \end{subfigure}
  \caption{Flow around a moving square, (a) Moving square test case geometry,
    (b) the body motion in time.}
  \label{fig:geom_tl}
\end{figure}

The geometric configuration is shown in \cref{fig:geom_tl-a} at time $t = 0$.
The body motion in time is shown in \cref{fig:geom_tl-b}. The domain size is
$L_y \times L_x = 5$ m $ \times\ 10$ m, with the square obstacle having a
length of 1m. The square motion is prescribed with a smooth acceleration in
the $x$-axis starting from rest until it reaches a steady maximum velocity
$U_{\text{square}} = 1.0\ \text{m/s}$. The boundary conditions are no-slip
and no-reflection on all walls of the rectangular domain and slip velocity on
the solid square. The simulation is conducted for a Reynolds number of $100$
with no gravity force and zero initial pressure field for the viscous
incompressible Newtonian fluid of density $1\,\text{kg m}^{-3}$.

\begin{figure}
  \centering
  \begin{subfigure}{0.47\textwidth}
    \includegraphics[width=\textwidth]{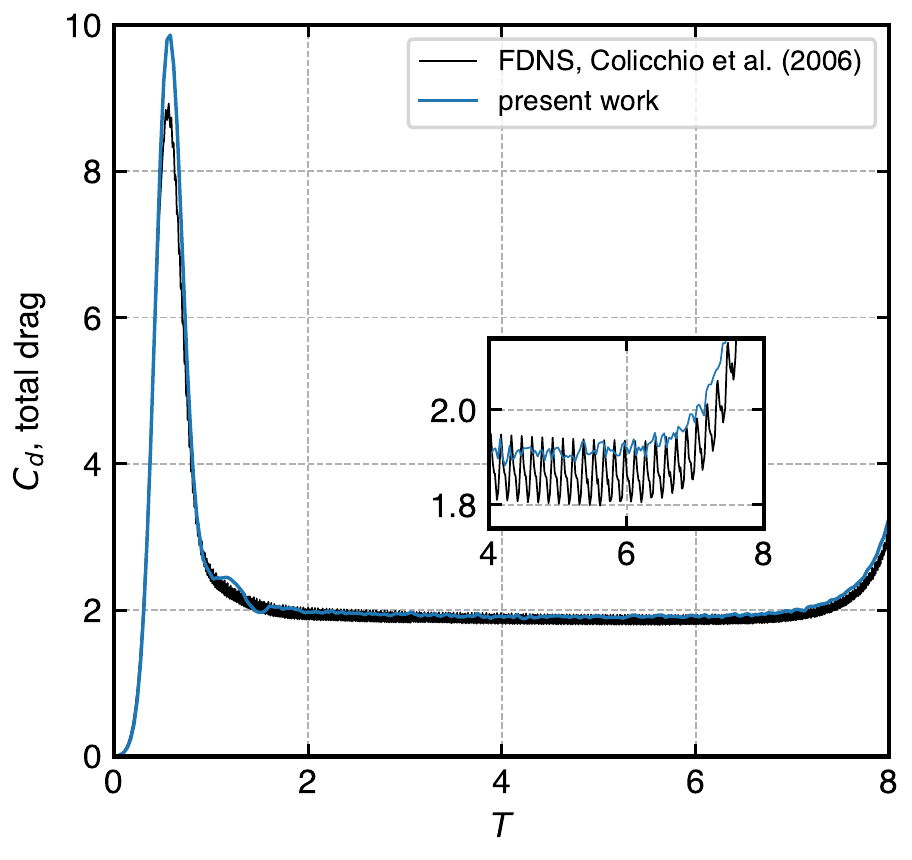}
    \subcaption{}
    \label{fig:re-100}
  \end{subfigure}
  \begin{subfigure}{0.47\textwidth}
    \includegraphics[width=\textwidth]{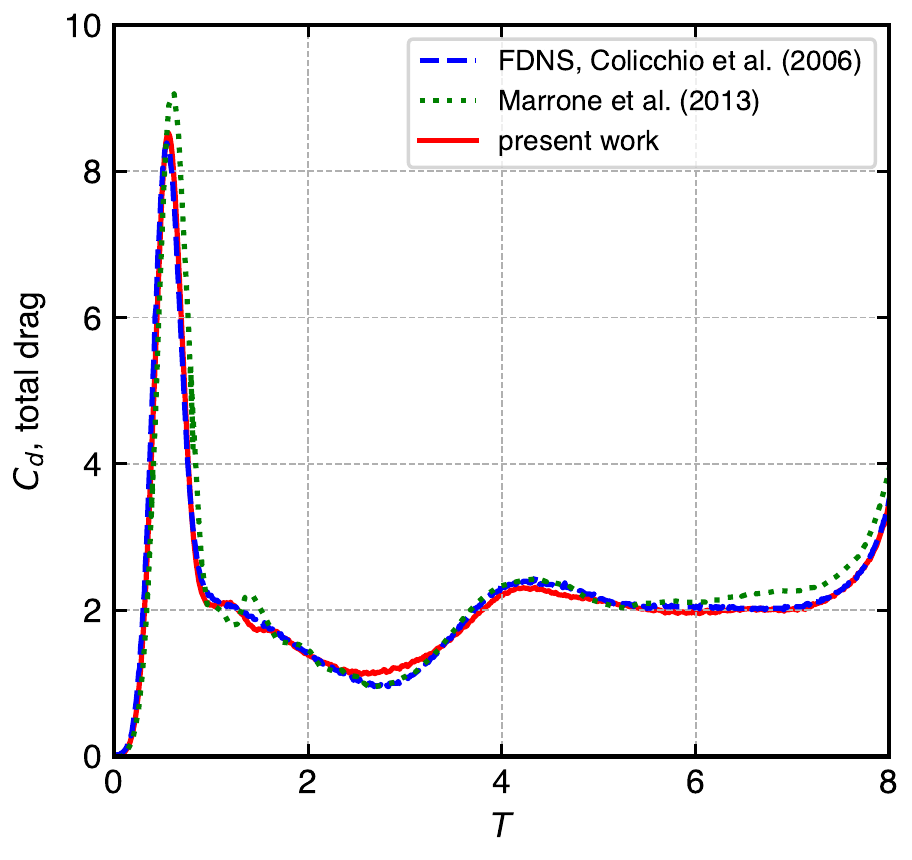}
    \subcaption{}
    \label{fig:re-600}
  \end{subfigure}
  \caption{Comparison of the total drag coefficient for a square cylinder at (a)
    $Re = 100$ and (b) $Re = 600$.  Simulated results of adaptive EDAC-SPH are
    compared with FDNS solver~\cite{colicchio2006fluid} and
    \citet{marrone2013accurate}.}%
  \label{fig:cdt-all-re}
\end{figure}

In \cref{fig:cdt-all-re}, we show the time evolution of the drag coefficient of
a moving square at (a) $Re = 100$ using $D/\Delta {x}_{\min} = 80$ and (b) at a
slightly higher $Re$ of $600$ using $D/\Delta {x}_{\min} = 200$. Simulated
results obtained from the present method are compared with the results
of the FDNS solver \cite{colicchio2006fluid} and
\citet{marrone2013accurate}. The comparison shows, especially with FDNS, very
good agreement with all the necessary features.

The simulation requires $44k$ fluid particles using adaptive resolution for the
converged solution. The initial spacing of particles on the domain is
$\Delta x_{\max} = 0.04$; the smallest spacing is $\Delta x_{\min} = 0.0125$
near the obstacle. The solid particles are packed using the minimum resolution
to create a mapping in the fluid-solid interface. If the $\Delta x_{\min}$ was
used in the entire domain, $D/\Delta x_{\min}$ would be $80$ which produces
$320k$ as the total number of particles. However, with an adaptive particle
resolution, we have used only $44k$ particles to produce accurate results. This
substantially reduces the number of particles by $7.273$ times when compared
with Marrone et al., \cite{marrone2013accurate} and by four times when compared
with the results by \cite{colicchio2006fluid}.

\begin{figure}[htp]
  \centering
  \includegraphics[width=0.8\textwidth]{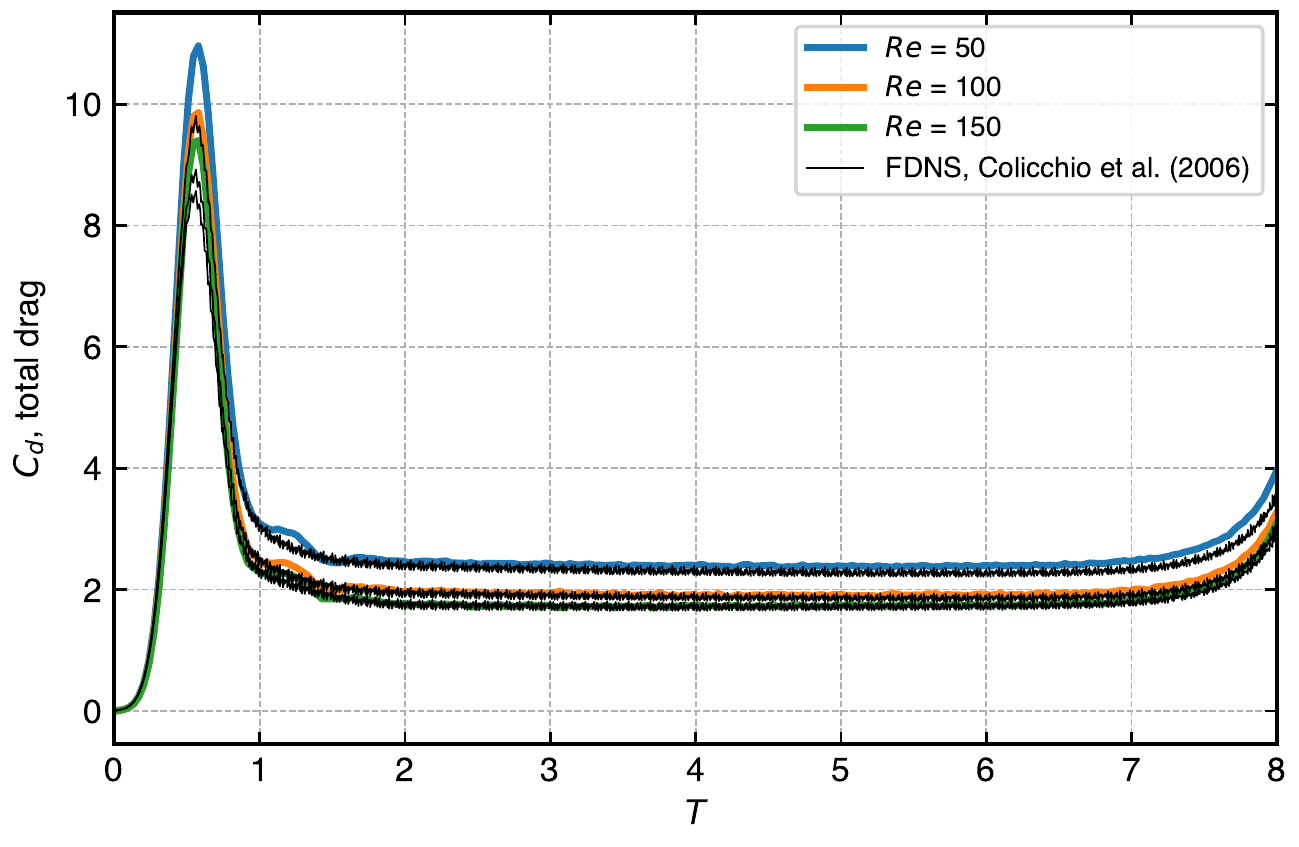}
  \caption{Comparison of results against the results of
    \citet{colicchio2006fluid}, with varying Reynolds number, for flow
    around a moving square.}%
  \label{fig:comprn}
\end{figure}
Figure \ref{fig:comprn} shows the variation of drag coefficient for flow around
a moving square at different Reynolds numbers. The comparison is made for three
Reynolds numbers of $50$, $100$, and $150$. The results of the adaptive EDAC-SPH
are in good agreement with \cite{colicchio2006fluid} results. However, due to
numerical reasons \cite{colicchio2006fluid} show the small amplitude and
high-frequency oscillations which are avoided in our case.

%%%%%================================================================================
\begin{figure}[!htp]
  \centering
  \begin{subfigure}{0.48\textwidth}
    \includegraphics[width=\textwidth]{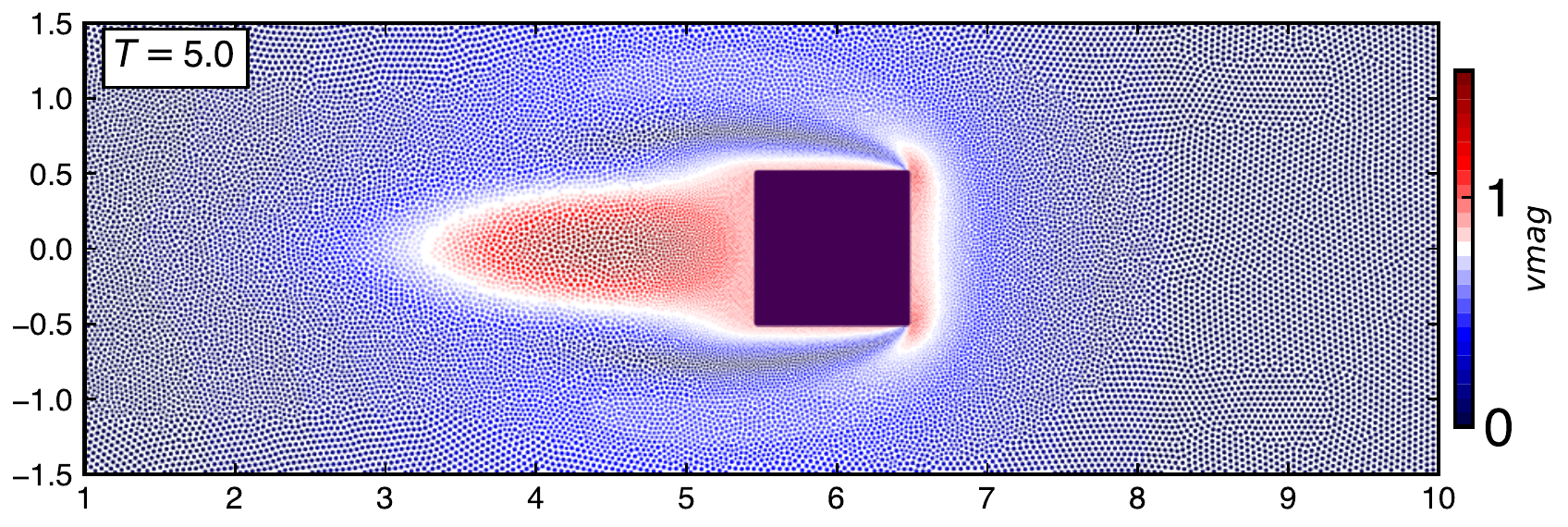}
  \end{subfigure}
  \begin{subfigure}{0.48\textwidth}
    \includegraphics[width=\textwidth]{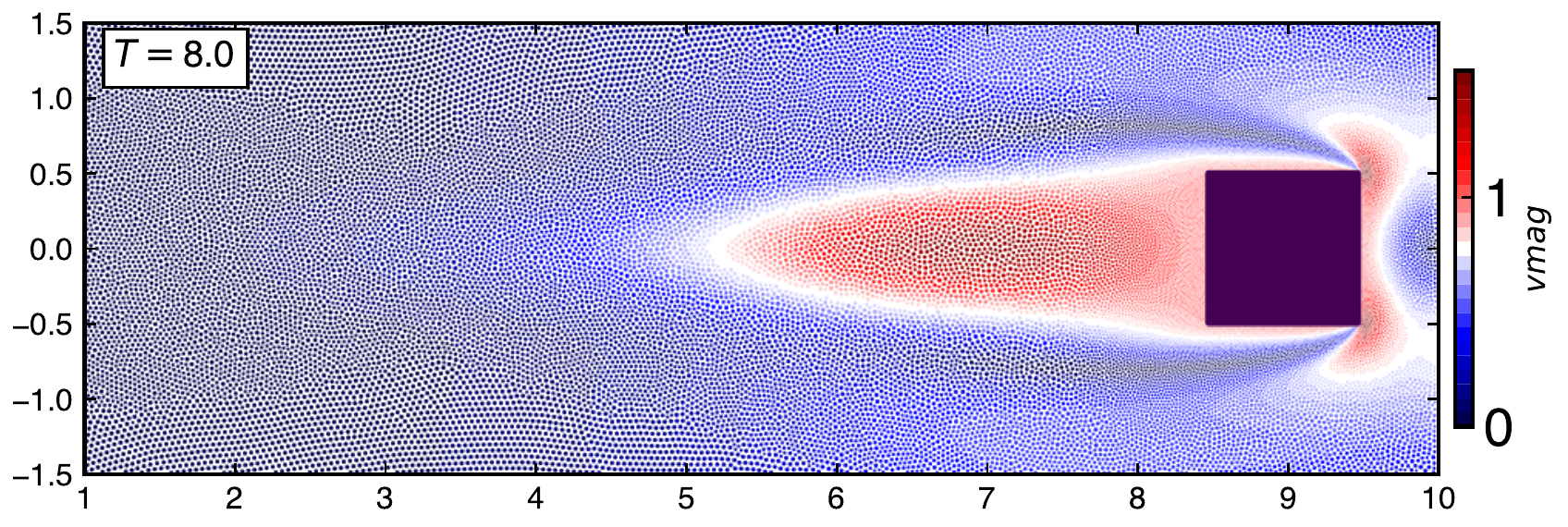}
  \end{subfigure}
  \\
  \begin{subfigure}{0.48\textwidth}
    \includegraphics[width=\textwidth]{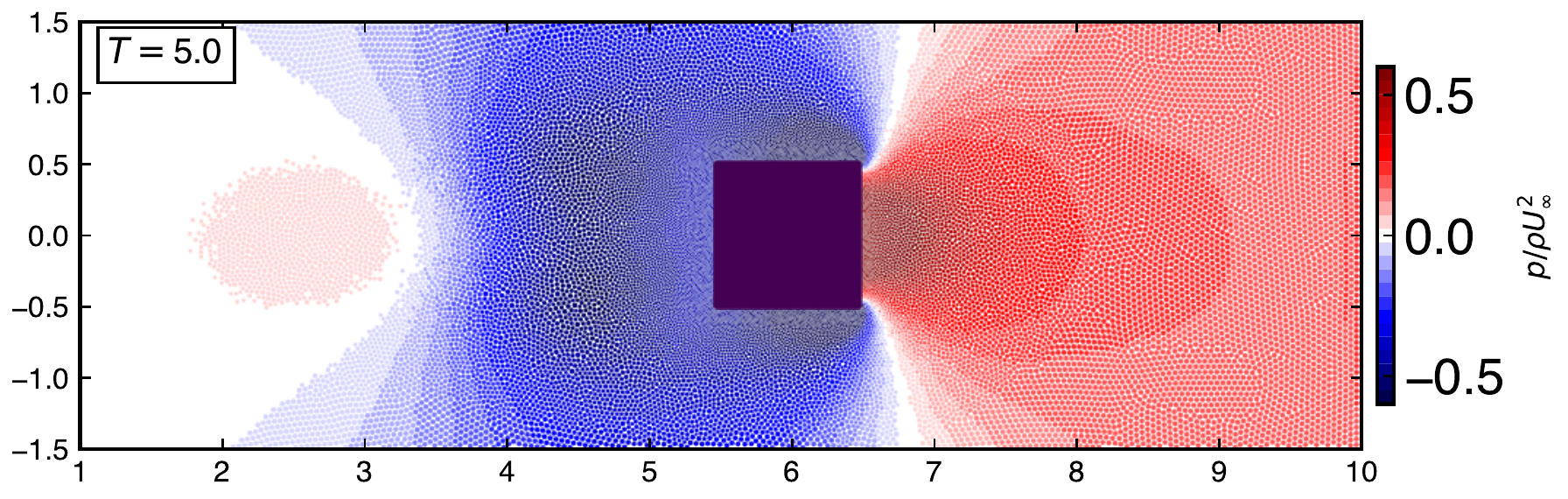}
  \end{subfigure}
  \begin{subfigure}{0.48\textwidth}
    \includegraphics[width=\textwidth]{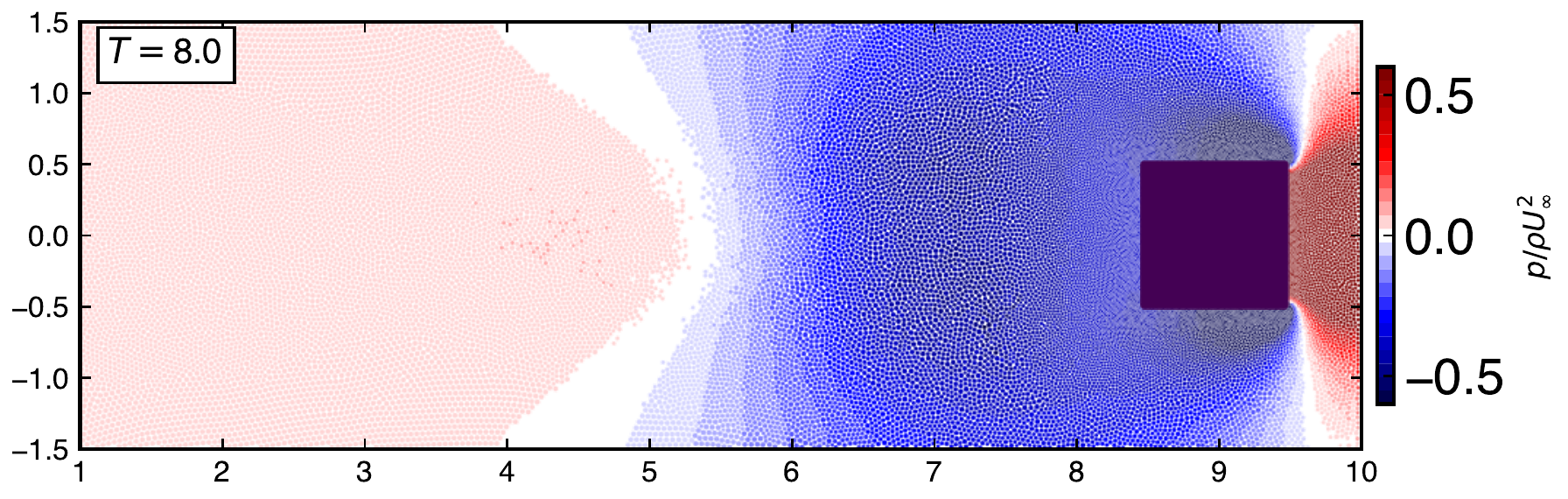}
  \end{subfigure}
  \\
  \begin{subfigure}{0.48\textwidth}
    \includegraphics[width=\textwidth]{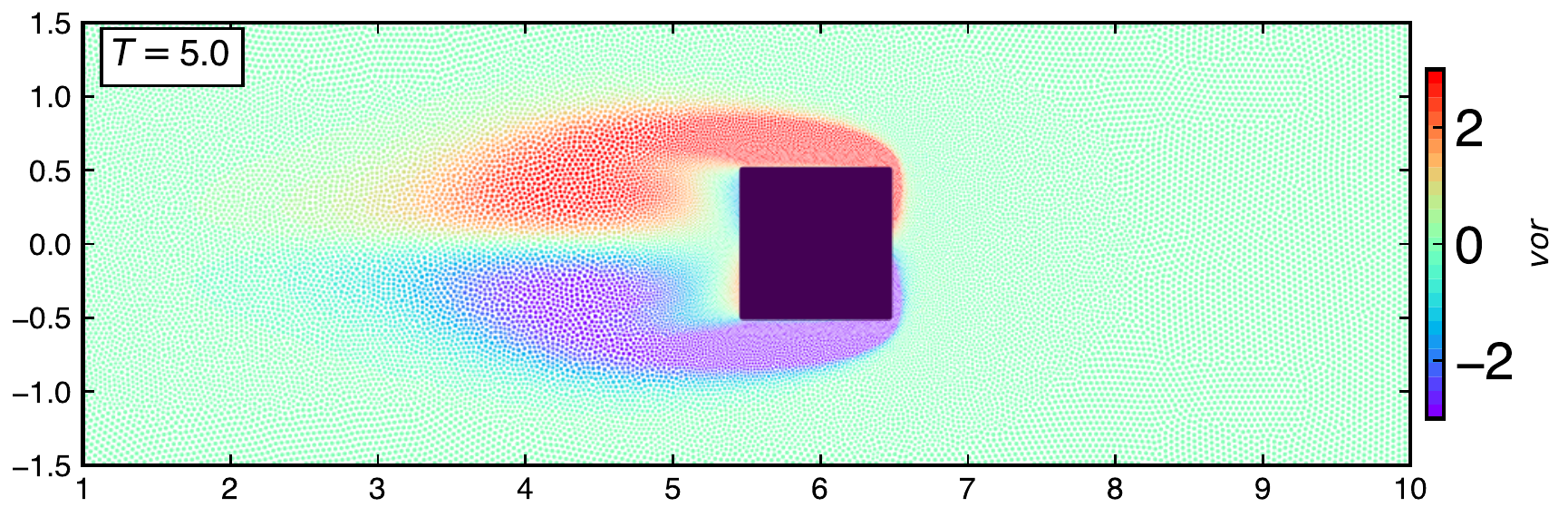}
  \end{subfigure}
  \begin{subfigure}{0.48\textwidth}
    \includegraphics[width=\textwidth]{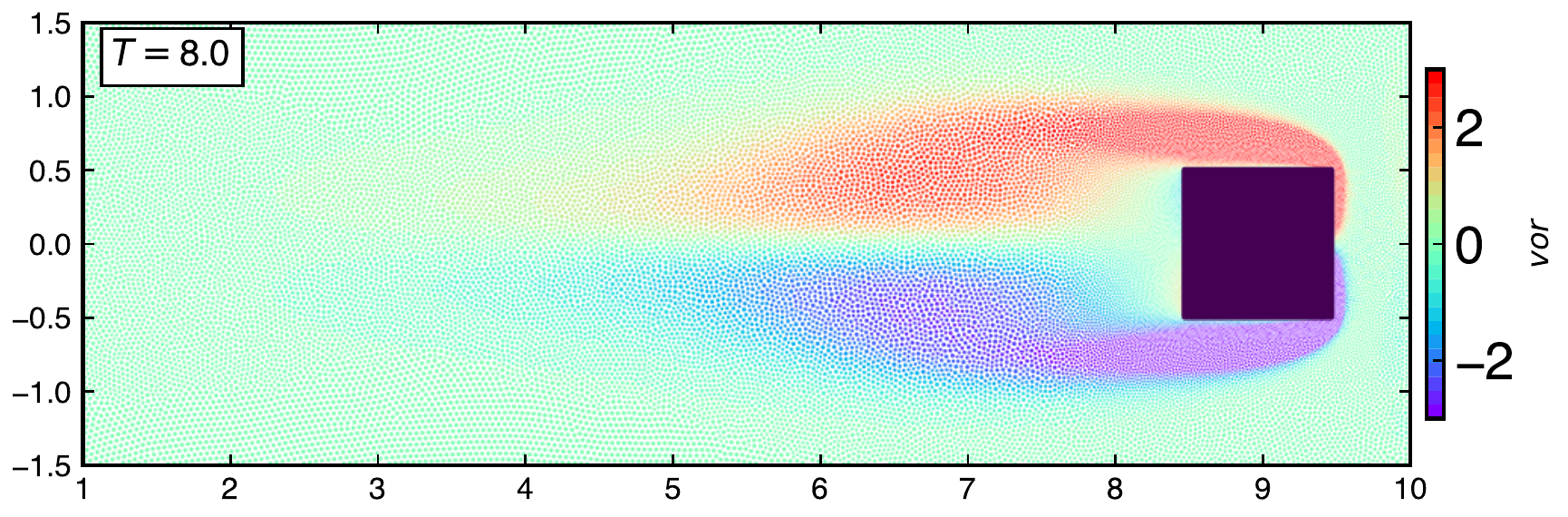}
  \end{subfigure}
  \caption{Particle distribution of moving square problem at $T = 5$ (left), and
    $T = 8$ (right) for $Re = 100$ showing: velocity magnitude (top), pressure
    (middle), and vorticity (bottom). The lowest resolution used is
    $D / \Delta x_{\max} = 25$ and the highest resolution used is
    $D / \Delta x_{\min} = 80$.}%
  \label{fig:ms-dist}
\end{figure}
%%%%%================================================================================

\Cref{fig:ms-dist} show the particle distribution with color indicating
pressure, vorticity and velocity magnitude taken at $T = 5$ and $T = 8$. The
qualitative observations illustrate the generation, diffusion, and convection of
the unsteady vorticity proving the capability of the adaptive EDAC-SPH scheme.

\subsection{Removal of background particles}%
\label{subsec:no-bg}

\citet{muta2021efficient} used a collection of particles called ``background''
particles to define the particle spacing $\Delta s$ and reference mass. But in
this work, we show that the background particles are not necessary to set this
reference mass. Updating the reference mass on the particles themselves
produces results closely matching those produced by the original algorithm using
background particles. This reduces memory usage by removing background
particles's storage, and a small reduction in the computational time. We perform
two simulations to establish the no background particle approach. We first
consider the flow past a circular cylinder at $Re = 1000$, the simulation
parameters and domain sizes are the same as defined in
\cite{muta2021efficient}. Furthermore, we compare the results with the
established vortex method results \cite{koumoutsakos1995,ramachandran2004}. We
then consider the flow past a moving square in a box at $Re = 150$. Refer to
\cref{subsec:ms-mc} for the simulation parameters. We compare the results to
the incompressible FDNS simulation results \cite{colicchio2006fluid}.  
In this case, the geometry is moving. In addition, we use solution adaptivity 
to refine the particles to the lowest resolution where the vorticity 
exceeds 5\% of the maximum vorticity in the simulation.

\begin{figure}[!ht]
  \centering
  \begin{subfigure}{0.49\textwidth}
  \includegraphics[width=\textwidth]{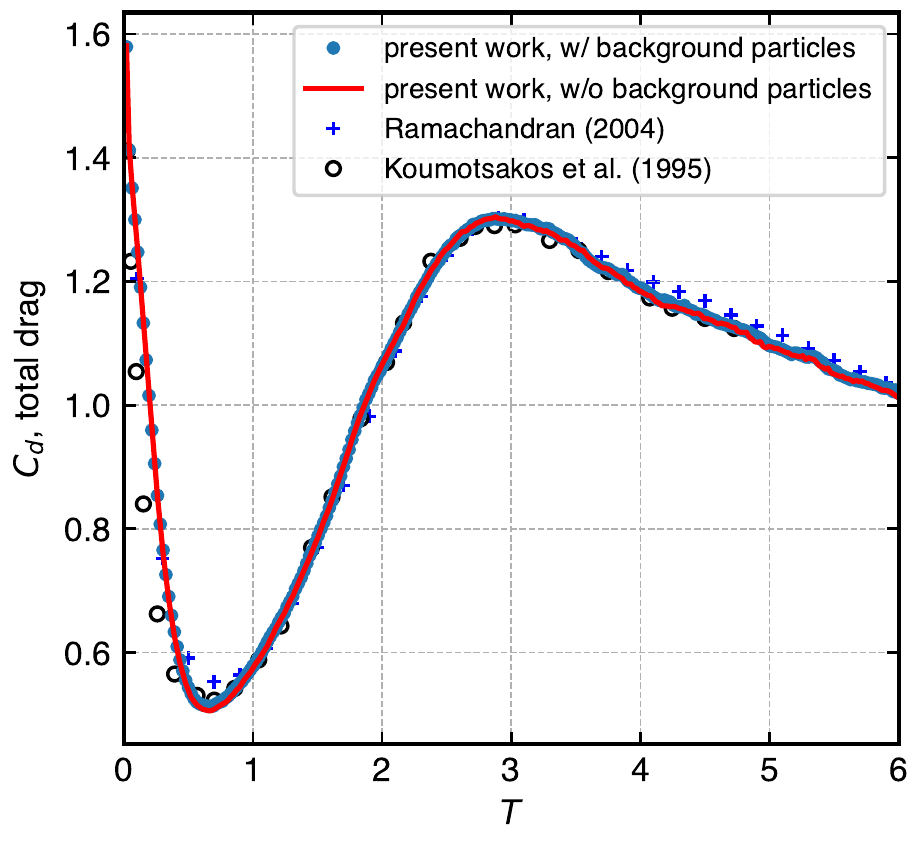}%
  \subcaption{}
  \label{fig:no-bg-fpc}
  \end{subfigure}
  \begin{subfigure}{0.49\textwidth}
  \includegraphics[width=\textwidth]{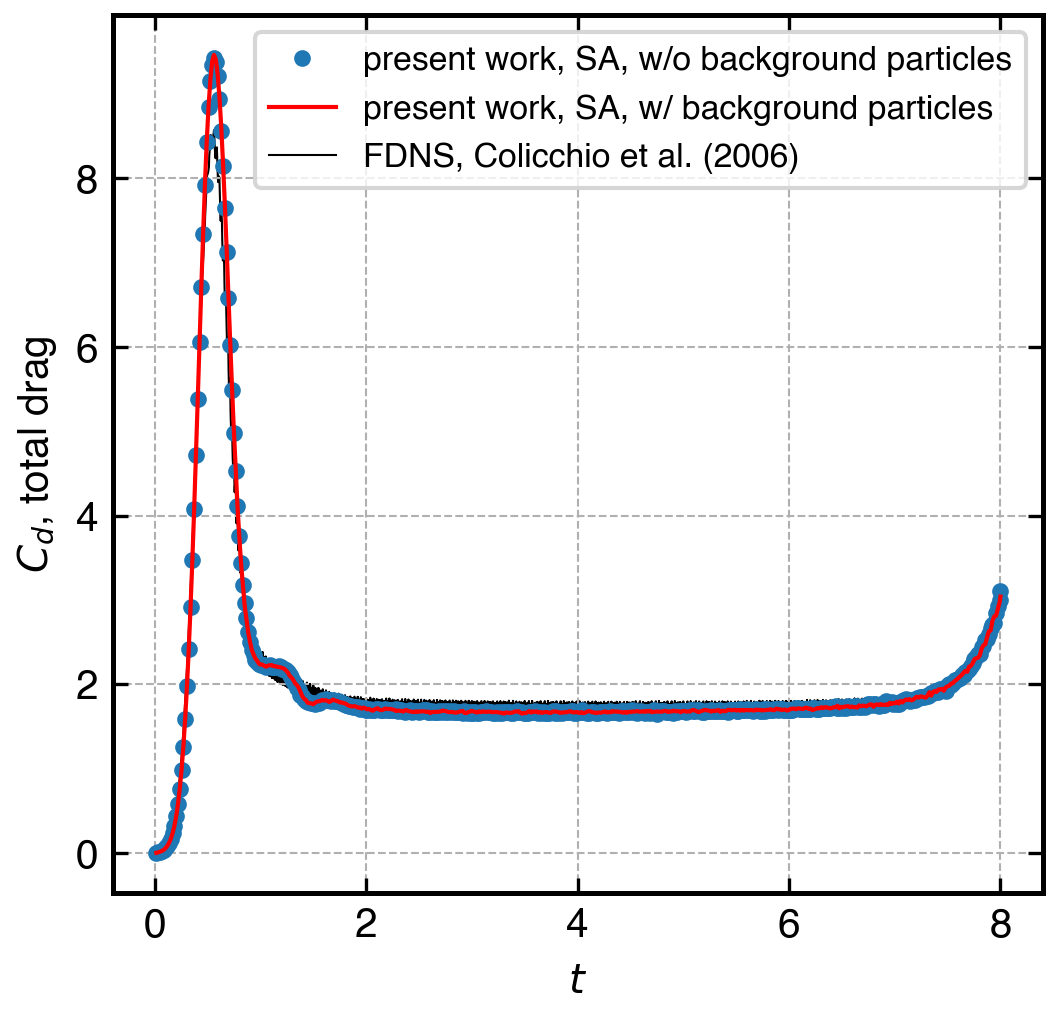}
  \subcaption{}
  \label{fig:no-bg-ms}
  \end{subfigure}
  \caption{(a) Comparison of the drag coefficient vs time with and without the use
    of background particles to that of the vortex method
    results\cite{koumoutsakos1995,ramachandran2004}. (b) Time history of the drag
    coefficient of a moving square simulated using solution adaptivity (SA) at $Re = 150$ 
    compared against FDNS results\cite{colicchio2006fluid}.}%
  \label{fig:no-bg-vs-bg}
\end{figure}

\begin{figure}[!ht]
  \centering
  \begin{subfigure}{0.49\textwidth}
  \includegraphics[width=\textwidth]{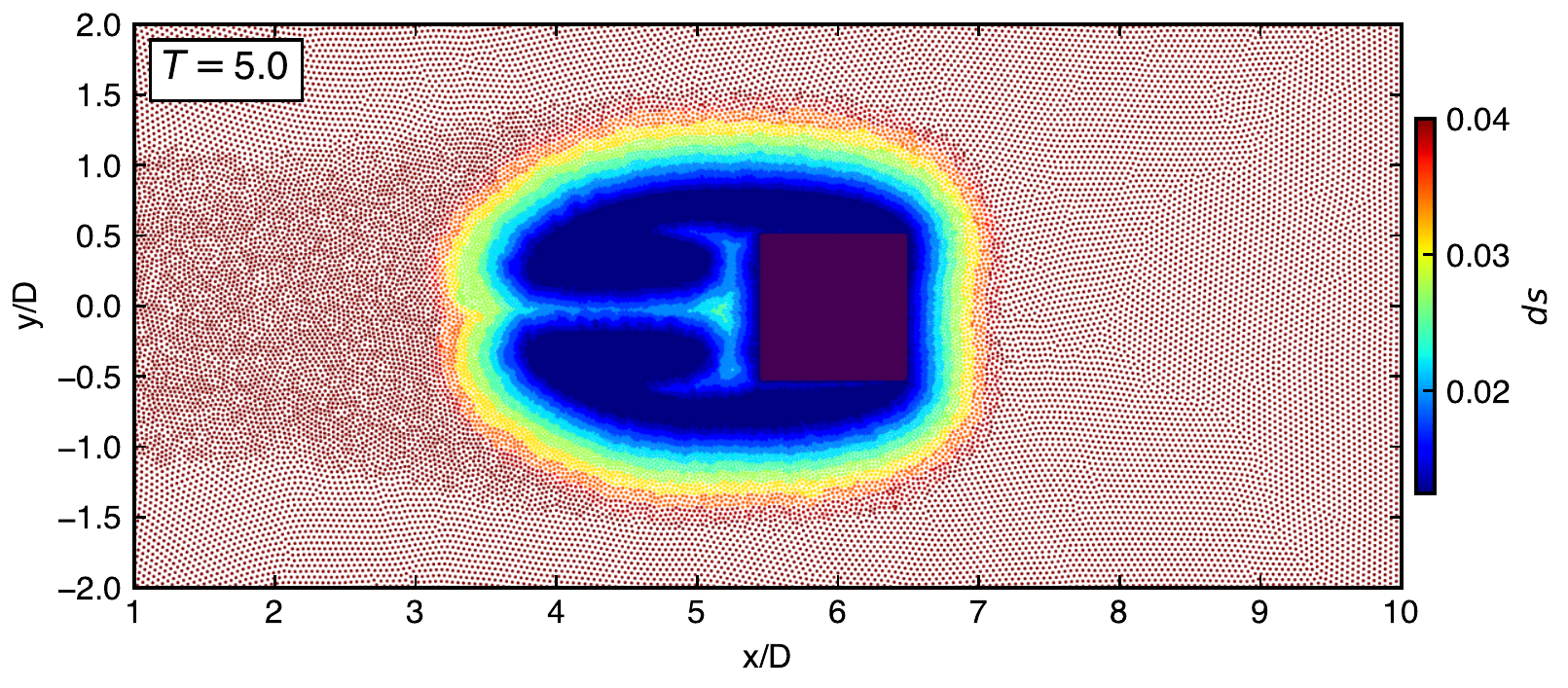}
  \subcaption{}
  \label{fig:no-bg-ds-pplot}
  \end{subfigure}
  \begin{subfigure}{0.49\textwidth}
  \includegraphics[width=\textwidth]{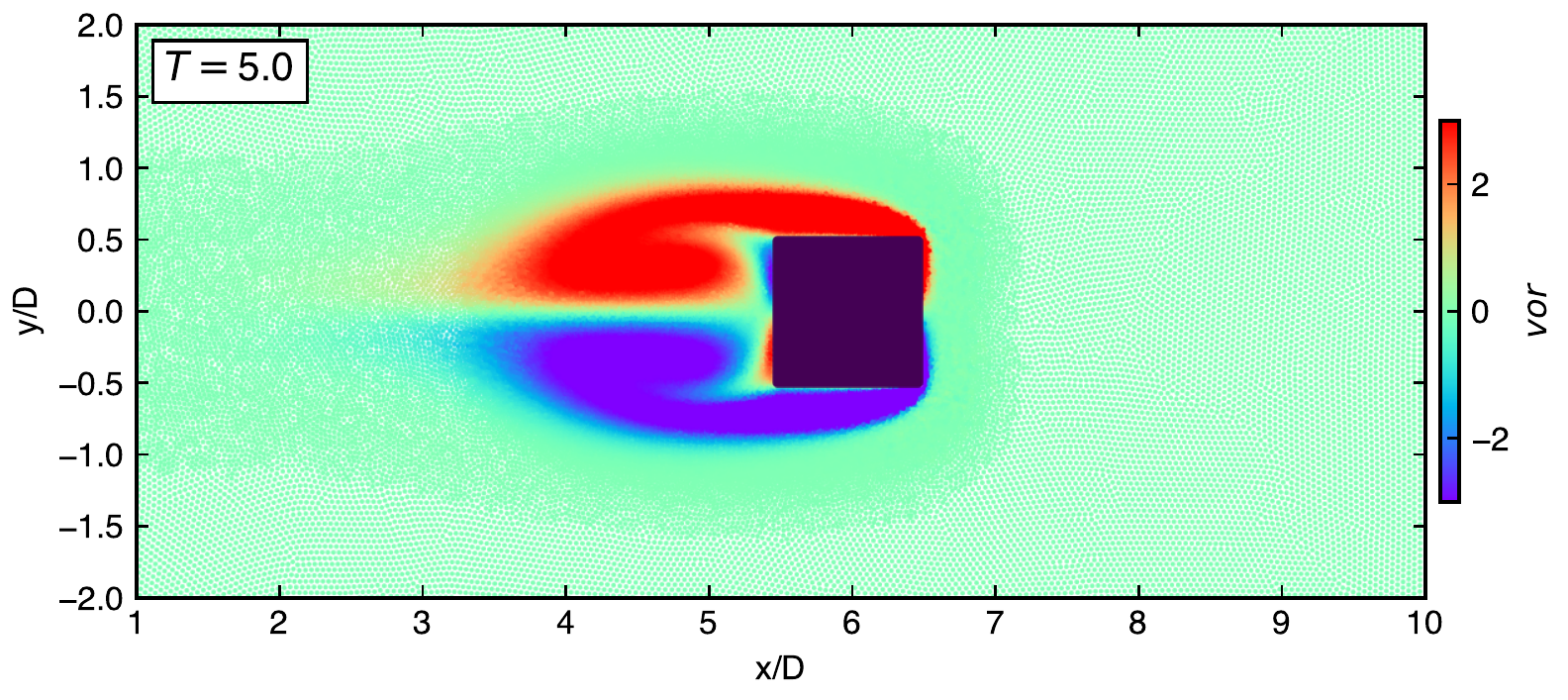}
  \subcaption{}
  \label{fig:no-bg-vor-pplot}
  \end{subfigure}
  \caption{(a) $\Delta s$ distribution (b) Vorticity distribution for the flow
    past a moving square at $Re = 150$, the highest resolution
    $D / \Delta x_{\min}$ is 80 and the lowest resolution $D / \Delta x_{\max}$
    is 25, and a $C_r$ spacing ratio of 1.12 is used.}%
  \label{fig:no-bg-pplots}
\end{figure}

\Cref{fig:no-bg-fpc} shows the time history of the drag coefficient of the
flow past cylinder problem without solution adaptivity. The results of adaptive 
EDAC-SPH are computed with and without use of background particles, and both 
results match very well with the reference data from vortex method simulations.
\Cref{fig:no-bg-ms} shows the time history of the
drag coefficient of the moving square problem using solution adaptivity.
This further confirms a good match when not using the background particles.
\Cref{fig:no-bg-ds-pplot} we show the distribution of the spacing $\Delta s$
without using background particles and in \cref{fig:no-bg-vor-pplot} we show
the vorticity distribution. It can be seen that the regions satisfying the
solution adaptivity criteria are refined to the highest resolution, and the
particles closest to the solid square particles are refined to the highest level.
%A good match can be seen between the adaptive EDAC-SPH results with and without background to the vortex method simulations. 
%

\FloatBarrier%
\subsection{Flow around an oscillating single cylinder}%
\label{subsec:fpc-moving}

The numerical examination of flow around an oscillating single cylinder is
helpful for the simulation and deeper analysis of two or more oscillating
cylinders. The study of flow over a cylinder is important for the study and
design of flow over a car, flow over aircraft, flow over airfoil blades, or
flow over a pillar for the design of bridges \cite{kang2003characteristics,
  bao2013flow, sumner2010two, stringer2014unsteady, can2020development}.

Bao et al., \cite{bao2013flow} classify the flow around the oscillatory motion
of cylinders into lock-on and unlock-on states which are also characteristic
of two cylinder flows. The classification depends on the shedding and
oscillating frequencies for which the transition occurs around a frequency
ratio of one.

We analyze the oscillatory motion of a single cylinder subject to cross-flow at
$Re = 100$. The sinusoidal motion of the cylinder is defined by
$Y = A \sin(\omega t)$, where $A$ is the amplitude, $\omega=2\pi f$ the angular
velocity, and $f$ the frequency of oscillation. Two amplitudes of oscillation
(low and high) and frequency ratio $(f_R)$ ranging from low to high are chosen
for the simulation. Here $f_R=f/f_{0}$, where $f_0$ is a natural vortex shedding
which is $0.165$ for a single cylinder oscillating at $Re=100$. The values are
$A = (0.25, 1.25)D$ and $f = (0.5, 1.5)f_{0}$.

\begin{figure}[!ht]
  \centering
  \begin{subfigure}{0.45\textwidth}
    \includegraphics[width=\textwidth]{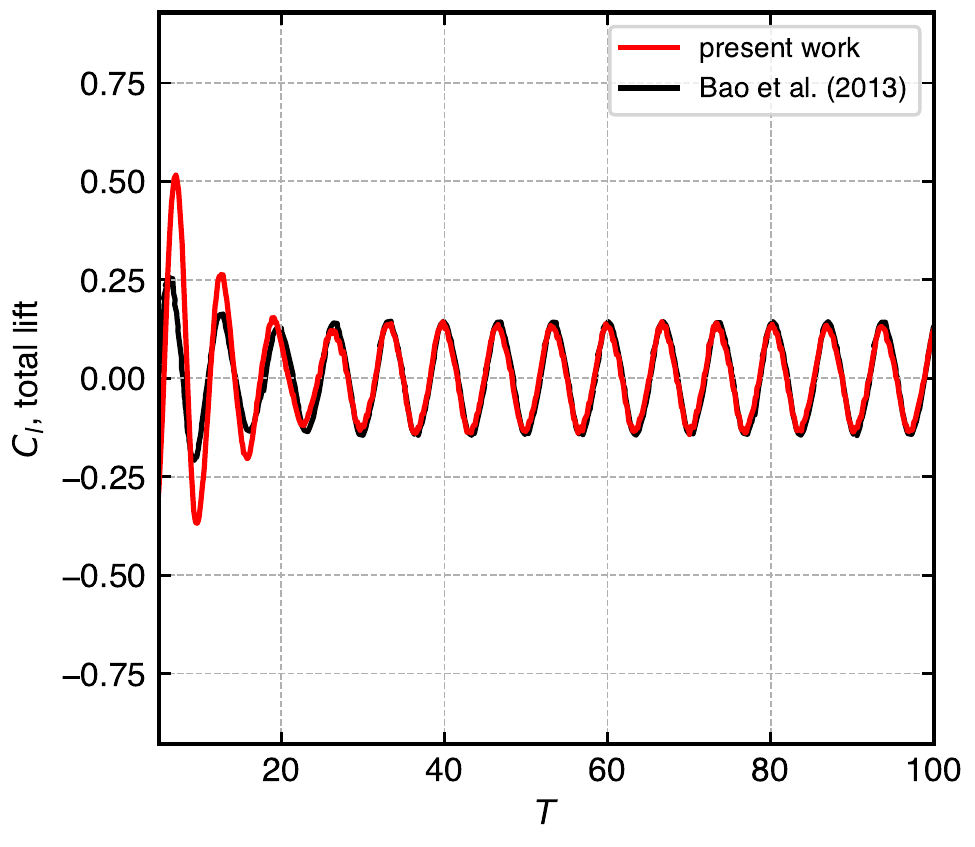}%
    \subcaption{}%
    \label{fig:bao_1cl-a}
  \end{subfigure}
  \begin{subfigure}{0.45\textwidth}
    \includegraphics[width=\textwidth]{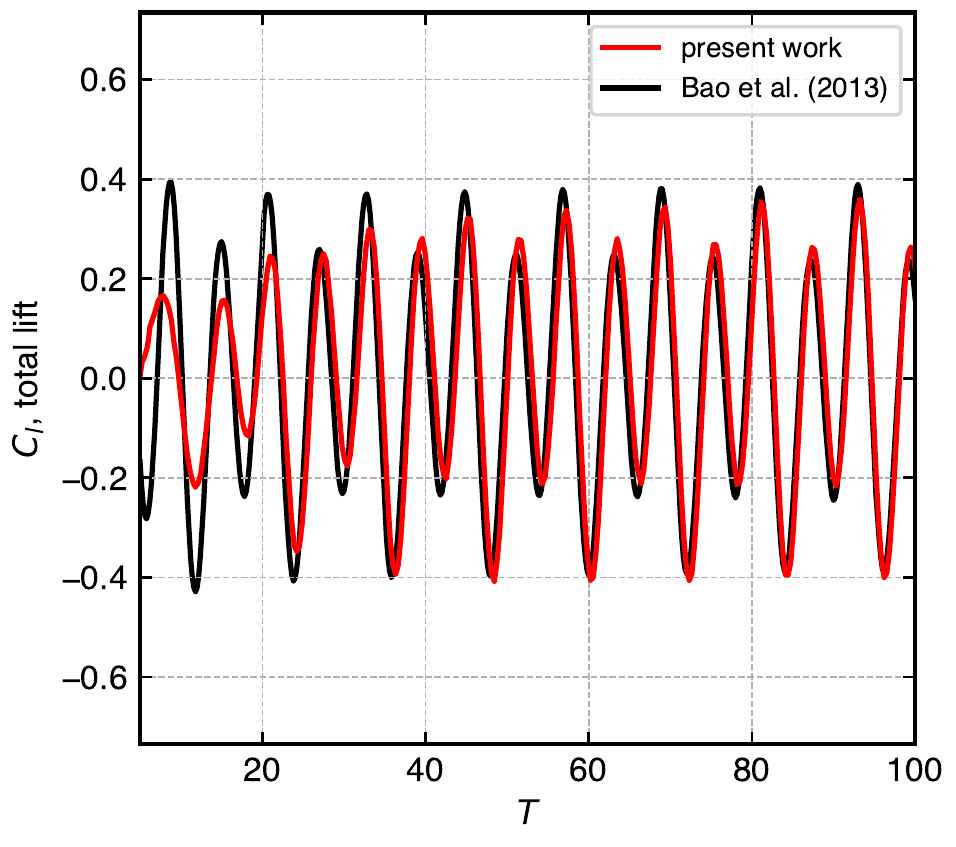}%
    \subcaption{}%
    \label{fig:bao_1cl-b}
  \end{subfigure}
  \\
  \begin{subfigure}{0.45\textwidth}
    \includegraphics[width=\textwidth]{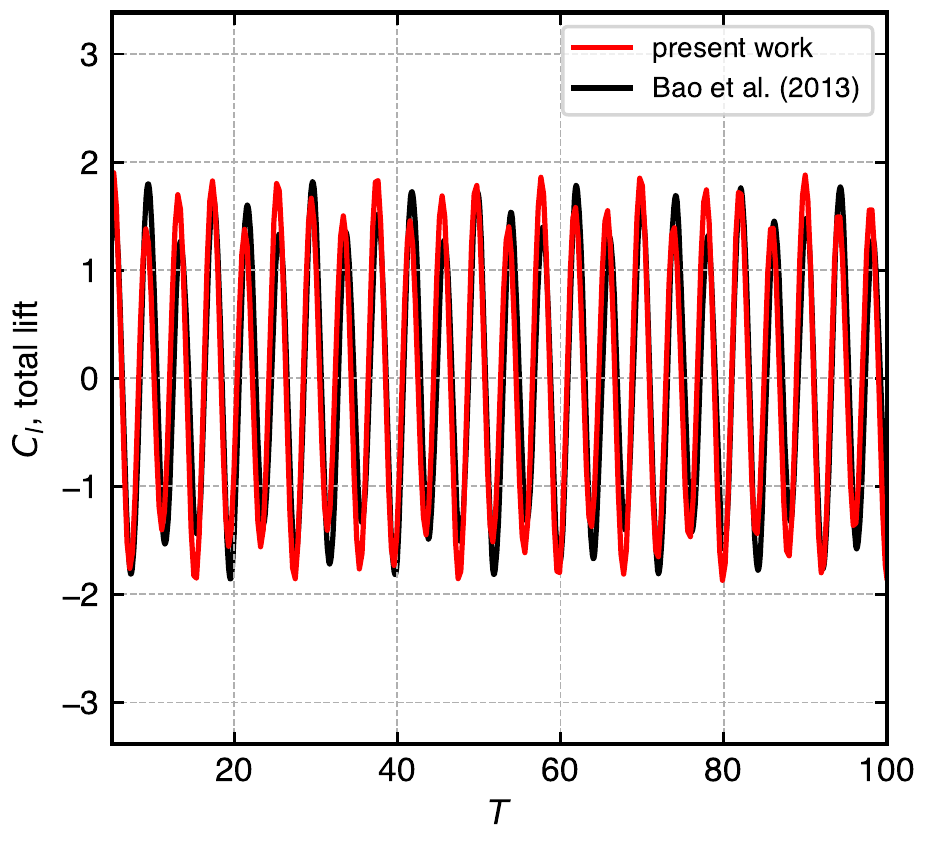}%
    \subcaption{}%
    \label{fig:bao_1cl-c}
  \end{subfigure}
  \begin{subfigure}{0.45\textwidth}
    \includegraphics[width=\textwidth]{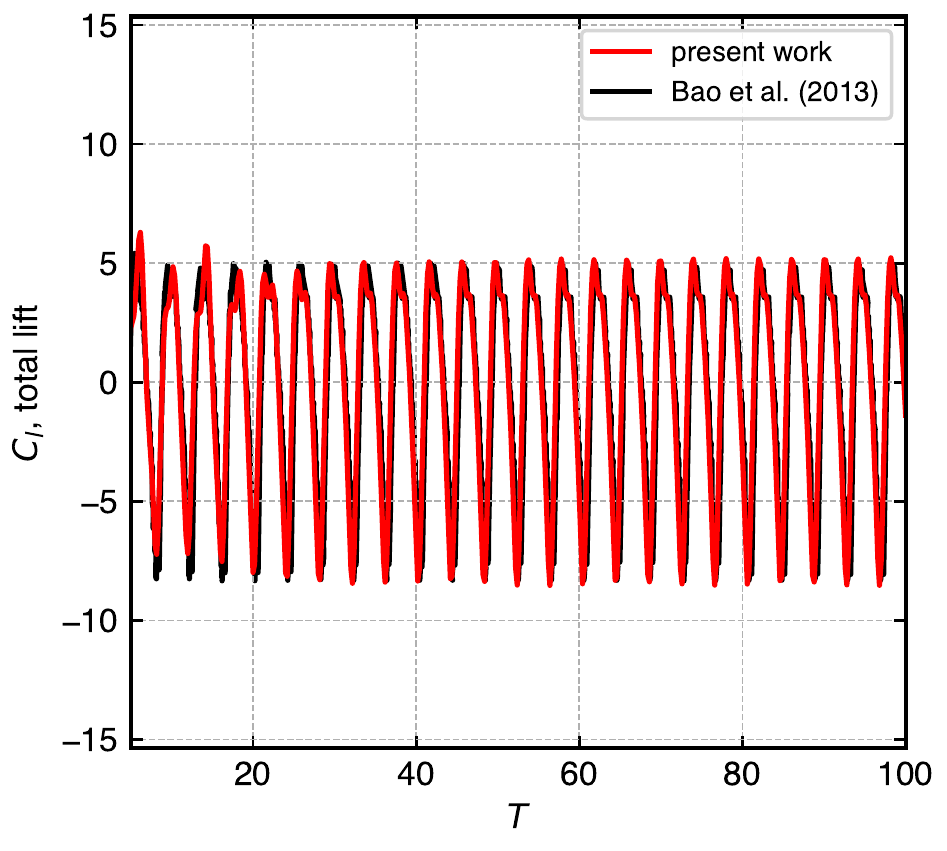}%
    \subcaption{}%
    \label{fig:bao_1cl-d}
  \end{subfigure}
  \caption{Time history of the lift coefficient for a single
    cylinder oscillating at $Re=100$ with different oscillating frequencies,
    Lock-on state: (a) ($A/D, f_R)=(0.25, 0.9)$; Unlock-on state: (b)
    ($A/D, f_R)=(0.25, 0.5)$, (c) ($A/D, f_R)=(0.25, 1.5)$, (d)
    ($A/D, f_R)=(1.25, 1.5)$, compared with the results of
    \citet{bao2013flow}.}%
  \label{fig:bao_1cl}
\end{figure}
\Cref{fig:bao_1cl} shows the time history of the fluctuating lift
coefficient for four combinations of amplitude and oscillation frequency. The
range includes lock-on and unlock-on states. In \cref{fig:bao_1cl-a}
$(A/D, f_R) = (0.25, 0.5)$, the simulated results initially show faster
transition and match with \cite{bao2013flow} capturing the peak-to-peak
amplitude and all trends. In \cref{fig:bao_1cl-b}
$(A/D, f_R) = (0.25, 1.1)$, our simulation depicts pure sinusoidal oscillation
from the beginning, matching very well for $t\geq 80$ seconds. The fluctuating
lift coefficient beats with a stronger magnitude for higher oscillating
frequencies. \Cref{fig:bao_1cl-c} $(A/D, f_R) = (0.25, 1.5)$, takes a
combination of the smallest amplitude ratio with the highest value of the
oscillating frequency, the results match very well from the beginning. In
\cref{fig:bao_1cl-d} $(A/D, f_R) = (1.25, 1.5)$, both parameters of the
amplitude and oscillating frequency are on the higher side. The results of this
simulation capture all the trends and peaks of the lift coefficients fluctuation
and are in good comparison with the reference data.

\subsection{Performance analysis}%
\label{subsec:perf}

In this section, we discuss the parallel performance of the adaptive EDAC-SPH
implemented using the framework of PySPH. To show that our adaptive algorithm is
fully parallel, we run the moving square test case with the highest resolution
of $\Delta x_{\min} = 0.0075$ and the lowest resolution of
$\Delta x_{\max} = 0.2$, giving us 57k particles. As opposed to the gradual
start in the moving square problem described in \cref{subsec:ms-mc} we move the
solid with a uniform velocity of 1 ms\textsuperscript{-1}. We proceed to test
this problem on a 6-Core Intel Core i7 CPU. We compare the time taken to run
1000 time-steps on a serial single-core single-thread execution with a parallel
(OpenMP) execution on the multi-core CPU with 4-threads and 6-threads.

\begin{table}[!ht]
  \centering
  \begin{tabular}{p{0.45\linewidth}lll}
    \toprule
    Component & 1-thread & 4-threads & 6-threads \\
    \midrule
    splitting \& merging & 24.74 s & 7.91 s & 6.64 s\\
    removing \& adding particles & 0.63 s & 0.61 s & 0.70 s \\
    EDAC scheme & 260.77 s & 78.44 s & 64.38 s\\
    \bottomrule
  \end{tabular}
  \caption{The time break-up for running 1000 time-steps of the moving square
    problem executed on a serial single-thread, and parallel 4-threads and
    6-threads on a 6-core CPU.}%
  \label{tab:perf-breakup}
\end{table}
\Cref{tab:perf-breakup} shows the time taken by various algorithms described in
this manuscript along with the time taken by the adaptive EDAC-SPH scheme. The
splitting and merging component constitutes the Update Spacing
\cref{alg:update-h}, splitting \cref{alg:splitting}, and merging
\cref{alg:merging}; the removing and adding of particles constitutes for the
time taken by the removal of the merged particles’ indices from the data
structure, and addition of new particles to the data structure; the EDAC scheme
constitutes the overall time taken by the SPH formulation.

\Cref{tab:perf-breakup} shows that the multi-thread implementation scales well
in parallel with a scale-up of 3.13x on 4-threads and 3.73x on 6-threads. The
time taken for removing and adding particles to the data structure constitutes
less than 2.5\% of the adaptive algorithm and this process is serial in
execution.  The present scheme again scales well with an increase in number of
threads. The scale-up is 3.32x on 4-threads and 4.05x on 6-threads. The parallel
execution of the adaptive algorithm constitutes 10--11\% of the adaptive EDAC-SPH
scheme.

The results of this section validate our formulation with the existing numerical
simulations.

\section{Complex motion demonstration}
\label{sec:demo}

Simulation of fluid flow around moving solid bodies is challenging and even
more challenging in complex motion scenarios such as the rotation of a complex
geometry and motions that combine pitching and translating. The adopted
adaptive EDAC-SPH method automatically refines the particles around solid bodies
performing any type of motion. Applications in complex moving cases are
demonstrated in this section. We consider the following examples for the
demonstration:
\begin{enumerate}
\item translating and pitching ellipse;
\item plunging ellipse with elliptical motion trajectory; and,
\item rotating S shape.
\end{enumerate}
An elliptic solid body and an S shape are used to illustrate the various
motions. The ellipse has a characteristic length of $1$ m and axis ratio
$0.4$. All the test cases use boundary conditions of the moving square problem
discussed in section \ref{subsec:ms-mc} for a stationary fluid.

\subsection{Translating and pitching ellipse}%
\label{subsec:trans-pitch}

The acceleration of the center of mass of the ellipse, $a_{\text{cm}}(t)$,
and the instantaneous pitch angle with respect to the horizontal, $\theta(t)$,
are given by,
\begin{equation}
  a_{\text{cm}}(t) = b_0 \exp\left(\frac{-{(t-b_1)}^2}{2c^2}\right) + d,
\label{eq:cmpx1a}
\end{equation}
\begin{equation}
  \theta(t) = \theta_0 \sin(\omega t + \psi)
  \label{eq:cmpx1b}
\end{equation}
where $b_0=2.8209512$, $b_1=0.525652151$, $c=0.14142151$, $d=-2.55580905e-08$,
and the pitch amplitude $\theta_0=10^{\circ}$ are the constants.

\begin{figure}[!htp]
  \centering
  \begin{subfigure}{0.5\textwidth}
    \includegraphics[width=\textwidth]{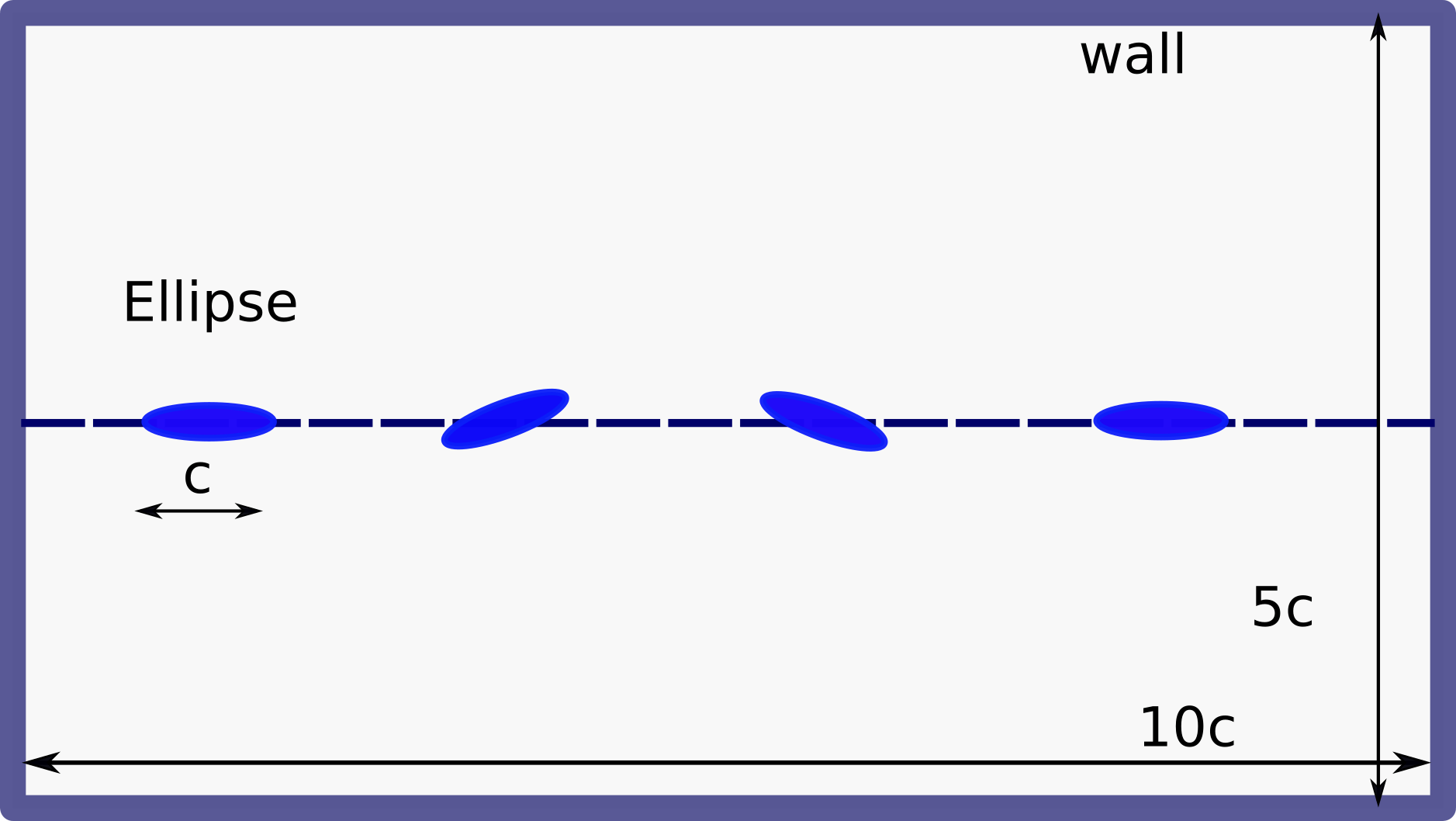}
  \end{subfigure}
  \caption{Translating pitching motion of an ellipse at four different times.}%
  \label{fig:elmove}
\end{figure}

We simulate the translating and pitching ellipse in a rectangular domain of size
$L_y \times L_x = 5c$ $ \times\ 10c$ at a Reynolds number of $550$. The ellipse
is started with a smooth acceleration in the $x$-axis starting from rest until
it reaches a steady maximum velocity of 1 m s\textsuperscript{-1}. The schematic
of the domain is shown in~\cref{fig:elmove}. The wall boundary condition is the
same as that used in the moving square problem of section
\ref{subsec:ms-mc}.
\begin{figure}[!htp]
  \begin{subfigure}{0.8\textwidth}
  \centering
   \hspace*{\fill}
  \begin{subfigure}{0.7\textwidth}
    \includegraphics[width=\textwidth]{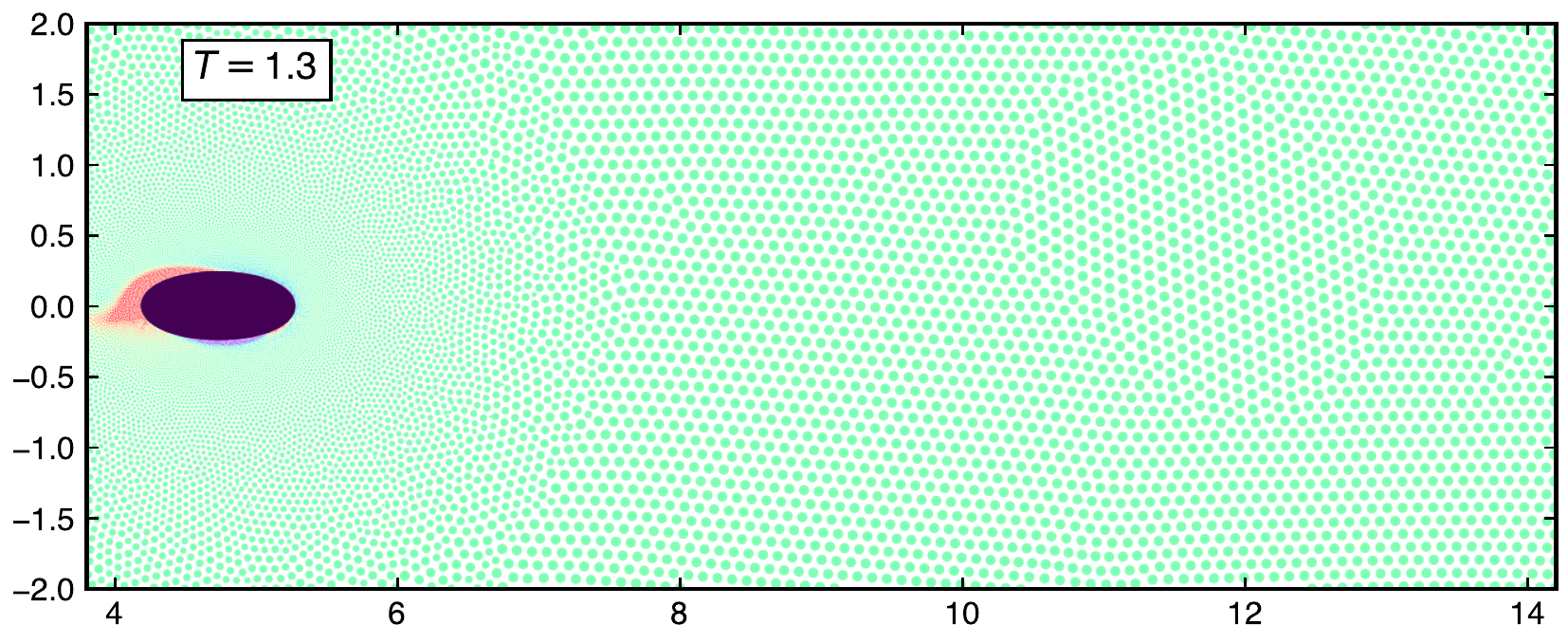}
    % \subcaption{}%
    \label{fig:ellpsetp:a}
  \end{subfigure}
 \\
   \hspace*{\fill}
  \begin{subfigure}{0.7\textwidth}
    \includegraphics[width=\textwidth]{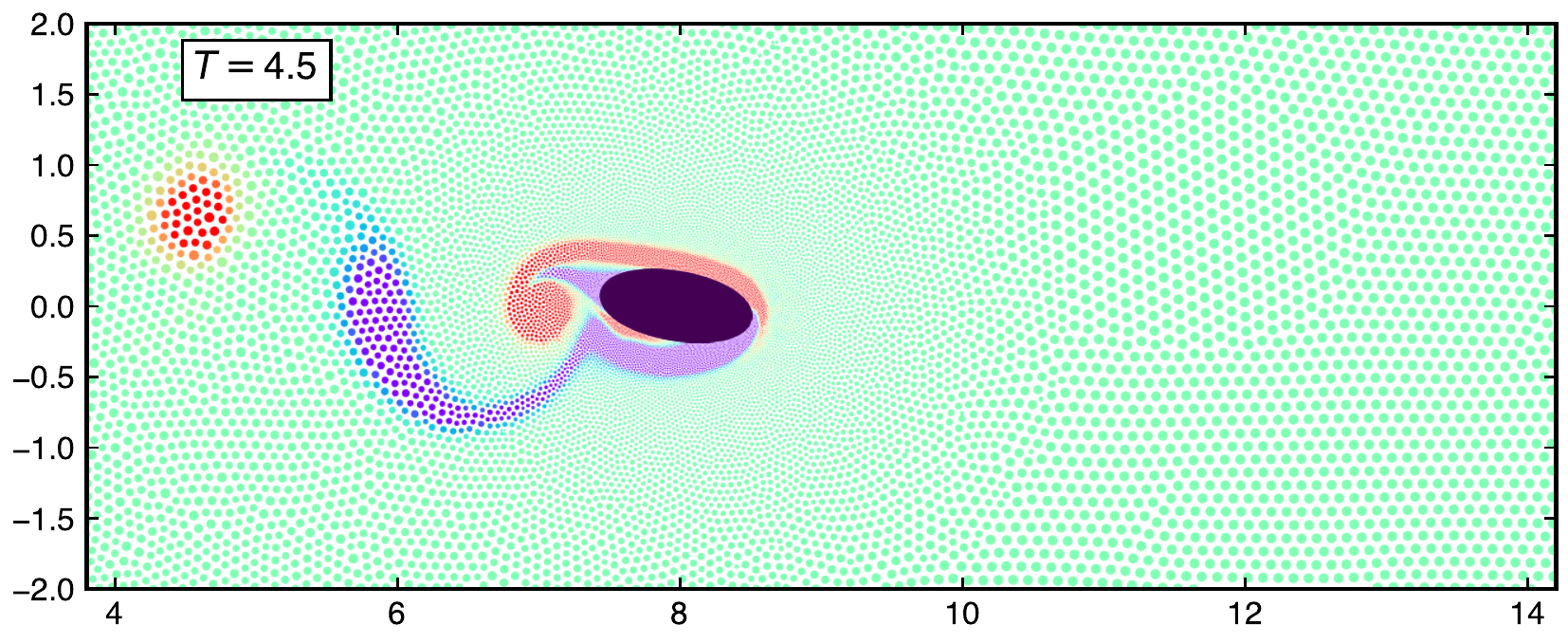}
    % \subcaption{}%
    \label{fig:ellpsetp:b}
  \end{subfigure}
  \\
   \hspace*{\fill}
  \begin{subfigure}{0.7\textwidth}
    \includegraphics[width=\textwidth]{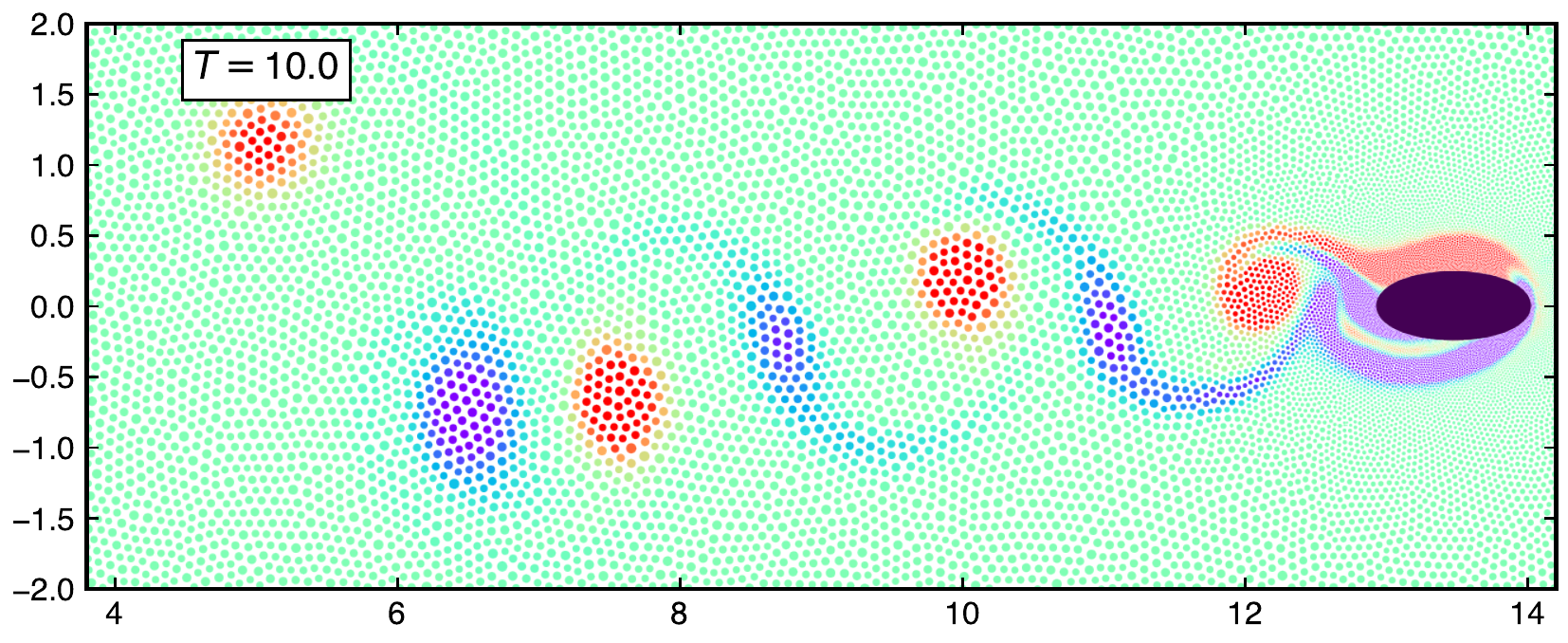}
    % \subcaption{}%
    \label{fig:ellpsetp:d}
  \end{subfigure}
  \end{subfigure}
  \begin{subfigure}{0.15\textwidth}
    \includegraphics[scale=0.6,left]{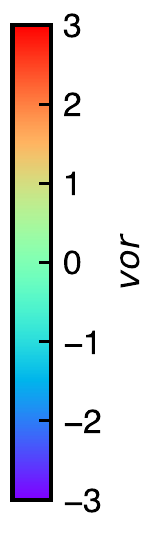}
  \end{subfigure}
  \caption{Vorticity distribution around a translating and pitching ellipse at
    three different times with the point size corresponding to the mass of the
    particle.}%
  \label{fig:translt}
\end{figure}
\Cref{fig:translt} shows the vorticity distribution around the ellipse at
different times. The top figure is at $T = 1.3$, where the ellipse is
$0^{\circ}$ from the horizontal. The middle figure is at $T = 4.5$, where is
ellipse is at $10^{\circ}$ from the horizontal and the bottom figure is at the
final time of $T = 8$.

\subsection{Plunging ellipse with elliptical motion trajectory}%
\label{subsec:flapping}

\begin{figure}[htp]
  \centering
  \begin{subfigure}{0.45\textwidth}
    \includegraphics[width=\textwidth]{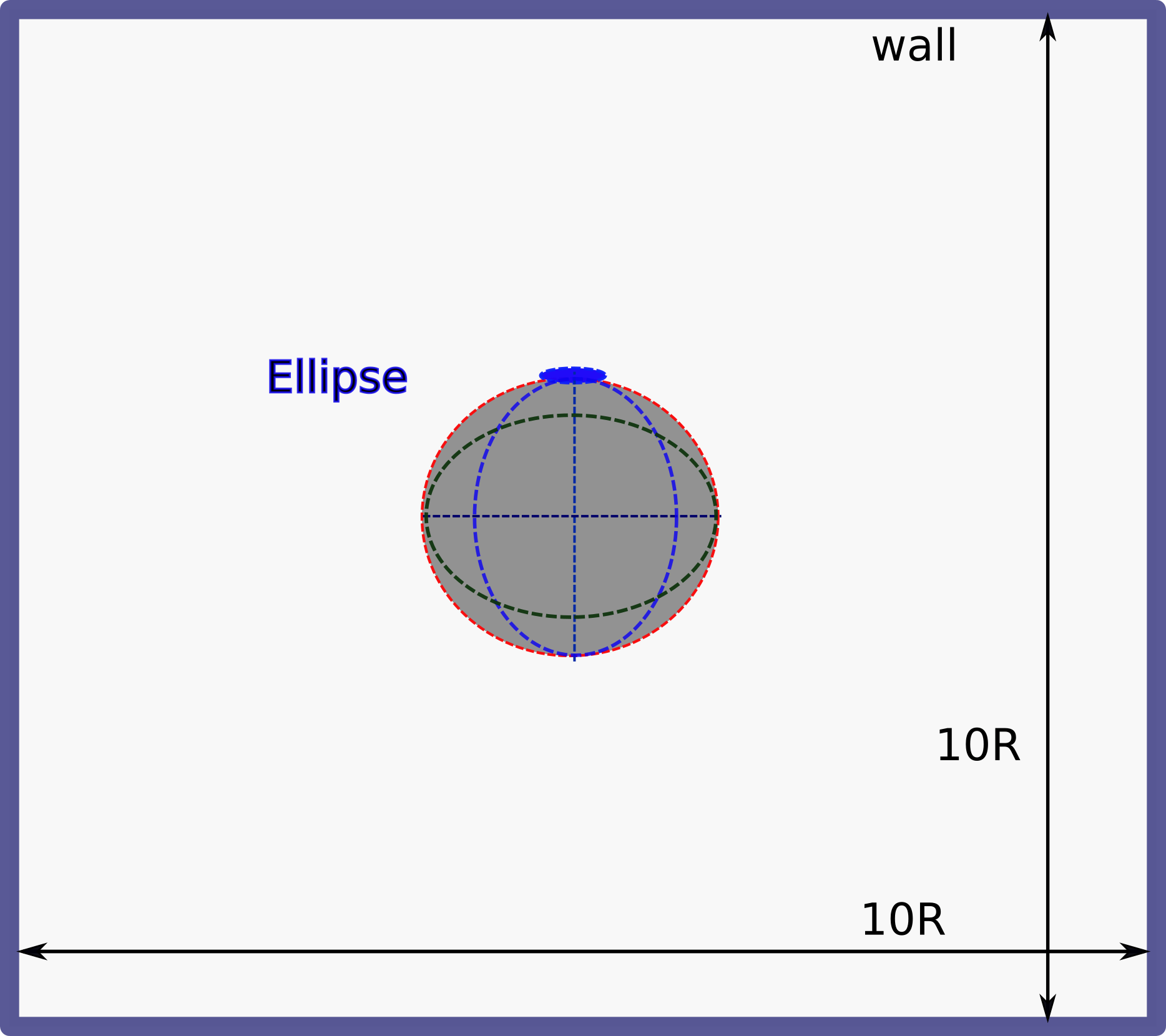}
    \subcaption{}
    \label{fig:el-flap1}
  \end{subfigure}
  \begin{subfigure}{0.45\textwidth}
    \includegraphics[width=\textwidth]{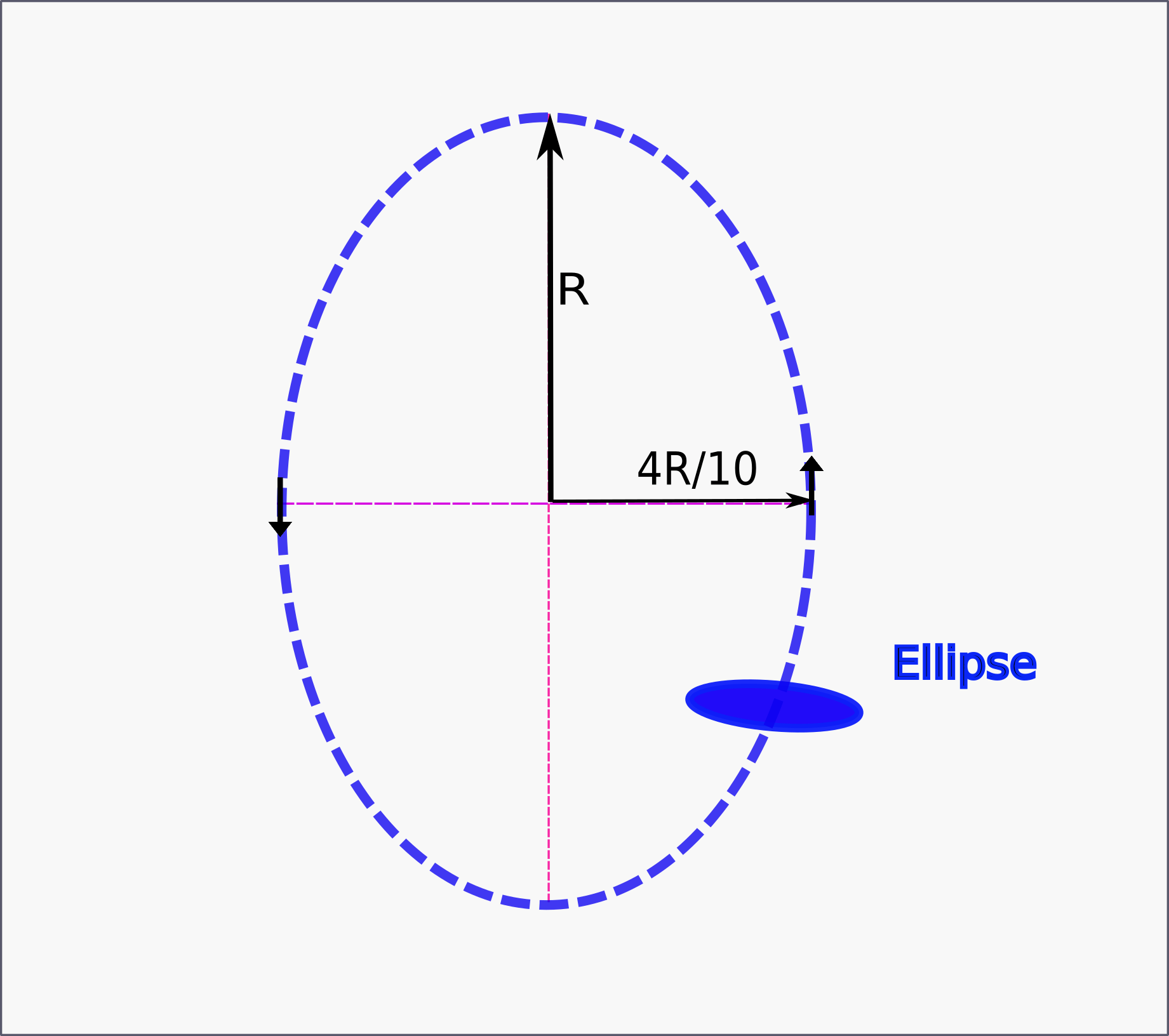}
    \subcaption{}
    \label{fig:el-flap2}
  \end{subfigure}
  \caption{Schematic diagram of the ellipse in plunging motion:
    (a) Domain size and boundary condition, (b) Enlarged view of the rotating area.}
  \label{fig:geom2}
\end{figure}

Kinematics of a flapping motion describes a vertical translation (heaving), a
horizontal oscillation, and pitching rotation. The amalgamation of these motions
constructs an elliptical trajectory (path) of the pitching ellipse. This kind of
kinematics is seen in the flying birds, aquatic animals, and has potential
applications in high-efficiency Micro Air Vehicles (MAVs), and other energy
harvesting vertical axis wind turbine (VAWT) designs
\cite{rozhdestvensky2003aerohydrodynamics, young2014review, esfahani2015fluid,
  wu2020review}. The pitching orientation and angle of attack can also vary
according to the type of motion undertaken for high-level performance. The
schematic diagram of ellipse motion is shown in \ref{fig:geom2}.
The motion equations for the vertical plunge, $(h(t)$, and the horizontal
sinusoidal motion, $w(t)$, of the ellipse are defined by,
\begin{equation}
  h(t) = h_0 \cos(\omega t),
  \quad
  w(t) = b h_0 \sin(\omega t),
\label{eq:cmpx2}
\end{equation}
where $h_0 = 2$, $\omega = 2\pi / 5$ is the angular velocity, and $b = 0.5$ is a
factor relating a horizontal motion with the vertical counterpart.

\begin{figure}[!htp]
  \begin{subfigure}{0.8\textwidth}
  \centering
   \hspace*{\fill}
  \begin{subfigure}{0.4\textwidth}
    \includegraphics[width=\textwidth]{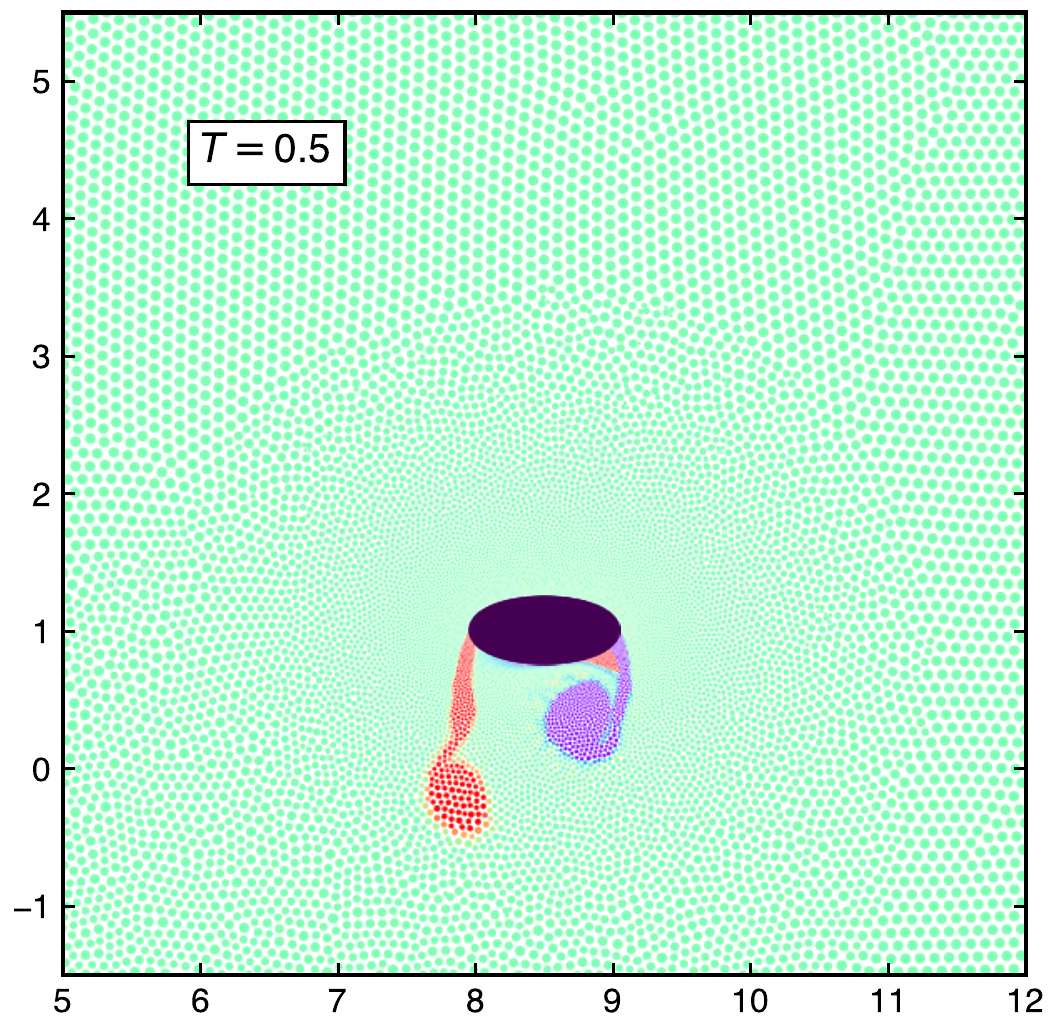}
    \subcaption{}\label{fig:plunge-a}
  \end{subfigure}
  \begin{subfigure}{0.4\textwidth}
    \includegraphics[width=\textwidth]{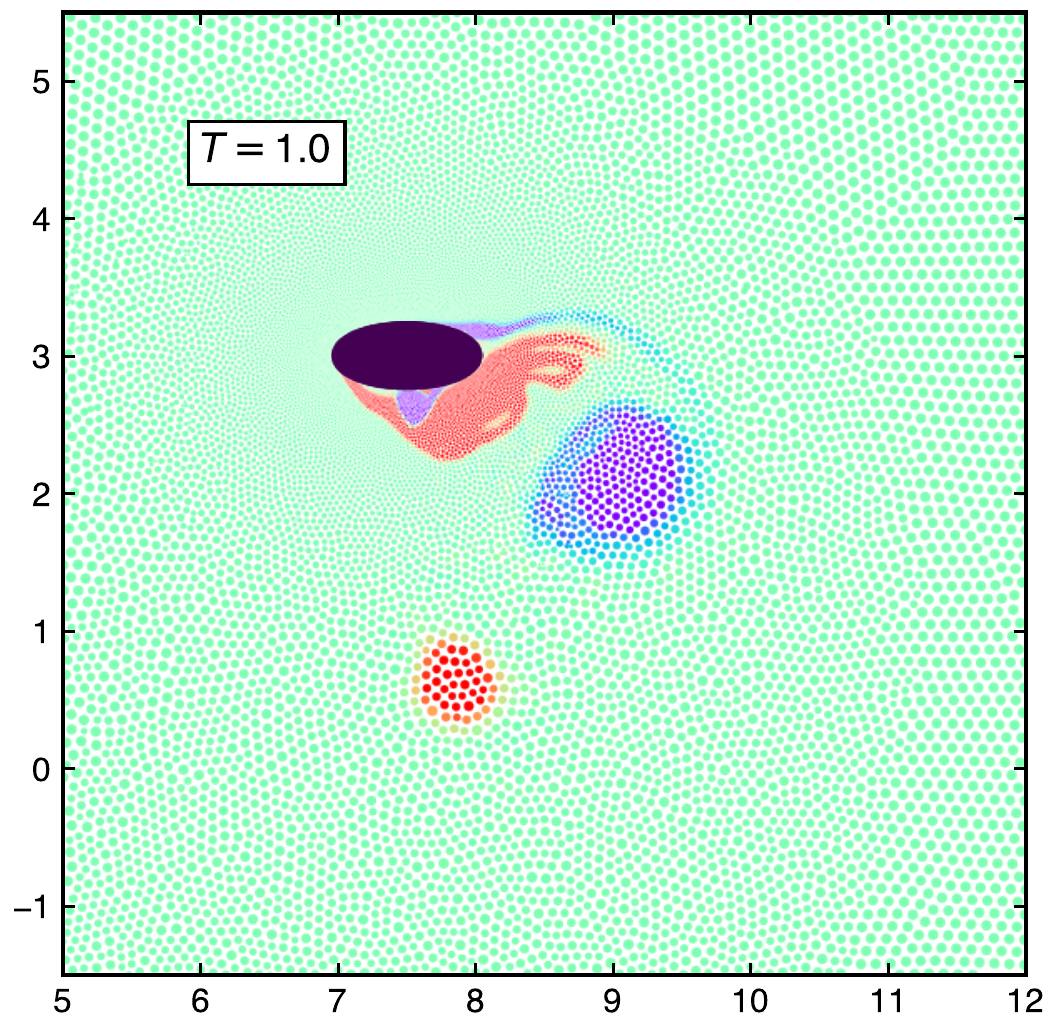}
    \subcaption{}\label{fig:plunge-b}
  \end{subfigure}
  \\
  \hspace*{\fill}
  \begin{subfigure}{0.4\textwidth}
    \includegraphics[width=\textwidth]{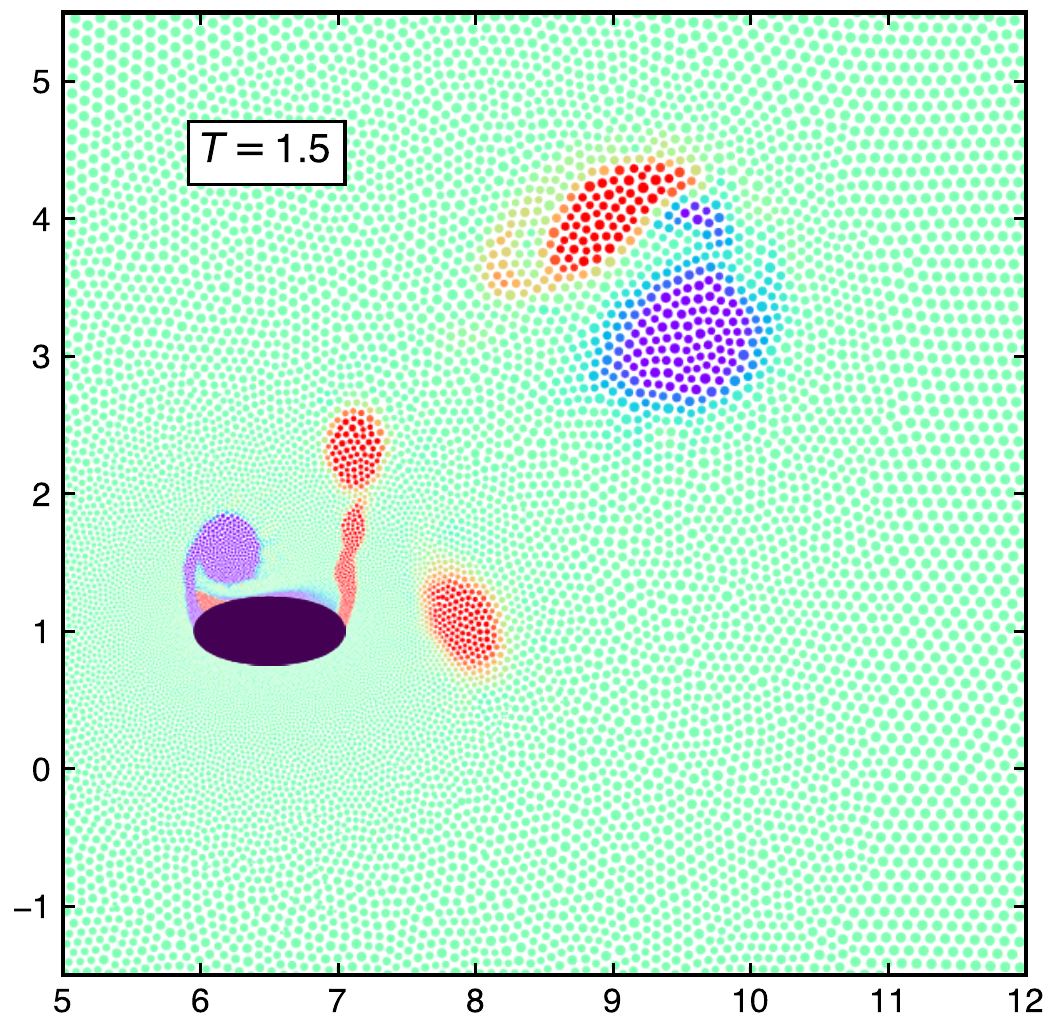}
    \subcaption{}\label{fig:plunge-c}
  \end{subfigure}
  \begin{subfigure}{0.4\textwidth}
    \includegraphics[width=\textwidth]{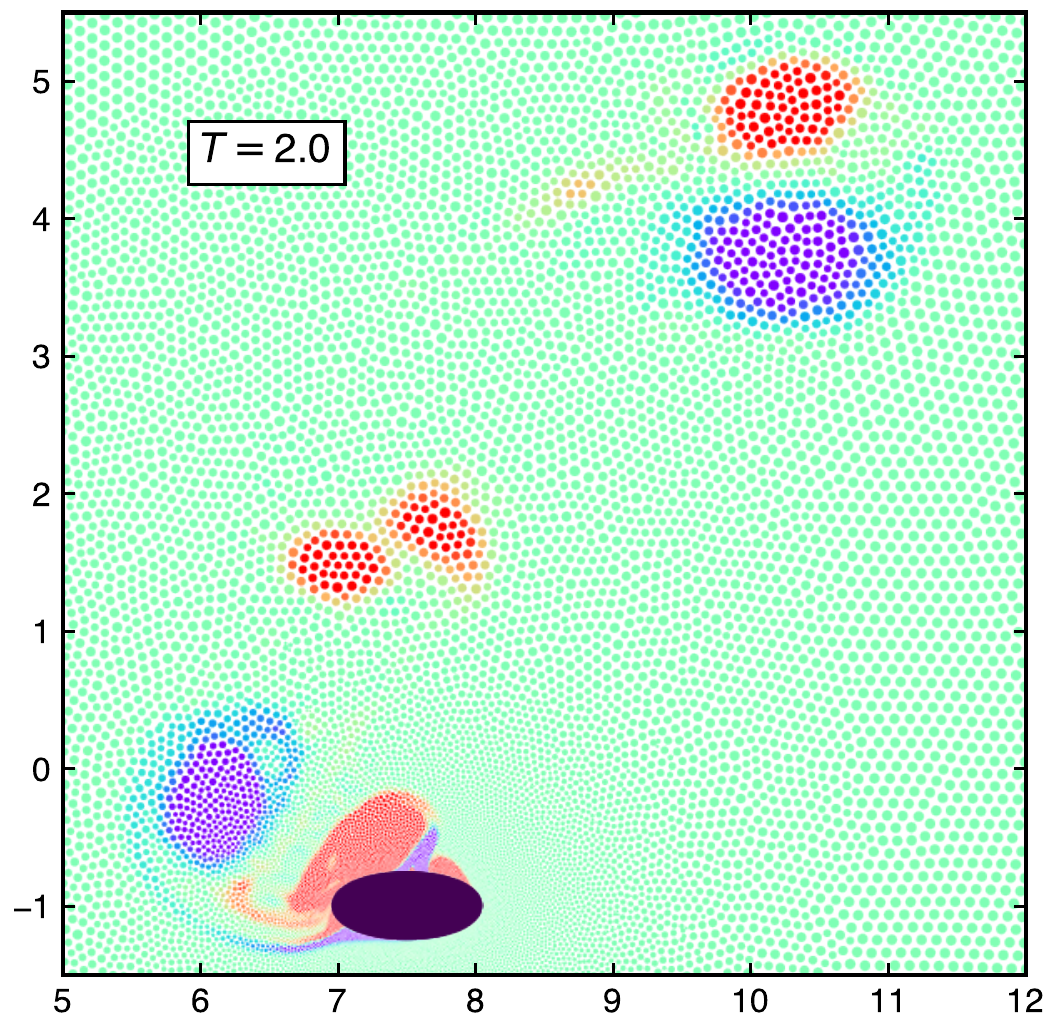}
    \subcaption{}\label{fig:plunge-d}
  \end{subfigure}
  \end{subfigure}
  \begin{subfigure}{0.09\textwidth}
    \includegraphics[scale=0.6,left]{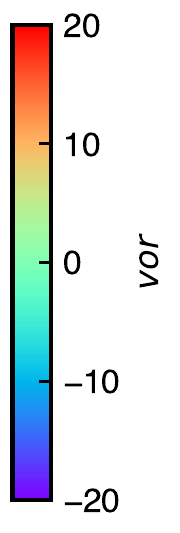}
  \end{subfigure}
  \caption{Vorticity distribution around the plunging ellipse at different time
    periods of two plunging cycles at $Re = 550$ at: (a) $T/4$, (b) $T/2$, (c)
    $3T/2$, and (d) $T$.  The results are computed using solution adaptivity,
    with lowest resolution of $D / \Delta x_{\max} = 25$, and highest resolution
    of $D / \Delta x_{\min} = 100$. The point size is proportional to the mass
    of the particle.}%
  \label{fig:plunge}
\end{figure}
\Cref{fig:plunge} shows the vorticity plots of a plunging ellipse undergoing the
mixed vertical-horizontal oscillatory motions forming an elliptical trajectory.
The motion of the solid ellipse is described using the equations in
\ref{eq:cmpx2}, of a stationary fluid with boundary conditions discussed in
\ref{subsec:ms-mc}.  The schematic diagram of the problem is illustrates in fig
\ref{fig:geom2} (a).  In \cref{fig:plunge-a} the ellipse is at the start of
upward stroke with backward motion. In \cref{fig:plunge-b} ellipse starts an
upward stroke with a forward motion at period $T/4$. At period $T/2$, the
ellipse starts the downward stroke with a forward motion. Similarly,
\cref{fig:plunge-c,fig:plunge-d} show the motion of the ellipse at different
time periods of $3T/2$, and $T$ respectively.

\subsection{Rotating S-shape}%
\label{subsec:sshape}

\begin{figure}[!htp]
  \centering
  \includegraphics[width=0.6\textwidth]{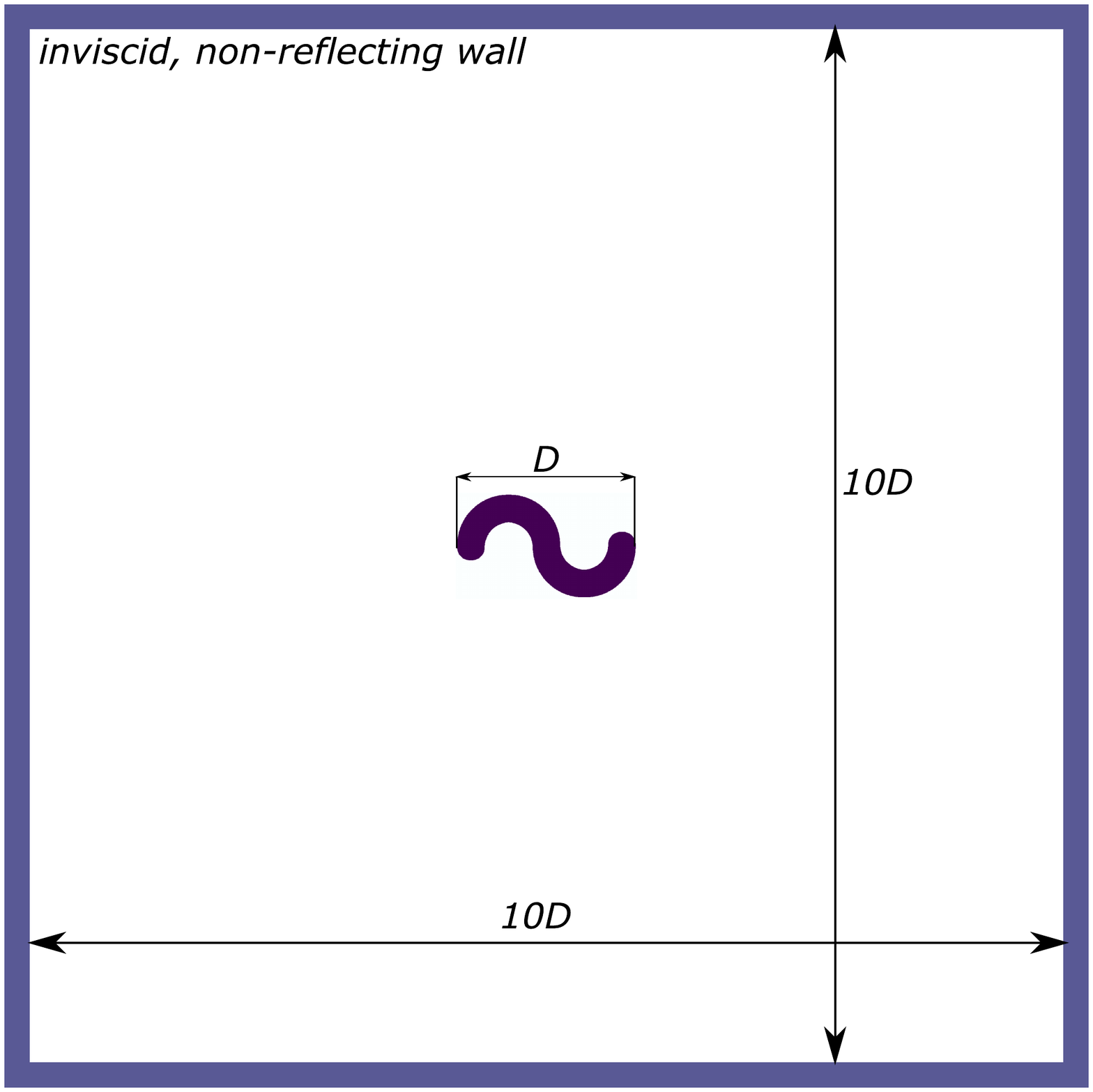}
  \caption{Schematic diagram of the rotating S-shape, where the outer diameter
    $D$ of the S-shape is 2 m.}%
  \label{fig:sshape-domain}
\end{figure}
\begin{figure}[!htp]
  \begin{subfigure}{0.8\textwidth}
   \hspace*{\fill}
  \centering
  \begin{subfigure}{0.4\textwidth}
    \includegraphics[width=\textwidth]{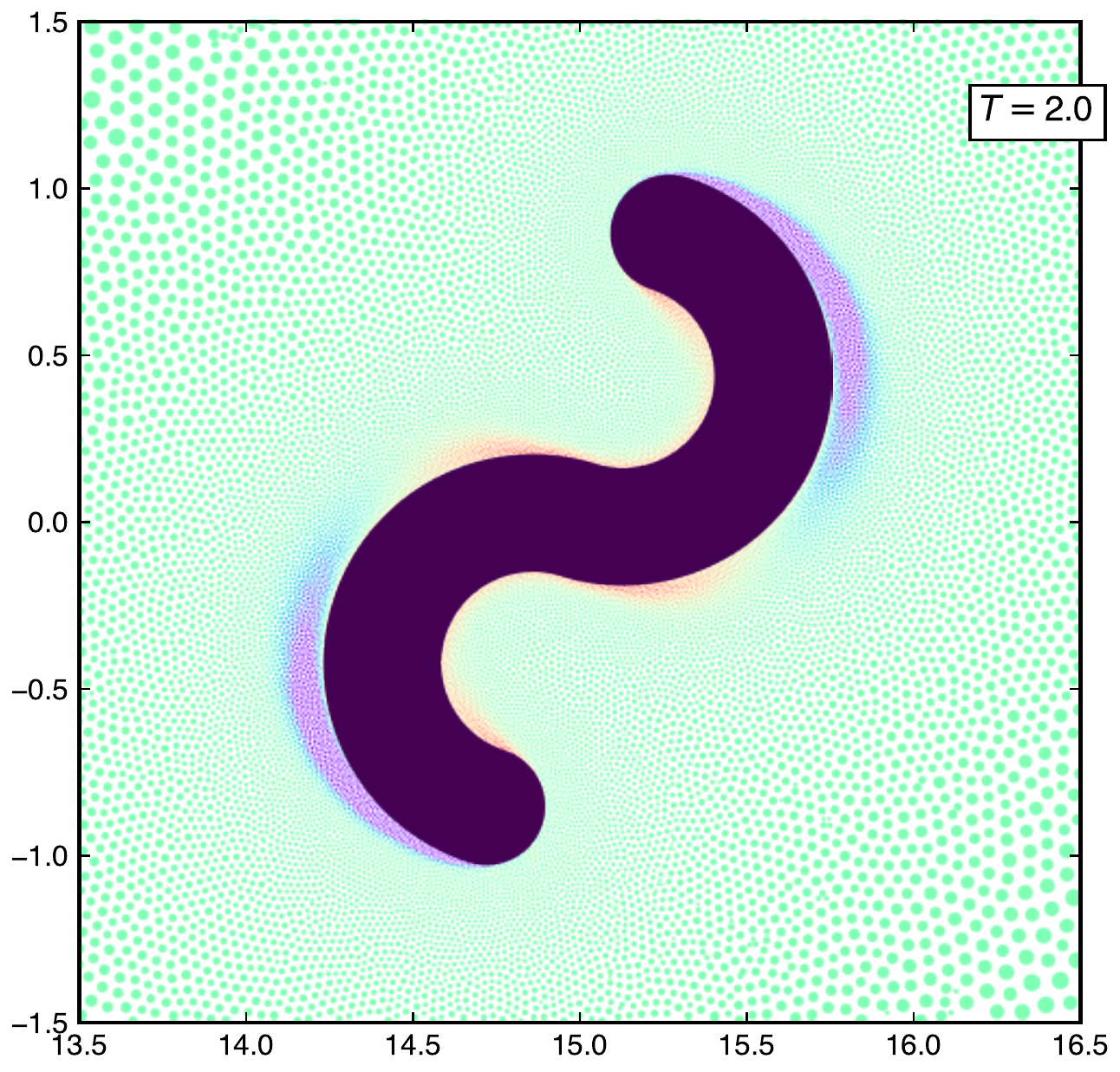}
    \subcaption{}\label{fig:sshape-a}
  \end{subfigure}
  \begin{subfigure}{0.4\textwidth}
    \includegraphics[width=\textwidth]{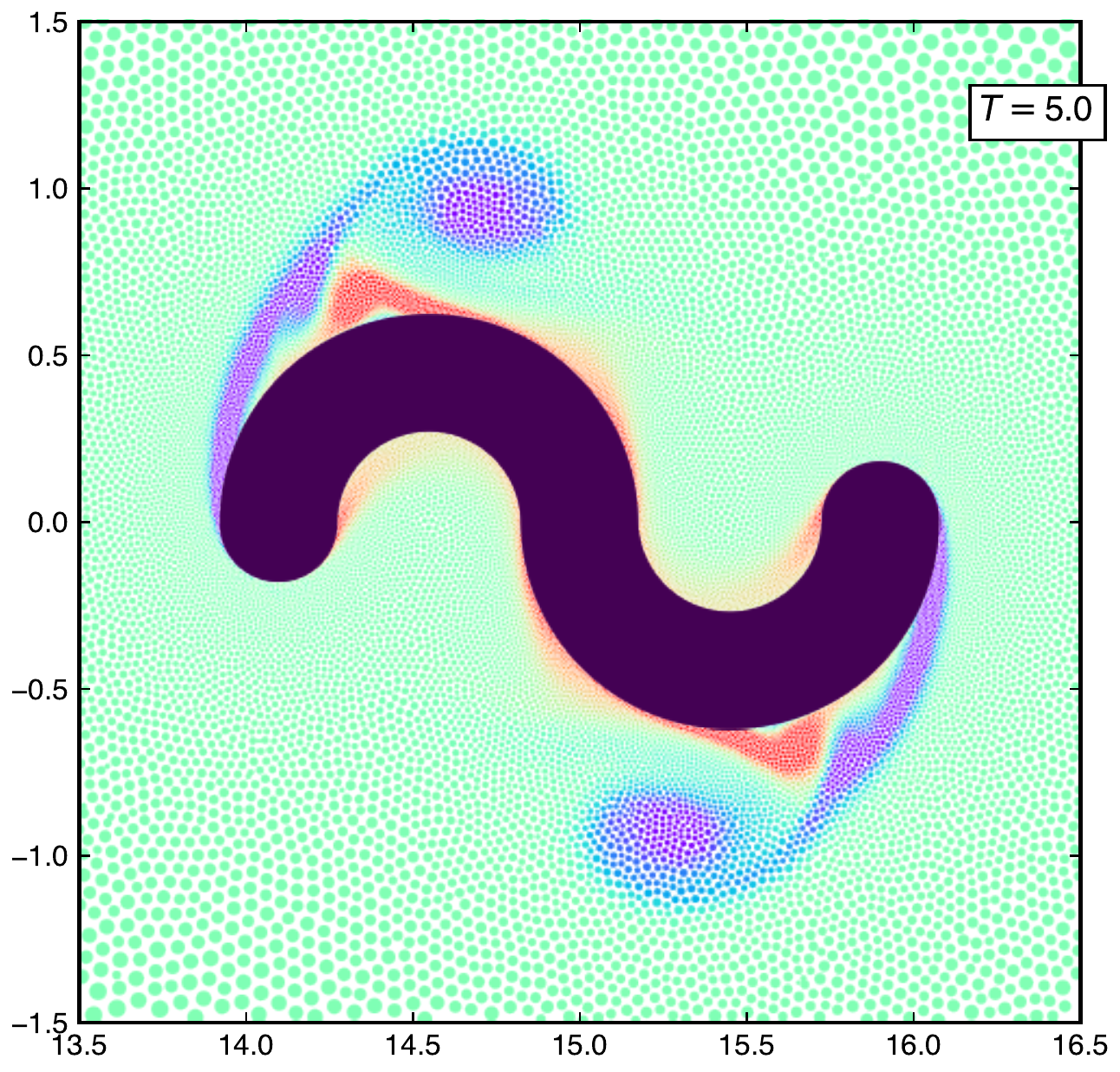}
    \subcaption{}\label{fig:sshape-b}
  \end{subfigure}
  \\
   \hspace*{\fill}
  \begin{subfigure}{0.4\textwidth}
    \includegraphics[width=\textwidth]{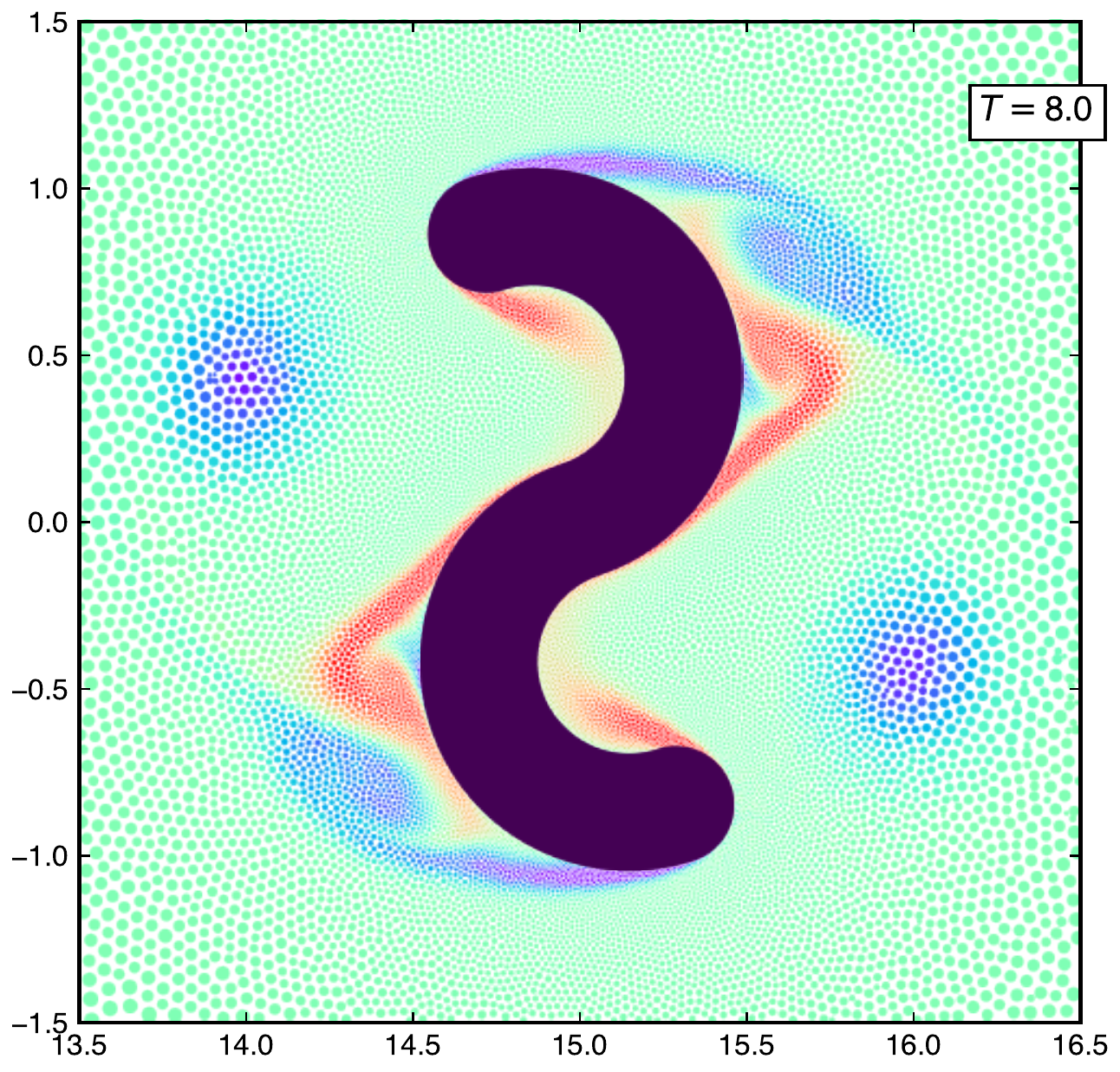}
    \subcaption{}\label{fig:sshape-c}
  \end{subfigure}
  \begin{subfigure}{0.4\textwidth}
    \includegraphics[width=\textwidth]{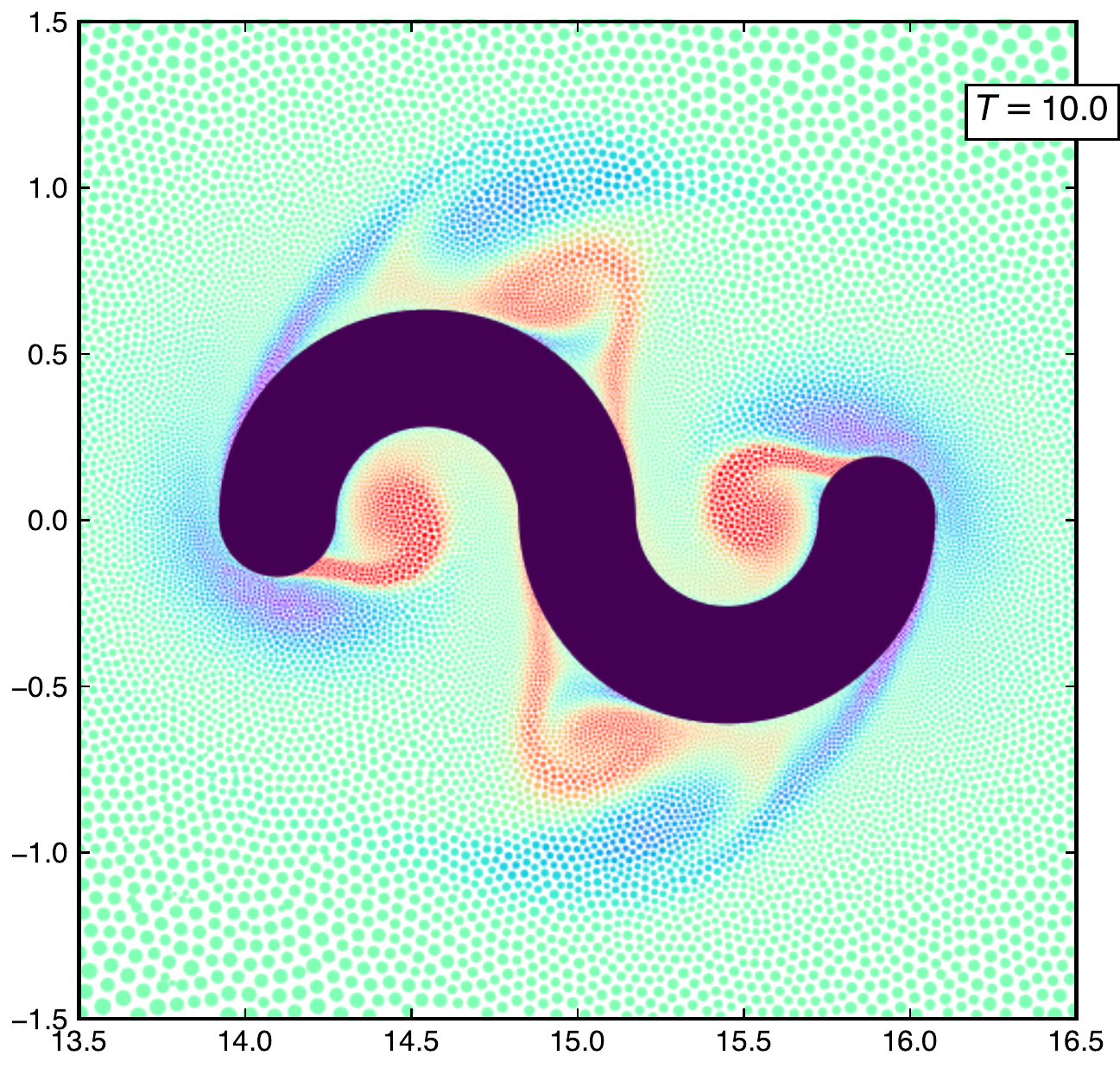}
    \subcaption{}\label{fig:sshape-d}
  \end{subfigure}
  \end{subfigure}
  \begin{subfigure}{0.09\textwidth}
    \includegraphics[scale=0.6,left]{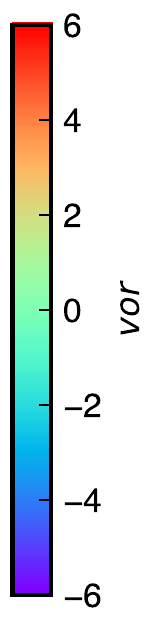}
  \end{subfigure}
  \caption{Vorticity distribution at different times around the
    counter-clockwise rotating S-shape at $Re = 2000$. The point size is
    proportional to the mass of the particle.}%
  \label{sshape:vor}
\end{figure}

We consider a counter-clockwise rotating S-shaped body at $Re = 2000$. The outer
diameter $D$ is 2 m and \cref{fig:sshape-domain} shows the domain schematic. The
minimum-resolution $D/\Delta x_{\min}$ is 10 and the maximum-resolution
$D/\Delta x_{\min}$ is 200. The body is rotated at a frequency of $0.1$
{sec}\textsuperscript{-1}. We simulate the problem for $t = 10$
sec. \Cref{sshape:vor} shows the vorticity distribution of the particles for the
rotating S-shape at $t = 2, 5, 8, 10$ secs.

This section demonstrates the capabilities of the adaptive particle refinement
in simulating complex motion and multiple bodies. This motivates further the use
of adaptive EDAC-SPH in a range of applications.

\FloatBarrier%
\section{Conclusions}%
\label{sec:conclusions}

In this work a modified version of the adaptive EDAC-SPH technique of
\citet{muta2021efficient} has been discussed and various applications have been
demonstrated. The adaptive particle refinement is automatic; the method is
efficient in terms of the number of particles that it creates and requires an
optimal number of neighbors. Besides the geometry-based automatic refinement, it
is possible to create static regions of refinement or use solution-based
adaptivity. Unlike the original method of \cite{muta2021efficient}, the proposed
method does not require any background particles making it more memory efficient
and easier to implement. The test cases presented in this paper clearly show the
applicability and computational efficiency of the method.  In particular the
original method did not benchmark the method for moving bodies or for multiple
bodies which we do in the current work. The results show good reliability of
using adaptive particle refinement in practical applications using SPH. The
proposed adaptive algorithm is executed before any SPH computations are
performed at every timestep. This makes it easy to incorporate into an existing
non-adaptive EDAC-SPH solver.

We show how the method can be used to simulate a variety of CFD problems that
involve stationary and moving geometries of varying complexity. The source code
is open-source and can be obtained from
\url{https://gitlab.com/pypr/asph_motion}. It is entirely written in Python and
can be made to run on different platforms, like, CPU (single and multi-core),
multi-CPU, as well as a GPGPU. This makes it easy for researchers to use and
extend. To our knowledge this is the only open-source code available for
adaptive SPH. In the future, we propose to apply this method to other practical
problems. The algorithms discussed work have not been tested for three
dimensional problems and this will be investigated in the future.

\section*{Acknowledgements}

We would like to thank the Aerospace Computational Engine (ACE) at the
Department of Aerospace Engineering, Indian Institute of Technology Bombay for
providing computational resources. We would like to thank Aditya Bhosale and
Miloni Atal for the parallel Octree nearest neighbor particle search algorithm
in PySPH, which was essential in the scale-up study done in this paper.

\bibliographystyle{model1a-num-names}
\bibliography{references}

\end{document}